\documentclass[a4,12pt]{article}
\usepackage[a4paper,
left=1.4in,right=1.4in,top=1.4in,bottom=1.4in,%
footskip=.25in]{geometry}
\usepackage[utf8]{inputenc}
\usepackage{authblk}
\title{Latent function-on-scalar regression models for observed sequences of binary data: a restricted likelihood approach}
\author[1]{Fatemeh  Asgari\thanks{ft.asgari@sci.ui.ac.ir}}
\author[1]{Mohammad Hossein Alamatsaz\thanks{alamatho@sci.ui.ac.ir}}
\author[2]{\newline Valeria Vitelli\thanks{valeria.vitelli@medisin.uio.no}}
\author[1]{Saeed Hayati\thanks{s.hayati@sci.ui.ac.ir}}
\affil[1]{\small{\textit{Department of Statistics, Faculty of Mathematics and Statistics,}}\\
	\small{\textit{University of Isfahan, Iran}}}
\affil[2]{\small{\textit{Oslo Center for Biostatistics and Epidemiology, Department of Biostatistics,}}\\
	\small{\textit{Institute of Basic Medical Sciences, University of Oslo, Norway}}}

\usepackage{eurosym}
\usepackage{amssymb}
\usepackage{amsfonts}
\usepackage{amssymb,amsmath,amsthm}
\RequirePackage[numbers]{natbib}
\RequirePackage[colorlinks,citecolor=blue,urlcolor=blue]{hyperref}

\usepackage{soul}
\usepackage{graphicx}
\usepackage{subcaption}
\usepackage{float}
\usepackage{nicefrac}
\usepackage{multirow}

\usepackage{tipa}
\def\code#1{\texttt{#1}}
\def\Ztipa{\text{\Large{\textcommatailz}}}
\def\ztipa{\text{\textcommatailz}}
\def\bZtipa{\text{\textbf{\Large{\textcommatailz}}}}

\DeclareMathOperator*{\argmin}{arg\,min}

\def\code#1{\texttt{#1}}

\usepackage[ruled,vlined,linesnumbered]{algorithm2e}

\makeatletter
\renewcommand{\algocf@captiontext}[2]{#1\algocf@typo. \AlCapFnt{}#2} 
\def\@algocf@capt@plain{top}
\renewcommand{\algocf@makecaption}[2]{%
	\addtolength{\hsize}{\algomargin}%
	\sbox\@tempboxa{\algocf@captiontext{#1}{#2}}%
	\ifdim\wd\@tempboxa >\hsize
	\hskip .5\algomargin%
	\parbox[t]{\hsize}{\algocf@captiontext{#1}{#2}}
	\else%
	\global\@minipagefalse%
	\hbox to\hsize{\box\@tempboxa}
	\fi%
	\addtolength{\hsize}{-\algomargin}%
}
\makeatother

\newtheorem{thm}{Theorem}
\newtheorem{rem}{Remark}
\newtheorem{prop}{Proposition}

\newcommand*{\defeq}{\mathrel{\vcenter{\baselineskip0.5ex \lineskiplimit0pt
			\hbox{\scriptsize.}\hbox{\scriptsize.}}}%
	=}
\makeatother

\begin{document}
\date{}
\maketitle
\vspace*{-1cm}
\begin{abstract}
In this paper, we study a functional regression setting where the random response curve is unobserved, and only its dichotomized version observed at a sequence of correlated binary data is available. We propose a practical computational framework for maximum likelihood analysis via the parameter expansion technique. Compared to existing methods, our proposal relies on the use of a complete data likelihood, with the advantage of being able to handle non-equally spaced and missing observations effectively. The proposed method is used in the Function-on-Scalar regression setting, with the latent response variable being a Gaussian random element taking values in a separable Hilbert space. Smooth estimations of functional regression coefficients and principal components are provided by introducing an  adaptive  MCEM algorithm that circumvents selecting the smoothing parameters. 
Finally, the performance of our novel method is demonstrated by various simulation studies and on a real case study. The proposed method is implemented in the \code{R} package \code{dfrr}.
\end{abstract}



\section{Introduction}\label{sec.1}
It is common that in some functional regression problems the response curve is observed as a sequence of correlated binary or multilevel data. This kind of situations can be handled via the family of generalized functional regression models. Bayesian methods are popular approaches for analyzing such data, see \citet{goldsmith2015}, \citet{meyer2015} and \citet{van2009} among others. However, Bayesian methods show some limitations, including heavy computations and lengthy fitting procedures. They also often lack flexibility in implementing necessary constraints, such as the orthogonality of the eigen-functions, which is vital for the identifiability of the principal components. Typically, the likelihood analysis of such models is supposed not to be practicable, because of computational burden and difficulty of handling irregular and missing data \citep{hall2008}. Nonetheless, this paper aims at providing a likelihood analysis framework for probit functional regression models that is flexible enough to handle non-equally, irregular, and hence missing data. 

This work is motivated by the Madras-dataset \citep[pp. 234-243]{diggle2002analysis}, including a sequence of binary variables whose values indicate the presence or absence of Schizophrenia related symptoms in a set of patients observed during a time period. This dataset has been previously studied and analyzed by longitudinal methods   \citep[see][]{schildcrout2007marginalized}. Lately,  the analysis of longitudinal data using methods from functional data analysis has been explored \citep{james2003,yao2005,zhou2018efficient}.
To get an insight into the relationship between functional and longitudinal data analysis we refer to \citet{zhao2004functional}.  

Alternative approaches to a likelihood analysis considered in  literature include non-parametric methods \citep{hall2008}, the use of approximating likelihood functions \citep{wang2014,sung2019,li2014hierarchical}, or the use of simplified and relaxed assumptions \citep{scheipl2016}. Specifically, \citet{hall2008} proposed a non-parametric estimation of the mean and kernel function of the underlying Gaussian process for a correlated sequence of binary data. They argued that the maximum likelihood approach is computationally demanding and unstable, because it would require a large number of parameters to ensure a sufficiently flexible parameterization of the underlying Gaussian process. They also claim that the maximum likelihood approach is hard to use in the irregular data case, where its implementation requires imposing additional assumptions that limit its flexibility. However, the non-parametric method proposed in \citet{hall2008} depends on the very limiting assumption that the variation of the underlying process about its mean is relatively small, and assumption that is not needed in the maximum likelihood approach here proposed. Moreover, a major drawback of their work is that it does not handle covariates in modeling the location parameter.

\citet{wang2014} proposed a generalized Gaussian process regression model for the binary case, in which location and the covariance operator are allowed depending on covariates. However, the approach is based on approximating the likelihood function by using a pre-specified kernel function that limits the flexibility of the model. A criticism of other likelihood-based approaches, like \citet{scheipl2016}, is that their method ignores the possible correlation structure of the residual functions over the domain, which is typically present with functional data. Additionally, a reliable approach to the estimation of the principal components is not provided.

The estimation of the covariance operator, or equivalently of the functional principal components, is crucial in the analysis of longitudinal and functional data \citep{james2000principal,li2010uniform, muller2005functional}. The functional principal components explain the pattern of variation of the observed random functions, and have many applications in dimension reduction \citep{Ramsay2005}. Although often the covariance function is specified a priori \citep{wang2014,sung2019}, our likelihood analysis method does not rely on prior specification of the covariance function, which can be considered as a free parameter.
In conclusion, the method proposed in this paper has several advantages over the non-parametric, Bayesian, and other likelihood-based methods currently developed in the literature: it does not depend on additional unnecessary assumptions, it is computationally faster and more stable than Bayesian approaches, it does not ignore the correlation structure of the residual functions over the domain, and it also treats the covariance function as a free parameter.

In this paper, we consider a Function-on-Scalar Regression (FoSR) model setting, where the response function is a latent Gaussian process, and we only observe its realization as a sequence of correlated binary observations. We propose a maximum likelihood approach, inspired by the idea of parameter expansion \citep{xu2010likelihood}, to carry out the estimation of regression coefficients and of the kernel function. Since we aim at automatically obtaining smooth functional parameters, functional regression coefficients and eigen-functions are estimated by introducing a novel method, namely the Adaptive Monte Carlo Expectation-Maximization (AMCEM) algorithm, which does not require the selection of the smoothing parameter. The AMCEM algorithm is a Monte Carlo EM algorithm in which the support of the distribution of the latent functions is restricted to a region consisting of functions that are smooth enough. It is adaptive in that the algorithm narrows the acceptance region step-by-step, so that the latent functions that fit the binary sequences become gradually smoother until the acceptance region converges to a non-empty set.

The rest of this paper is organized as follows. In Section \ref{seq.2}, the dichotomized FoSR model is introduced and the identifiability issue of the parameters is thoroughly investigated. In Section \ref{sec.3}, the estimation method and the AMCEM algorithm are described.
The performance of the method is demonstrated by various simulation studies, and by showing the results of the analysis of the Madras dataset, in Section \ref{sec.5}. Section \ref{sec.6} is devoted to discussion and conclusion. Proofs are collected in the Appendix.

\section{The Functional Probit Regression Model}\label{seq.2}

This article is concerned with functional data of the form ${\ztipa_{i}\in L^2[0,1]}$; $i=1,2,\ldots,N$. Let $\Ztipa_i$ be a Gaussian variable with a mean function $\boldsymbol{\beta}^\top \mathbf{x}_i$ and covariance operator $T$, where $\mathbf{x}_i=(1,x_{i1},\ldots,x_{i(q-1)})^\top$ is a $q$-dimensional vector of covariates, and $\boldsymbol{\beta}=(\beta_0,\beta_1,\ldots,\beta_{(q-1)})^{\top}$ is a $q$-dimensional vector of functions in $L^2[0,1]$, representing the unknown regression coefficients. Suppose that the functions $\ztipa_i$ are latent and unobserved, that they are affected by measurement error, and only observed in a dichotomized and sparse version. That is, for each $\ztipa_i$ and some points $0\leq t_{i1},t_{i2},\ldots,t_{iM_i}\leq 1$, observations are of the form:
\begin{equation}\label{eq.2.1}
y_{ij}=\text{I}\left(\ztipa_i(t_{ij})+\epsilon_{it_{ij}}>0\right),\qquad i=1,2,\ldots,N,\qquad j=1,2,\ldots,M_i ,
\end{equation}
where $\text{I}(\cdot)$ is the indicator function. For describing the latent functions we thus consider a Function-on-Scalar Regression (FoSR) model of the form
\begin{align}\label{eq.2.2.0}
\mathcal{W}_i(t)=\Ztipa_i(t)+\epsilon_{it},
\end{align}
where
\begin{align}
\Ztipa_i(t)=\boldsymbol{\beta}^\top(t) \mathbf{x}_i+\varepsilon_i(t).\label{eq.2.2}
\end{align}
The residual function  $\varepsilon_i\in L^2[0,1]$ is a Gaussian variable with mean function zero and covariance operator $T=\sum_{j\geq1}\nu_{j}\psi_{j}\otimes\psi_{j}$, where $\otimes$ denotes the tensor product and $\{\psi_j\in L^2[0,1], j\in\Bbb{N}\}$ is a complete orthonormal basis of $\overline{\text{Image}}(T)$. This means that $T$ has kernel function $K(s,t)=\sum_{j\geq 1}\nu_{j}\psi_{j}(s)\psi_{j}(t)$.
The error term $\epsilon_{it}$ is independent of the residual function, and for each $i=1,2,\ldots,N$ and $t\in[0,1]$, it is independently distributed as a normal random variable with mean zero. The term $\epsilon_{it}$ denotes the measurement error, while the residual function $\varepsilon_{i}(t)$ explains the complex and smooth structure of dependence along the domain within each sample. The variance of $\epsilon_{it}$ together with the parameters $\beta$ and $T$  are not identifiable. The identifiability issues associated to the model parameters is discussed in Section 
\ref{sec.2}.

Let $$\boldsymbol{\mu}_{i}=\left(\Bbb{E}\mathcal{W}_{i}(t_{ij})\right)_{1\leq j\leq M_{i}}=\left(\boldsymbol{\beta}^{\top}(t_{ij})\mathbf{x}_i\right)_{1\leq j\leq M_{i}},$$ and $$\boldsymbol{\Sigma}_{i}=\left[\text{cov}\left(\mathcal{W}_{i}\left(t_{ij}\right),\mathcal{W}_{i}\left(t_{ik}\right)\right)\right]_{1\leq j,k\leq M_{i}},$$ where $\boldsymbol{\beta}(t)=\left(\beta_0(t),\beta_1(t),\ldots,\beta_{(q-1)}(t)\right)^{\top}$.
The probability density function (pdf) of the binary response vector $\mathbf{Y}_i=(Y_{i1},Y_{i2},\ldots,Y_{iM_{i}})^\top$ given the parameters $\boldsymbol{\beta}$ and $T$ is 
\begin{equation}\label{eq.2.77}
f(\mathbf{y}_i\mid \boldsymbol{\beta},T)=\int_{C_{i1}}\ldots\int_{C_{iM_{i}}}{\phi_{M_{i}}(\mathbf{u};\boldsymbol{\mu}_i,\boldsymbol{\Sigma}_{i})d\mathbf{u} },
\end{equation}
where $C_{ij}$; $j=1,2,\ldots,M_{i}$; are equal to the interval $(-\infty,0]$ if $y_{ij}=0$ and $(0,+\infty)$ if $y_{ij}=1$. Moreover, $\phi_{M_{i}}(\mathbf{u};\boldsymbol{\mu}_i,\boldsymbol{\Sigma}_{i})$ is the pdf of a $M_{i}$-variate Gaussian distribution with mean vector $\boldsymbol{\mu}_i$ and variance-covariance matrix $\boldsymbol{\Sigma}_{i}$. 

\subsection{Identifiability Issues}\label{sec.2}

The parameters of the FoSR model (\ref{eq.2.2}) are not identifiable unless we add a constraint like $K(t,t)=1$ for all $t\in [0,1]$. In fact, consider the transformation $\Ztipa_i^{'}=C\Ztipa_i$, where  ${C:L^2[0,1]\to L^2[0,1]}$ is a linear operator and,  for all $g\in L^2[0,1]$, $C$ is defined as
\begin{equation*}
(Cg)(t)=f(t)g(t),
\end{equation*}
in which $f$ is a positive and bounded function in $ L^2[0,1]$.  $\Ztipa_i^{'}$ is then a $L^2[0,1]$-valued Gaussian variable  with mean function $m^{'}=\boldsymbol{\beta}^{'^{\top}}\mathbf{x}_i$ with $\boldsymbol{\beta}^{'}(t)=f(t)\boldsymbol{\beta}(t)$, and covariance operator $T^{'}=CTC$ having kernel function
\begin{align}\label{p1.2}
K^{'}(s,t)=E[(\Ztipa^{'}(s)-m^{'}(s))(\Ztipa^{'}(t)-m^{'}(t))]&=f(s)f(t)K(s,t).
\end{align}
Let  $\mathbf{F}_{i}=\text{diag}\{f(t_{ij})\}_{1\leq j \leq M_{i}}$. Since $f$ is positive, by using the change of variable $\mathbf{u}^{'}=\mathbf{F}_{i}\mathbf{u}$ in equation (\ref{eq.2.77}), we obtain
\begin{equation*}
f(\mathbf{y}_i\mid\boldsymbol{\mu}_{i},\boldsymbol{\Sigma}_{i})=f(\mathbf{y}_i\mid\mathbf{F}_{i}\boldsymbol{\mu}_{i},\mathbf{F}_{i}\boldsymbol{\Sigma}_{i}\mathbf{F}_{i}),
\end{equation*}
where $\mathbf{F}_{i}\boldsymbol{\mu}_{i}=\left(\Bbb{E}\Ztipa^{'}_{i}(t_{ij})\right)_{1\leq j\leq M_{i}}$ and 
$$\mathbf{F}_{i}\boldsymbol{\Sigma}_{i}\mathbf{F}_{i}=\left[\text{cov}\left(\Ztipa^{'}_{i}\left(t_{ij}\right),\Ztipa^{'}_{i}\left(t_{ik}\right)\right)+\delta_{jk}f^{2}(t_{ij})\text{var}(\epsilon_{it_{ij}})\right]_{1\leq j,k\leq M_{i}}.$$
Thus, $\left(\boldsymbol{\mu}_{i},\boldsymbol{\Sigma}_{i}\right)$, and hence $\left(\boldsymbol{\beta},T,\text{var}(\epsilon_{it})\right)$  are not  identifiable.

Although $(\boldsymbol{\beta},T,\text{var}(\epsilon_{it}))$ are not identifiable, a function of these parameters of the form $(\boldsymbol{\alpha},R)$ is identifiable, where $\boldsymbol{\alpha}(t)=K(t,t)^{-1/2}\boldsymbol{\beta}(t)$ and $R=LTL$. $L$ is a linear mapping from $L^2[0,1]$ to $L^2[0,1]$ and, for any $g\in L^2[0,1]$, we have $(Lg)(t)=K(t,t)^{-1/2}g(t)$.
\begin{rem}
	\label{re.1}
	Note that $\boldsymbol{\alpha}$ and $R$ are respectively the mean function and covariance operator of the standardized version of the Gaussian process $\Ztipa$ in  (\ref{eq.2.2}), given by 
	\begin{equation*}
	L\Ztipa_{i}=(L\boldsymbol{\beta})^{\top}\mathbf{x}_{i}+L\varepsilon_{i},
	\end{equation*}
	and the kernel of the covariance operator  $R$ is given by
	\begin{align*}
	K^{*}(s,t)=K(t,t)^{-\frac{1}{2}}K(s,t)K(s,s)^{-\frac{1}{2}}.
	\end{align*}
\end{rem}

\begin{prop}
	\label{t1}
	Consider the FoSR model given by (\ref{eq.2.2.0})-(\ref{eq.2.2}), and let its functional response  be dichotomized as in (\ref{eq.2.1}). Then,  the parameters $\boldsymbol{\alpha}$ and $R$ are identifiable.
\end{prop}
The relationship between the covariance operators of the standardized and non-standardized versions of the underlying Gaussian process is provided in the next proposition.
\begin{thm}\label{p2}
	Let $L^{'}:L^2[0,1]\to L^2[0,1]~~g(\cdot)\mapsto K(\cdot,\cdot)^{1/2}g(\cdot)$. Then, we have $T=L^{'}RL^{'}$ and $R=LTL$ with $||R||\leq 1$.
\end{thm}

For the sake of identifiability of the variance of the measurement error term, we consider the following standardized version of (\ref{eq.2.2.0}), 
\begin{align}\label{eq.es}
L\mathcal{W}_i(t)&=L\Ztipa_i(t)+L\epsilon_{it}.
\end{align}
We assume for simplicity that $L\epsilon_{it}\sim\mathcal{N}(0,\sigma^2)$, independently  for each $i=1,2\ldots,N$ and $t\in[0,1]$. This assumption is equivalent to that of the variance of $\epsilon_{it}$ being proportional to the variance of $\Ztipa_{i}$ at $t$ in the unstandardized original model (\ref{eq.2.2.0}).

\section{Estimation Method}\label{sec.3}

First, we present some necessary notations in this section.
Let $\{e_k\}_{k\geq 1}$ be an arbitrary complete orthonormal basis for the separable Hilbert space $L^2[0,1]$ with the inner product $\langle\cdot,\cdot\rangle$ and the corresponding norm $\parallel\cdot\parallel$, and define $b_{lk}\defeq\left\langle\beta_{l},e_{k}\right\rangle$, $\theta_{jk}\defeq\left\langle\psi_j,e_k\right\rangle$ and $\omega_{ij}\defeq\left\langle\varepsilon_i,\psi_j\right\rangle$, where $\psi_j$ is the $j$\textsuperscript{th} eigen-function of $T$. Consider the Karhunen-Lo\`eve expansion  $\varepsilon_i=\sum_{j\geq 1}\omega_{ij}\psi_j$, where  $\omega_{ij}$ are independent Gaussian univariate random variables with mean zero and variance $\nu_j$. For the sake of identifiablitiy of the eigen-functions, we assume that $\nu_1> \nu_2> \ldots > 0$.
All separable Hilbert spaces are isomorphic to the space of square summable sequences, thus an equivalent form of model (\ref{eq.2.2}) is given by
\begin{align}\label{15}
\left\langle \Ztipa_i,e_k\right\rangle&=\left\langle\boldsymbol{\beta},e_k\right\rangle^{\top}\mathbf{x}_i+\left\langle\varepsilon_i,e_k\right\rangle\nonumber\\
&=\left\langle\boldsymbol{\beta},e_k\right\rangle^{\top}\mathbf{x}_i+\sum_{j\geq 1}\omega_{ij}\left\langle\psi_j,e_k\right\rangle,~~~k\geq 1.
\end{align}
For any fixed $J\in \Bbb{N}$, let $\boldsymbol{\theta}_j=(\theta_{j1},\theta_{j2},\ldots,\theta_{jJ})^\top$, $j=1,\ldots,p$ and $\mathbf{b}_l=(b_{l1},b_{l2},\ldots,b_{lJ})^\top$, $l=0,2,\ldots,q-1,$ be the $J$-dimensional vectors of unknown parameters. Let also $\boldsymbol{\omega}_i=(\omega_{i1},\omega_{i2}\ldots,\omega_{iJ})^\top$ be a $J$-dimensional multivariate normally distributed random variable with mean vector zero and variance-covariance matrix $\mathbf{A}=\text{diag}(\nu_j)_{1\leq j \leq J}$.
Define $\mathbf{B}=(\mathbf{b}_0,\mathbf{b}_1,\ldots,\mathbf{b}_{q-1})^\top$ a $q\times J$ matrix and $\boldsymbol{\Theta}=(\boldsymbol{\theta}_1,\boldsymbol{\theta}_2,\ldots,\boldsymbol{\theta}_J)^\top$ a $p\times J$ matrix with orthonormal rows, and $\mathbf{Z}_{i}=[\left\langle \Ztipa_{i},e_{j}\right\rangle]_{1\leq j\leq J}$ the $J$-dimensional vector of Fourier coefficients.  
Now, from (\ref{15}), we can write
\begin{equation}\label{eq.2.6}
\mathbf{Z}_i=\mathbf{B}^\top\mathbf{x}_i+\boldsymbol{\Theta}^\top \boldsymbol{\omega}_i .
\end{equation}
The random vector $\mathbf{Z}_i$ has a $J$-variate Gaussian distribution with mean vector $\mathbf{B}^\top\mathbf{x}_i$ and variance-covariance matrix 
$\boldsymbol{\Sigma}_{\theta}=\boldsymbol{\Theta}^{\top}\mathbf{A}\boldsymbol{\Theta}$.

In Section \ref{EM}, a truncated version of the model (\ref{15}) as presented in (\ref{eq.2.6}) is used to provide Maximum Likelihood Estimation (MLE) of the location parameter, and  Restricted Maximum Likelihood (REML) estimation of the covariance function. However, we first propose in Section \ref{REMLsec} a framework related to the REML approach that facilitates providing unbiased estimation of covariance operators in infinite-dimensional  Hilbert spaces.

\subsection{Unbiased Estimation of the Covariance Operator}\label{REMLsec}
The REML approach is a special case of maximum likelihood estimation that uses a likelihood function provided from a transformed  data, so that nuisance parameters have no effects on estimation. 
Although a likelihood function in the infinite-dimensional separable Hilbert spaces cannot be defined, an approach similar to REML can be used to provide an unbiased estimation of covariance function in such spaces.

First, note that some of the key results that hold for Gaussian random vectors in finite-dimensional spaces are also valid in the Gaussian case in infinite-dimensional separable Hilbert spaces.
\begin{prop}\label{p3}
	Let $X_1$ and $X_2$ be two jointly Gaussian functional variables in a separable Hilbert space $H$  with mean functions $m_1$ and $m_2$ and covariance operators $C_1$ and $C_2$, respectively. Then, $X_1$ and $X_2$ are independent if  $C_{12}=C_{21}=0$, where $C_{12}$ and $C_{21}$ are the cross-covariance operators.
\end{prop}
Assume $\Ztipa_{1},\Ztipa_{2},\ldots,\Ztipa_{N}$ are  $N$ independent realizations of the  FoSR model in (\ref{eq.2.2}). Let ${\bZtipa=(\Ztipa_1,\Ztipa_2,\ldots,\Ztipa_{N})^\top}$, and $\mathbf{X}=(\mathbf{x}_{1},\mathbf{x}_{2},\ldots,\mathbf{x}_{N})^{\top}$ be a $N\times q$ matrix of scalar  covariates. Then 
\begin{equation*}
\bZtipa=\mathbf{X}\boldsymbol{\beta}+\boldsymbol{\varepsilon},
\end{equation*}
where, with regard to Proposition \ref{p3}, $\boldsymbol{\varepsilon}=(\varepsilon_1,\varepsilon_2,\ldots,\varepsilon_{N})^\top$ is a Gaussian random vector with a mean vector of zero function elements and a variance-covariance matrix $\mathbf{I}_{N}\otimes T$, where $\otimes$ denotes the Kronecker product and $\mathbf{I}_{N}\otimes T$ is a  diagonal matrix of operators whose diagonal  elements are  $T$.

Consider the matrix $\mathbf{A}=\mathbf{I}_{N}-\mathbf{X}(\mathbf{X}^\top \mathbf{X})^{-1}\mathbf{X}^\top$ with $\text{rank}(\mathbf{A})=N-q$. Since $\mathbf{A}$ is not a full rank matrix, the eigen-value decomposition of matrix $\mathbf{A}$ is
\begin{equation*}
\mathbf{A}=\mathbf{U}\boldsymbol{\Lambda}\mathbf{U}^\top ,
\end{equation*}
where $\mathbf{U}$ is a $N\times (N-q)$ matrix whose $i$th column is the eigen-vector $\mathbf{u_i}$ of $\mathbf{A}$, and $\boldsymbol{\Lambda}$ is the diagonal matrix $\boldsymbol{\Lambda}=\text{diag}(\nu_{i})_{1\leq i\leq (N-q)}$ whose diagonal elements are the corresponding positive eigen-values. Thus, noting that
\begin{align*}
\mathbf{U}^\top \mathbf{X}&=[(\mathbf{U}^\top \mathbf{U}\boldsymbol{\Lambda})^{-1}\mathbf{U}^\top \mathbf{U}\boldsymbol{\Lambda}]\mathbf{U}^\top \mathbf{X}\\
&=\boldsymbol{\Lambda}^{-1}\mathbf{U}^\top \mathbf{A}\mathbf{X}=0,
\end{align*}
we have that $\bZtipa^{*} =\mathbf{U}^\top \bZtipa =\mathbf{U}^\top \boldsymbol{\varepsilon}$ is a Gaussian random vector with mean vector of zero functions. From  Proposition \ref{p4} (proof given in the appendix) below, it follows that the elements of $\bZtipa^{*}$ are independent Gaussian functional variables with mean function zero and covariance operator $T$. 
\begin{prop}\label{p4}
	Suppose $H$ is a separable Hilbert space and  $\mathbf{X}=(X_{1},\ldots,X_{n})^\top$ is a $n$-dimensional vector of $H$-valued  independent Gaussian  elements with mean functions zero and covariance operators $C$. let $\mathbf{Q}=(\mathbf{q}_1 ,\mathbf{q}_2 ,\ldots,\mathbf{q}_{m})$ be a $n\times m$ matrix whose columns are orthogonal unit vectors, that is $\mathbf{Q}^\top \mathbf{Q}=\mathbf{I}_{m}$. If $\mathbf{Y}=\mathbf{Q}^\top \mathbf{X}$, then  $\mathbf{Y}=(Y_{1},Y_{2},\ldots,Y_{m})^\top$ is a $m$-dimensional vector whose elements are independent Gaussian functional variables  with mean functions zero and covariance operators $C$.
\end{prop}
Let $\Ztipa^{*}_{k}=\mathbf{u}^{\top}_{k}\bZtipa$, then $\bZtipa^{*}=\left(\Ztipa^{*}_{1},\Ztipa^{*}_{2},\ldots,\Ztipa^{*}_{(N-q)}\right)^{\top}$. From Proposition \ref{p4}, $\Ztipa^{*}_{k}$ are $N-q$ independent and identically distributed Gaussian random elements with mean function zero and covariance operators $T$. Hence, an unbiased estimate of $T$ can be obtained by the $N-q$ independent random samples $\Ztipa^{*}_{k}$ as
\begin{equation}\label{16}
\hat{T}=\frac{1}{N-q}\sum_{k=1}^{N-q}\Ztipa^{*}_{k}\otimes \Ztipa^{*}_{k},
\end{equation}
where $\otimes$ stands for the tensor product\footnote{Throughout the paper, we indicate with $\otimes$ both the tensor product and the Kronecker product, and we have so far explicitly mentioned which of the two was used. Hereafter, we will not explicitly specify the meaning of the notation $\otimes$, if it will be possible to infer it with no ambiguity from the mathematics.}. 

In the next section, we describe the EM algorithm, we provide the ML estimator of $\boldsymbol{\beta},$ and we describe the REML estimator of $T$, similarly to what already proposed in equation (\ref{16}).

\subsection{Model inference via Expectation-Maximization}\label{EM}

The likelihood function depends on the covariance function through the matrix $\boldsymbol{\Sigma}_{\theta}$, thus the estimation of $\boldsymbol{\Theta}$ and $A$ and hence $T$ are obtained by eigen-decomposition of  $\boldsymbol{\Sigma}_{\theta}$. Consequently, to estimate the parameters of the FoSR model in (\ref{eq.2.2.0}),  it is sufficient to estimate the Fourier coefficients of the functional parameters $\mathbf{B}$, the covariance matrix $\boldsymbol{\Sigma}_{\theta}$ and $\sigma^2$.  
Smooth estimations of $\boldsymbol{\beta}$ and $\psi_{j}$, $j\geq 1$ are given by introducing the AMCEM algorithm. In our estimation method, $J$ in model (\ref{eq.2.6}) can be chosen large enough to reflect (\ref{15}) with a high precision.

The EM algorithm is an iterative method to derive maximum likelihood estimates of parameters when the model depends on unobserved latent variables  \citep{dempster1977maximum}. 
The maximization step of the EM algorithm concerning the standardized parameters does not yield a closed-form solution in the case of our model. Inspired by the parameter expansion technique proposed by \citet{liu1998parameter}, and \citet{liu1999parameter}, we consider expanding the parameter space and estimating unstandardized parameters. The standardized parameters, which are uniquely identifiable, are then computed from the estimated unstandardized parameters.

Let $\mathbf{e}=\left(e_1,e_2,\ldots,e_J\right)^\top$ be the  vector of the first $J$ basis functions.  Let $\mathbf{E}_{i}=\left[e_{j}\left(t_{ik}\right)\right]_{1\leq j\leq J, 1\leq k\leq M_i}$ be the $J\times M_{i}$ matrix of basis functions evaluated at $t_{i1},t_{i2},\ldots,t_{iM_{i}}$.
According to (\ref{eq.2.6}), we can rewrite the models (\ref{eq.2.2}) and (\ref{eq.2.2.0}) in a truncated form as 
\begin{equation*}
\mathbf{Z}_i=\mathbf{B}^{\top}\mathbf{x}_{i}+\boldsymbol{\varepsilon}_i,
\end{equation*}
and 
\begin{equation*}
\mathbf{W}_i=\mathbf{E}_i^\top \mathbf{Z}_i+\boldsymbol{\epsilon}_i ,
\end{equation*}
where $\mathbf{Z}_i=(Z_{i1} ,Z_{i2} ,\ldots,Z_{iJ})^\top$ is the vector of latent Fourier coefficients.     The residual vector $\boldsymbol{\varepsilon}_i=(\varepsilon_{i1},\varepsilon_{i2},\ldots,\varepsilon_{iJ})^\top$ is Gaussian distributed  with mean vector zero and variance-covariance matrix $\boldsymbol{\Sigma}_{\theta}$. Further, 
$\boldsymbol{\epsilon}_i=(\epsilon_{it_{i1}},\epsilon_{it_{i2}},\ldots,\epsilon_{it_{iM_{i}}})^\top$ is the vector of independent measurement error terms, whose standardized version given by (\ref{eq.es}) is Gaussian distributed  with mean vector zero and variance-covariance matrix $\sigma^2\times\mathbf{I}_{M_{i}}$. Finally, define 
$
\mathbf{W}_i=(W_{i1} ,W_{i2} ,\ldots,W_{iM_{i}})^\top.
$

Let us assume that the sequence of binary data for subject $i$, $i=1,\ldots,N$, $\mathbf{Y}_{i}=(Y_{i1},Y_{i2},\ldots,Y_{iM_{i}})^\top$, is generated according to  the truncated model $$Y_{ij}=\text{I}\left(\sum_{k=1}^{J}Z_{ik}e_k\left(t_{ij}\right)+\epsilon_{it_{ij}}>0\right)=\text{I}\left(W_{ij}>0\right).$$
Let $\mathbf{Y}=\left\{\mathbf{Y}_{1},\mathbf{Y}_{2},\ldots,\mathbf{Y}_{N}\right\}$ and $\mathbf{W}=\left\{\mathbf{W}_{1},\mathbf{W}_{2},\ldots,\mathbf{W}_{N}\right\}$ be two sets of random vectors and  ${\mathbf{Z}=\left[\mathbf{Z}_{1},\mathbf{Z}_{2},\ldots,\mathbf{Z}_{N}\right]^{\top}}$  be a $N\times J$ random matrix, collecting the subject-specific i.i.d. model variables, and let $\mathbf{y}$, $\mathbf{w}$ and $\mathbf{z}$ be the respective corresponding observations. Assuming that  $\boldsymbol{\gamma}=(\mathbf{B},\boldsymbol{\Sigma}_{\theta},\sigma^2)$ is the model set of parameters, the complete-data log-likelihood function is given by 
\begin{align*}
\ell_{\text{com}}(\boldsymbol{\gamma}\mid \mathbf{y},\mathbf{z},\mathbf{w})=\log f(\mathbf{z},\mathbf{w}\mid \boldsymbol{\gamma})+\log f(\mathbf{y}\mid \mathbf{w},\boldsymbol{\gamma}),
\end{align*}
where $f(\mathbf{z},\mathbf{w}\mid \boldsymbol{\gamma})$ denotes the joint pdf of $\mathbf{Z}$ and $\mathbf{W}$ given the parameters, and $f(\mathbf{y}\mid \mathbf{w},\boldsymbol{\gamma})$ is the pdf of $\mathbf{Y}$ given $\mathbf{W}$ and the parameter set.

The E-step at iteration $m+1$ of the EM algorithm given the parameter estimate $\boldsymbol{\gamma}^{(m)}$ at iteration $m$, involves evaluating the expectation $$Q(\boldsymbol{\gamma}\mid \mathbf{y},\boldsymbol{\gamma}^{(m)})=\Bbb{E}_{\mathbf{Z},\mathbf{W}\mid \mathbf{Y}=\mathbf{y},\boldsymbol{\gamma}^{(m)}}[\ell_{\text{com}}(\boldsymbol{\gamma}\mid \mathbf{y},\mathbf{Z},\mathbf{W})],$$ where
the latter expression is the expectation of the log-likelihood function given the observed variable $\mathbf{Y}$. 
Considering that 
$$Q(\boldsymbol{\gamma}\mid \mathbf{y},\boldsymbol{\gamma}^{(m)})=\Bbb{E}_{\mathbf{W}\mid \mathbf{Y}=\mathbf{y},\boldsymbol{\gamma}^{(m)}}\left\{\Bbb{E}_{\mathbf{Z}\mid \mathbf{W},\boldsymbol{\gamma}^{(m)}}\left[\ell_{\text{com}}\left(\boldsymbol{\gamma}\mid \mathbf{y},\mathbf{Z},\mathbf{W}\right)\right]\right\},$$ the  idea behind our smoothing procedure for the functional parameters is to substitute $Q(\boldsymbol{\gamma}\mid \mathbf{y},\boldsymbol{\gamma}^{(m)})$ by 
\begin{equation*}
Q_{\boldsymbol{\lambda}}(\boldsymbol{\gamma}\mid \mathbf{y},\boldsymbol{\gamma}^{(m)})=\Bbb{E}_{\mathbf{W}\mid \mathbf{Y}=\mathbf{y},\boldsymbol{\gamma}^{(m)}}\left\{\Bbb{E}_{\mathbf{Z}\mid \mathbf{W},V_{\boldsymbol{\lambda}},\boldsymbol{\gamma}^{(m)}}\left[\ell_{\text{com}}\left(\boldsymbol{\gamma}\mid \mathbf{y},\mathbf{Z},\mathbf{W}\right)\right]\right\},
\end{equation*} 
in which $\boldsymbol{\lambda}=(\lambda_{1},\lambda_{2},\ldots,\lambda_{N})^{\top}$ and $V_{\boldsymbol{\lambda}}$ is the event $\mathbf{Z}_{i}^{\top}\mathbf{P}^{(n)}\mathbf{Z}_{i}\leq \lambda_{i},~\forall i=1,2,\ldots,N$, where $\mathbf{P}^{(n)}=[\langle e_{j}^{(n)},e_{k}^{(n)}\rangle]_{1\leq j,k\leq J}$ is the penalty matrix associated with the basis functions $e_{j}(t)$, and $e_{j}^{(n)}(t)$ is the n\textsuperscript{th} derivative of $e_{j}(t)$. Throughout this paper, we set $n=2$. The parameter $\boldsymbol{\lambda}$ controls the smoothness of the latent random functions, yielding to a smooth estimation of the eigen-functions.

Note that
\begin{align*}
&\Bbb{E}_{\mathbf{W}\mid \mathbf{Y}=\mathbf{y},\boldsymbol{\gamma}^{(m)}}\left\{\Bbb{E}_{\mathbf{Z}\mid \mathbf{W},V_{\lambda},\boldsymbol{\gamma}^{(m)}}\left[\log 
f(\mathbf{y}\mid \mathbf{W},\boldsymbol{\gamma})\right]\right\}\\
&~=\Bbb{E}_{\mathbf{W}\mid \mathbf{Y}=\mathbf{y},\boldsymbol{\gamma}^{(m)}}\left\{\Bbb{E}_{\mathbf{Z}\mid \mathbf{W},V_{\lambda},\boldsymbol{\gamma}^{(m)}}\left[\log\prod_{i=1}^{N}\prod_{j=1}^{M_{i}}I(W_{ij}\in C_{ij})\right]\right\}\\
&~=\Bbb{E}_{\mathbf{W}\mid \mathbf{Y}=\mathbf{y},\boldsymbol{\gamma}^{(m)}}\left\{\log\prod_{i=1}^{N}\prod_{j=1}^{M_{i}}I(W_{ij}\in C_{ij})\right\}=0,
\end{align*}
and thus $Q_{\boldsymbol{\lambda}}(\boldsymbol{\gamma}\mid \mathbf{y},\boldsymbol{\gamma}^{(m)})$ can be rewritten as
\begin{align}\label{3}\nonumber
Q_{\boldsymbol{\lambda}}(\boldsymbol{\gamma}\mid \mathbf{y},\boldsymbol{\gamma}^{(m)})=&\Bbb{E}_{\mathbf{W}\mid \mathbf{Y}=\mathbf{y},\boldsymbol{\gamma}^{(m)}}\left\{\Bbb{E}_{\mathbf{Z}\mid \mathbf{W},V_{\lambda},\boldsymbol{\gamma}^{(m)}}\left[\log f\left( \mathbf{Z},\mathbf{W}\mid\boldsymbol{\gamma}\right)\right]\right\}\\\nonumber
=&\Bbb{E}_{\mathbf{W}\mid \mathbf{Y}=\mathbf{y},\boldsymbol{\gamma}^{(m)}}\left\{\Bbb{E}_{\mathbf{Z}\mid \mathbf{W},V_{\lambda},\boldsymbol{\gamma}^{(m)}}\left[\log f\left(\mathbf{Z}\mid \mathbf{B},\boldsymbol{\Sigma}_{\theta}\right)\right]\right\}\\
&+\Bbb{E}_{\mathbf{W}\mid \mathbf{Y}=\mathbf{y},\boldsymbol{\gamma}^{(m)}}\left\{\Bbb{E}_{\mathbf{Z}\mid \mathbf{W},V_{\lambda},\boldsymbol{\gamma}^{(m)}}\left[\log f\left(\mathbf{W}\mid \mathbf{Z},\sigma^{2}\right)\right]\right\},
\end{align}
where $f\left(\mathbf{z}\mid \mathbf{B},\boldsymbol{\Sigma}_{\theta}\right)$ stands for the pdf of $\mathbf{Z}$ and $f\left(\mathbf{w}\mid \mathbf{z},\sigma^{2}\right)$ is the pdf of $\mathbf{W}$ given $\mathbf{Z}$.

To implement the M-step, we derive closed-form REML estimation of $\boldsymbol{\Sigma}_{\theta}$ and ML estimations of $\mathbf{B}$ and $\sigma^2$.
For the REML estimation of $\boldsymbol{\Sigma}_{\theta}$, we substitute the restricted log-likelihood function to the  log-likelihood function in  (\ref{3}).  Differentiating  with respect to $\boldsymbol{\Sigma}_{\theta}$ and setting the derivative equal to zero yields
\begin{equation}\label{eq:MstepSigma}
\mathbf{\Sigma}_{\theta}^{(m+1)}=\left(\frac{1}{N-q}\right)\sum_{k=1}^{N-q}\Bbb{E}_{\mathbf{W}\mid \mathbf{Y}=\mathbf{y},\boldsymbol{\gamma}^{(m)}}\left\{\Bbb{E}_{\mathbf{Z}\mid \mathbf{W},V_{\lambda},\boldsymbol{\gamma}^{(m)}}\left(\mathbf{Z}^{*}_{k}\mathbf{Z}^{*^{\top}}_{k}\right)\right\},
\end{equation}
where $\mathbf{Z}^{*}_{k}=\mathbf{Z}^{\top}\mathbf{u}_{k}$.
Estimation of eigen-values and smoothed eigen-functions is then obtained by a simple eigen-decomposition of $\boldsymbol{\Sigma}_{\boldsymbol{\theta}}^{(m+1)}$. Consequently, $T^{(m+1)}=\sum_{j=1}^{J}\nu_{j}^{(m+1)}\psi_{j}^{(m+1)}\otimes\psi_{j}^{(m+1)}$, where $\psi_{j}^{(m+1)}=\boldsymbol{\theta}_{j}^{(m+1)^{\top}}\mathbf{e}$ and the sequence $(\nu_{j}^{(m+1)},\boldsymbol{\theta}_{j}^{(m+1)})$ are the eigen-values and eigen-vectors of $\boldsymbol{\Sigma}_{\boldsymbol{\theta}}^{(m+1)}$, respectively.

By setting the first-order derivative of (\ref{3}) with respect to $\mathbf{B}$ equal to zero, the  update of the estimation of $\mathbf{B}$ is obtained as follows:
\begin{equation}\label{eq:MstepB}
\mathbf{B}^{(m+1)}=\left(\mathbf{X}^{\top}\mathbf{X}\right)^{-1} \mathbf{X}^{\top}\Bbb{E}_{\mathbf{W}\mid \mathbf{Y}=\mathbf{y},\boldsymbol{\gamma}^{(m)}}\left\{\Bbb{E}_{\mathbf{Z}\mid \mathbf{W},V_{\lambda},\boldsymbol{\gamma}^{(m)}}(\mathbf{Z} )\right\}.
\end{equation}
Finally, the estimation of $\sigma^2$ at iteration $m+1$ is updated by
\begin{equation}\label{eq:Mstepsigma2}
\frac{1}{\sum_{i=1}^{N}M_{i}}\sum_{i=1}^{N}\Bbb{E}_{\mathbf{W}\mid \mathbf{Y}=\mathbf{y},\boldsymbol{\gamma}^{(m)}}\hspace*{-0.3em}\left\{\Bbb{E}_{\mathbf{Z}\mid \mathbf{W},V_{\lambda},\boldsymbol{\gamma}^{(m)}}\left(\mathbf{W}_{i}-\mathbf{E}_{i}^{\top}\mathbf{Z}_{i}\right)^{\hspace*{-0.3em}\top}\hspace*{-0.3em}
\mathbf{K}_{i}^{-1^{(m+1)}}\hspace*{-0.3em}\left(\mathbf{W}_{i}-\mathbf{E}_{i}^{\top}\mathbf{Z}_{i}\right)\right\},
\end{equation}
in which $\mathbf{K}_{i}^{(m+1)}=\text{diag}\left(K(t_{ij},t_{ij})^{(m+1)}\right)_{1\leq j\leq M_{i}}$, where
$K^{(m+1)}$ is the kernel of the covariance operator $T^{(m+1)}$.  

Let us define the matrix $\mathbf{D}$ as  
\begin{equation*}
\mathbf{D}=\left[\int{\left(K^{(m+1)}(s,s)\right)^{-\frac{1}{2}}e_{i}(s)e_{j}(s)ds}\right]_{1\leq i,j\leq J}.
\end{equation*}
Then, the standardized  regression coefficient $\boldsymbol{\alpha}$ and the  standardized covariance operator $R$ can be updated by 
\begin{align}\label{25}
\boldsymbol{\alpha}^{(m+1)}&=\mathbf{B}^{(m+1)}\mathbf{D}\mathbf{e},\\\label{26}
R^{(m+1)}&=\sum_{j=1}^{J}\varrho_{j}^{(m+1)}\varphi_{j}^{(m+1)}\otimes\varphi_{j}^{(m+1)},
\end{align}
in which  $\varphi_{j}^{(m+1)}=\boldsymbol{\vartheta}_{j}^{(m+1)^{\top}}\mathbf{e}$, where $(\varrho_{j}^{(m+1)},\boldsymbol{\vartheta}_{j}^{(m+1)})$ are respectively the eigen-values and eigen-vectors of $\mathbf{D}\boldsymbol{\Sigma}_{\boldsymbol{\theta}}^{(m+1)}\mathbf{D}^{\top}$. 
If in each iteration we replace $\mathbf{B}^{(m+1)}$ with $\mathbf{B}^{(m+1)}\mathbf{D}$ and $\boldsymbol{\Sigma}^{(m+1)}_{\boldsymbol{\theta}}$ with $\mathbf{D}\boldsymbol{\Sigma}_{\boldsymbol{\theta}}^{(m+1)}\mathbf{D}^{\top}$, then  the estimation of $\sigma^2$ is updated by
\begin{equation}
\sigma^{2^{(m+1)}}=\frac{1}{\sum_{i=1}^{N}M_{i}}\sum_{i=1}^{N}\Bbb{E}_{\mathbf{W}\mid \mathbf{Y}=\mathbf{y},\boldsymbol{\gamma}^{(m)}}\left\{\Bbb{E}_{\mathbf{Z}\mid \mathbf{W},V_{\lambda},\boldsymbol{\gamma}^{(m)}}\left(\mathbf{W}_{i}-\mathbf{E}_{i}^{\top}\mathbf{Z}_{i}\right)^{\hspace*{-0.3em}\top}\hspace*{-0.5em}
\left(\mathbf{W}_{i}-\mathbf{E}_{i}^{\top}\mathbf{Z}_{i}\right)\right\}.
\end{equation}

\subsection{The AMCEM Algorithm}

In the E-step of the EM algorithm described above, it is necessary to compute the sufficient statistics in equations (\ref{eq:MstepSigma}), (\ref{eq:MstepB}) and (\ref{eq:Mstepsigma2}), specifically:
\begin{align*}
&\Bbb{E}_{\mathbf{W}\mid \mathbf{Y}=\mathbf{y},\boldsymbol{\gamma}^{(m)}}\left\{\Bbb{E}_{\mathbf{Z}\mid \mathbf{W},V_{\lambda},\boldsymbol{\gamma}^{(m)}}\left(\mathbf{Z}^{*}_{k}\mathbf{Z}^{*^{\top}}_{k}\right)\right\},\\
&\Bbb{E}_{\mathbf{W}\mid \mathbf{Y}=\mathbf{y},\boldsymbol{\gamma}^{(m)}}\left\{\Bbb{E}_{\mathbf{Z}\mid \mathbf{W},V_{\lambda},\boldsymbol{\gamma}^{(m)}}\left(\mathbf{Z}_{i} \right)\right\},\\
&\Bbb{E}_{\mathbf{W}\mid \mathbf{Y}=\mathbf{y},\boldsymbol{\gamma}^{(m)}}\left\{\Bbb{E}_{\mathbf{Z}\mid \mathbf{W},V_{\lambda},\boldsymbol{\gamma}^{(m)}}\left(\mathbf{W}_{i}-\mathbf{E}_{i}^{\top}\mathbf{Z}_{i}\right)^{\top}
\mathbf{K}_{i}^{-1^{(m+1)}}\left(\mathbf{W}_{i}-\mathbf{E}_{i}^{\top}\mathbf{Z}_{i}\right)\right\}.
\end{align*}
None of the above expectations have closed-form expressions, thus a Monte Carlo method needs to be employed to compute the integrations numerically. For the outer expectation, a Gibbs sampling algorithm can be developed to generate samples from the truncated multivariate normal distribution, and for the inner expectation, an accept-and-reject scheme can be used. The AMCEM algorithm is ``adaptive'' because the sequence of smoothing parameters decreases at each step, inducing the latent functions fitted to the data to be smoother at each step proceeding along the algorithm. As a result, the AMCEM algorithm yields estimations of functional parameters which are as smooth as possible. 

Consider that $\mathbf{W}_{i}$ is $M_{i}$-variate Gaussian distributed with mean vector $\boldsymbol{\mu}_{i}=\mathbf{E}_{i}^{\top}\mathbf{B}^{\top}\mathbf{x}_{i}$ and variance-covariance matrix $\boldsymbol{\Sigma}_{i}=\mathbf{E}_{i}^{\top}\boldsymbol{\Sigma}_{\theta}\mathbf{E}_{i}+\sigma^{2}\mathbf{K}_{i}$,
and thus $f(\mathbf{w}_{i}\mid \mathbf{Y}_{i}=\mathbf{y}_{i},\boldsymbol{\gamma}^{(m)})$ is a normal distribution with mean $\boldsymbol{\mu}_{i}^{(m)}=\mathbf{E}_{i}^{\top}\mathbf{B}^{(m)^{\top}}\mathbf{x}_{i}$ and variance-covariance matrix ${\boldsymbol{\Sigma}_{i}^{(m)}=\mathbf{E}_{i}^{\top}\boldsymbol{\Sigma}_{\theta}^{(m)}\mathbf{E}_{i}+\sigma^{2^{(m)}}\mathbf{K}_{i}^{(m)}}$, truncated to the region $C_{i1}\times C_{i2}\times\ldots\times C_{iM_{i}}$. In order to apply the Gibbs sampling procedure, the full conditional distributions need to be specified. The conditional pdf $f(w_{ij}\mid \mathbf{W}_{i,-j}=\mathbf{w}_{i,-j},Y_{ij}=y_{ij},\boldsymbol{\gamma}^{(m)})$ is a univariate Gaussian density function $\mathcal{N}(\tau_{ij},\sigma^{2}_{ij})$ truncated to $C_{ij}$ with parameters
\begin{align}\label{30}
\tau_{ij}&=\mu_{ij}^{(m)}+\boldsymbol{\Sigma}_{i}^{(m)}[j,-j]\boldsymbol{\Sigma}_{i}^{(m)^{-1}}[-j,-j](\mathbf{w}_{i,-j}-\boldsymbol{\mu}^{(m)}_{i,-j}),\\\label{31}
\sigma^{2}_{ij}&={\Sigma_{i}^{(m)}}_{(jj)}-\boldsymbol{\Sigma}_{i}^{(m)}[j,-j]\boldsymbol{\Sigma}_{i}^{(m)^{-1}}[-j,-j]\boldsymbol{\Sigma}_{i}^{(m)^{\top}}[j,-j].
\end{align}
In  equations (\ref{30}) and (\ref{31}), $\mu_{ij}^{(m)}$ is the $j$\textsuperscript{th} element of $\boldsymbol{\mu}_{i}^{(m)}$, $\boldsymbol{\mu}^{(m)}_{i,-j}$ is the vector $\boldsymbol{\mu}_{i}^{(m)}$  excluding the $j$\textsuperscript{th} element, ${\Sigma_{i}^{(m)}}_{(jj)}$ is the $j$\textsuperscript{th} diagonal element of the matrix $\boldsymbol{\Sigma}_{i}^{(m)}$, $\boldsymbol{\Sigma}_{i}^{(m)}[j,-j]$ is the $j$\textsuperscript{th} row of $\boldsymbol{\Sigma}_{i}^{(m)}$ excluding the $j$\textsuperscript{th} element, and finally $\boldsymbol{\Sigma}_{i}^{(m)}[-j,-j]$ is the matrix $\boldsymbol{\Sigma}_{i}^{(m)}$ excluding the $j$\textsuperscript{th} row and  $j$\textsuperscript{th} column.
To generate samples from $\mathcal{N}(\tau_{ij},\sigma^{2}_{ij})$ truncated to $C_{ij}$, a simple method of inverse transform sampling can be employed. 
Let $\Phi\left(x;\tau_{ij},\sigma^{2}_{ij}\right)$ be the cumulative distribution function  of $\mathcal{N}(\tau_{ij},\sigma^{2}_{ij})$, and define $p=\Phi\left(0;\tau_{ij},\sigma^{2}_{ij}\right)$. Let $u$ be a random sample from the uniform distribution $U\left(0,p\right)$ if $y_{ij}=0$, and from $U\left(p,1\right)$ otherwise. A random sample from $\mathcal{N}(\tau_{ij},\sigma^{2}_{ij})$ truncated to $C_{ij}$ is obtained by $\Phi^{-1}\left(u;\tau_{ij},\sigma^{2}_{ij}\right)$. 

To compute the inner expectations, consider that the conditional distribution of  $\mathbf{Z}_{i}$ given $\mathbf{W}_{i}$ and $\boldsymbol{\gamma}^{(m)}$, for $1\leq i\leq N$, is the Gaussian distribution $\mathcal{N}_{J}(\boldsymbol{\eta}_{i},\boldsymbol{\Delta}_{i})$ with
\begin{align*}
\boldsymbol{\eta}_{i}&=\mathbf{B}^{(m)^{\top}}\mathbf{x}_{i}+\boldsymbol{\Sigma}_{\theta}^{(m)}\mathbf{E}_{i}\boldsymbol{\Sigma}_{i}^{(m)^{-1}}(\mathbf{w}_{i}-\boldsymbol{\mu}_{i}^{(m)}),\\
\boldsymbol{\Delta}_{i}&=\boldsymbol{\Sigma}_{\theta}^{(m)}-\boldsymbol{\Sigma}_{\theta}^{(m)}\mathbf{E}_{i}\boldsymbol{\Sigma}_{i}^{(m)^{-1}}\mathbf{E}_{i}^{\top}\boldsymbol{\Sigma}_{\theta}^{(m)}.
\end{align*}
Thus, the inner expectation can be computed numerically by generating random samples from the given conditional distribution, truncated to the region specified by $V_{\boldsymbol{\lambda}}$. Random sample generation from this truncated multivariate normal distribution can be achieved by implementing an accept-and-reject scheme.
The rejection rate of this procedure depends on $\lambda_{i}$, with poor specification of $\lambda_i$ possibly leading to a high rejection rate. In the adaptive MCEM algorithm,  $\lambda_{i}$ is chosen such that the rejection rate can be controlled at a prespecified level. The smoothing parameter $\lambda_{i}$ is decreasingly updated at each step of the Gibbs sampler. This approach avoids the selection of the smoothing parameter for the latent variables and eigen-functions, thus providing an automatically tuned smooth estimation.

\begin{algorithm}[!htp]\label{al.amcem}
	\SetAlgoLined
	\let\oldnl\nl
	\newcommand{\nonl}{\renewcommand{\nl}{\let\nl\oldnl}}\nonl
	
	$~$\\\nonl
	At iteration $m+1$:
	$~$\\
	$\delta^{(m+1)}\leftarrow \mathop{\argmin}\limits_{\delta}AV^{(m)}(\delta)$\label{em.0}
	
	\For{$i$ \textnormal{in} $1:N$}
	{
		
		\uIf{$t\leq 2$\label{em.1}}
		{

			$\lambda_{i}^{(m+1)}\leftarrow\infty$
			
		}\Else{
			
			$\lambda_{i}^{(m+1)}\leftarrow \text{Quantile}\left(\left\{h_{i}\left[m,k\right],~~ k=1,2,\ldots,K\right\},1-\delta^{(m+1)}\right)$ \label{em.2}
			
		}

		Draw $K$ samples $\mathbf{w}_{i}^{(k)}\sim \mathcal{N}\left(\boldsymbol{\mu}_{i}^{(m)},\boldsymbol{\Sigma}_{i}^{(m)}\right)$, $k=1,2,\ldots,K$, truncated to the region $C_{i1}\times C_{i2}\times\ldots\times C_{iM_{i}}$, using Gibbs sampler
		
		$\boldsymbol{\Delta}_{i}\leftarrow\boldsymbol{\Sigma}_{\theta}^{(m)}-\boldsymbol{\Sigma}_{\theta}^{(m)}\mathbf{E}_{i}\boldsymbol{\Sigma}_{i}^{(m)^{-1}}\mathbf{E}_{i}^{\top}\boldsymbol{\Sigma}_{\theta}^{(m)}
		$
		
		\For{$k$ \textnormal{in} $1:K$}
		{

			$\boldsymbol{\eta}_{i}\leftarrow\mathbf{B}^{(m)^{\top}}\mathbf{x}_{i}+\boldsymbol{\Sigma}_{\theta}^{(m)}\mathbf{E}_{i}\boldsymbol{\Sigma}_{i}^{(m)^{-1}}(\mathbf{w}_{i}^{(k)}-\boldsymbol{\mu}_{i}^{(m)})$

			Draw $\mathbf{x}\sim\mathcal{N}(\boldsymbol{\eta}_{i},\boldsymbol{\Delta}_{i})$

			$h\leftarrow \mathbf{x}^{\top}\mathbf{P}^{(n)}\mathbf{x}$

			\While{$h>\lambda_{i}^{(m+1)}$\label{em.3}}{

				Draw $\mathbf{x}\sim\mathcal{N}(\boldsymbol{\eta}_{i},\boldsymbol{\Delta}_{i})$

				$h\leftarrow \mathbf{x}^{\top}\mathbf{P}^{(n)}\mathbf{x}$
				
			}
			
			$h_{i}[m+1,k]\leftarrow \mathbf{x}^{\top}\mathbf{P}^{(n)}\mathbf{x}$

			$\mathbf{z}_{i}^{(k)}\leftarrow \mathbf{x}$

		}

	}
	\nonl
	Update $\mathbf{b}^{(m+1)}$, $\boldsymbol{\Sigma}_{\theta}^{(m+1)}$ and $\sigma^{2^{(m+1)}}$ from the generated sample $(\mathbf{w}_{i}^{(k)},\mathbf{z}_{i}^{(k)})$, $i=1,2,\ldots,N$ and $k=1,2,\ldots,K$.

	\caption{Adaptive Monte Carlo EM}
\end{algorithm}

We give here some further specifications concerning the AMCEM algorithm sketched above:
\begin{itemize}
	\renewcommand{\labelitemi}{\scriptsize$-$}
	\item Line \ref{em.0}, initialization of iteration $m + 1$: the update of $\delta^{(m+1)}$, the rate of rejection, makes use of the validation function $AV^{(m)}(\delta)$ of the AMCEM algorithm evaluated on the parameter values at the previous iteration $(m)$, as defined in (\ref{eq:AV}) below. 
	\item Line \ref{em.2}: the set $\{h_{i}[m,k],~k=1,2,\ldots,K\}$ is a random sample from the distribution  $\mathbf{Z}_{i}^{\top}P^{(n)}\mathbf{Z}_{i}$ given $\mathbf{Y}_{i}$ and $\boldsymbol{\gamma}^{(m)}$. Thus the $100(1-\delta)$\textsuperscript{th} percentile  of $\{h_{i}[m,k],~k=1,2,\ldots,K\}$ is an approximation of   the $(1-\delta)$\textsuperscript{th} quantile of the random variable $\mathbf{Z}_{i}^{\top}P^{(n)}\mathbf{Z}_{i}\mid \mathbf{Y}_{i},\boldsymbol{\gamma}^{(m)}$. Hence, the rejection rate of the algorithm can be controlled to be close to an a priori specified level $\delta$.
	Note that, at the second iteration of the EM algorithm (see line \ref{em.1}), the parameters differ significantly from their initial values, and hence the distribution of  $\mathbf{Z}_{i}^{\top}P^{(n)}\mathbf{Z}_{i}\mid \mathbf{W}_{i},\boldsymbol{\gamma}^{(1)}$ changes dramatically relative to the distribution of $\mathbf{Z}_{i}^{\top}P^{(n)}\mathbf{Z}_{i}\mid \mathbf{W}_{i},\boldsymbol{\gamma}^{(0)}$ at the first iteration, possibly causing the procedure at line \ref{em.2} to yield sub-optimal results. We therefore choose to employ this procedure only from the third iteration of the algorithm.
	\item Line \ref{em.3}: obtaining smooth functions that fit well the sequence of binary data is guaranteed by the implemented accept-and-reject procedure. 
\end{itemize}



The validation function in the AMCEM algorithm for a given $0<\delta<1$ is defined by  
\begin{equation}\label{eq:AV}
\frac{1}{\sum_{i=1}^{N}M_{i}}\sum_{i=1}^{N}\Bbb{E}_{\mathbf{W}\mid \mathbf{Y}=\mathbf{y},\boldsymbol{\gamma}^{(m)}}\left\{\Bbb{E}_{\mathbf{Z}\mid \mathbf{W},V_{\delta},\boldsymbol{\gamma}^{(m)}}\left(\mathbf{W}_{i}-\mathbf{E}_{i}^{\top}\mathbf{Z}_{i}\right)^{\top}
\hat{\mathbf{K}}_{i,\delta}^{-1}\left(\mathbf{W}_{i}-\mathbf{E}_{i}^{\top}\mathbf{Z}_{i}\right)\right\}.
\end{equation}
$V_{\delta}$ is the event $\mathbf{Z}_{i}^{\top}\mathbf{P}^{(n)}\mathbf{Z}_{i}\leq \lambda_{i},~\forall i=1,2,\ldots,N$ with 
$$\lambda_{i}= \text{Quantile}\left(\left\{h_{i}\left[m,k\right],~~ k=1,2,\ldots,K\right\},1-\delta\right),$$
and $\hat{\mathbf{K}}_{i,\delta}$ is the evaluation of the covariance function estimated by the generated samples. 
Since when increasing the smoothing level the fitted latent functions cause the inflation of $\sigma^2$, the validation function  prevents the algorithm from over-smoothing.

The sequence $\lambda_{i}^{(m)}$ is decreasing with respect to $m$, which causes the imputed missing functions to get  smoother as the AMCEM proceeds. The decreasing sequence $\lambda_{i}^{(m)}$ is bounded from below, thus implying its convergence. Furthermore, the convergence of $\lambda_{i}^{(m)}$ corresponds to the convergence of the rejection rate $\delta^{(m)}$ to zero. Since $\delta^{(m)}$ converges to zero, $V_{\delta}$ includes the support of $\mathbf{Z}$ given $\boldsymbol{\gamma}^{(m)}$ and $\mathbf{Y}$.  

With the random samples generated at iteration $m+1$, the parameters $\mathbf{B}$, $\boldsymbol{\Sigma}_{\theta}$ and $\sigma^2$ are updated as follows:
\begin{align*}
\mathbf{\Sigma}_{\theta}^{(m+1)}=&\left(\frac{1}{N-q}\right)\sum_{i=1}^{N-q}\frac{1}{K}\sum_{k=1}^{K}\mathbf{z}_{i}^{*(k)}\mathbf{z}_{i}^{*(k)^{\top}},\\
\mathbf{B}^{(m+1)}=&\left(\mathbf{X}^{\top}\mathbf{X}\right)^{-1}\mathbf{X}^{\top}\left(\frac{1}{K}\sum_{k=1}^{K}\mathbf{Z}^{(k)}\right),\\
\sigma^{2^{(m+1)}}=&\frac{1}{\sum_{i=1}^{N}M_{i}}\sum_{i=1}^{N}\frac{1}{K}\left(\mathbf{w}_{i}^{(k)}-\mathbf{E}_{i}^{\top}\mathbf{z}_{i}^{(k)}\right)^{\top}
\mathbf{K}_{i}^{-1^{(m+1)}}\left(\mathbf{w}_{i}^{(k)}-\mathbf{E}_{i}^{\top}\mathbf{z}_{i}^{(k)}\right),
\end{align*}
in which $\mathbf{z}_{i}^{*(k)}=\mathbf{Z}^{(k)^{\top}}\mathbf{u}_{i}$, where $\mathbf{Z}^{(k)}=\left[\mathbf{z}_{1}^{(k)},\mathbf{z}_{2}^{(k)},\ldots,\mathbf{z}_{N}^{(k)}\right]^{\top}$ is a $N\times J$ matrix and $\mathbf{u}_{i}$ is the $i$\textsuperscript{th} column of $\mathbf{U}$.


\subsection{Further Comments}\label{sec.4}

In the model description given in Section \ref{sec.3}, the set of basis $\{e_{i}\}_{i\geq 1}$ was considered to be orthonormal. This assumption can be relaxed without any noticeable changes in our arguments, that is  $\{e_{i}\}_{i\geq 1}$ can be chosen as any arbitrary basis, such as for example B-splines. The only differences are in estimating the standardized parameters in equations (\ref{25}) and (\ref{26}), and extracting the eigen-functions from the matrix $\boldsymbol{\Sigma}_{\theta}$. The new standardized estimates of $\boldsymbol{\alpha}$ and $R$ are given by
\begin{align*}
\boldsymbol{\alpha}^{(m+1)}&=\mathbf{B}^{(m+1)}\mathbf{D}\boldsymbol{\Omega}^{-1}\mathbf{e},\\
R^{(m+1)}&=\sum_{j=1}^{J}\varrho_{j}^{(m+1)}\varphi_{j}^{(m+1)}\otimes\varphi_{j}^{(m+1)},
\end{align*}
where $\boldsymbol{\Omega}=\left[\left\langle e_{i},e_{j}\right\rangle\right]_{1\leq i,j\leq J}$ is a $J\times J$ matrix  and $\varphi_{j}^{(m+1)}=\boldsymbol{\vartheta}_{j}^{(m+1)^{\top}}\boldsymbol{\Omega}^{-1/2}\mathbf{e}$, where $(\varrho_{j}^{(m+1)},\boldsymbol{\vartheta}_{j}^{(m+1)})$ are respectively the eigen-values and eigen-vectors of $\boldsymbol{\Omega}^{-1/2}\mathbf{D}\boldsymbol{\Sigma}_{\theta}^{(m+1)}\mathbf{D}^{\top}\boldsymbol{\Omega}^{-1/2}$. 

In our approach for estimating the functional parameters as described in Section (\ref{EM}), we proposed the smooth estimation of both functional regression coefficients and eigen-functions. As already mentioned, the smooth estimation of eigen-functions is obtained by introducing the novel AMCEM algorithm, which automatically tunes the smoothing parameters. It is worth noticing that the method, in addition to the eigen-functions, can also provide smooth estimations of the regression coefficients.

Prediction is often a crucial task in statistical modeling. Two different scenarios can be considered for prediction in our modeling framework. First, suppose that only the covariate values are given for a new individual. The estimation of the individual trajectory for the new case on its whole domain is then given by $\hat{\mathcal{W}}=\hat{\Bbb{E}}[\mathcal{W}\mid \mathbf{X}=\mathbf{x}]=(\hat{\mathbf{B}}\mathbf{e})^{\top}\mathbf{x}$. Another possible prediction scenario entails that, in addition to the value of the covariates, the value of $Y(t)$ is also observed for some time points $t_{1},t_{2},\ldots,t_{M}$. The individual trajectory for this new case can be then estimated by 
\begin{align}\nonumber
\hat{\mathcal{W}}&=\hat{\Bbb{E}}[\mathcal{W}
\mid\mathbf{X}=\mathbf{x},\mathbf{Y}=\mathbf{y}]\\
&=\mathbf{e}^{\top}\left[\hat{\mathbf{B}}^{\top}\mathbf{x}+\hat{\boldsymbol{\Sigma}}_{\theta}\mathbf{E}(\mathbf{E}^{\top}\hat{\boldsymbol{\Sigma}}_{\theta}\mathbf{E})^{-1}\left(\Bbb{E}_{\mathbf{W}\mid \mathbf{X}=\mathbf{x},\mathbf{Y}=\mathbf{y}}(\mathbf{W})-(\hat{\mathbf{B}}\mathbf{E})^{\top}\mathbf{x}\right)\right],
\end{align}
where
\begin{align*}
\mathbf{Y}&=(Y({t_1}),Y({t_2}),\ldots,Y({t_M})),\\
\mathbf{E}&=\left[e_{j}\left(t_{k}\right)\right]_{1\leq j\leq J, 1\leq k\leq M},\\
\mathbf{W}&=\left(\mathcal{W}(t_{1}),\mathcal{W}(t_{2}),\ldots,\mathcal{W}(t_{M})\right)^{\top}.
\end{align*}
The expectation $\Bbb{E}(\mathbf{W}\mid \mathbf{X}=\mathbf{x},\mathbf{Y}=\mathbf{y})$ can be computed by MCMC integration approximation, using Gibbs sampling.

\section{Applications}\label{sec.5}

In this section, we evaluate the performance of the AMCEM algorithm via simulation studies and a real-world example. In the simulation studies, we examine the effect of: sample size, the number of points sampled per curve, the regularity/irregularity of the design, the magnitude of the variance of the measurement error, and the complexity of the covariance structure. For the regular/irregular designs, four scenarios are taken into account: (R) regular designs with equally spaced time points, (RT) regular designs with equally spaced right-truncated time points,  (RM) regular designs with equally spaced and missing at random time points, (IRS) irregular designs with stochastic time points.

\subsection{Simulation Studies}

In our simulation study, we consider the  FoSR model (\ref{eq.2.2.0}) with $q=1$, given by
\begin{align}\label{eq.2.2.0.0}
\mathcal{W}_i(t)=\beta_{0}(t)+\beta_{1}(t)x_i+\varepsilon_i(t)+\epsilon_{it},~~~\varepsilon_i (t)=\sum_{j\geq 1}\sqrt{\nu_j}\varepsilon_{ij}\psi_{j}(t),
\end{align}
in which $\varepsilon_{ij}$ are  identically and independently distributed standard normal random variables. The parameters $\nu_j$ and $\psi_j$ are eigen-values and eigen-functions of  the covariance operator of $\varepsilon_i$, and also $\epsilon_{it}\stackrel{iid}{\sim}\mathcal{N}\left(0,\sigma^2 K\left(t,t\right)\right)$, where $K(t,t)=\sum_{j\geq 1}\nu_j\psi_{j}^{2}(t)$. A dichotomized version of this model is observed as 
\begin{equation*}
y_{ij}=\text{I}\left(\mathcal{W}(t_{ij})>0\right),\qquad i=1,2,\ldots,N,\qquad j=1,2,\ldots,M_i .
\end{equation*}

In the data generating procedure, we consider two different sets of functional regression coefficients, detailed below. For what concerns the covariance structure of model (\ref{eq.2.2.0.0}), we assume that only the first $p$ eigen-values are non-zero. The eigen-values are chosen as $\nu_{j}=r\rho^{j-1}$ for fixed $r>0$ and ${0< \rho<1}$, and the  parameter $\rho$ determines the decay rate of the eigen-values. If $\rho$ tends to zero (one), the sequence of eigen-values decreases fast (slowly), thus leading to smoother  (less smooth) residual functions, and to more (less) correlated binary variables in the sequence. 
For the eigen-functions, we
set  $\psi_{2j}\left(t\right)=\sqrt{2}\sin\left(2\pi jt\right)$
and $\psi_{2j+1}\left(t\right)=\sqrt{2}\cos\left(2\pi \left(j+1\right)t\right)$
for $j\geq 1$. 

The number and location of time points per curve are chosen according to four different scenarios related to the designs (R), (RT), (RM), and (IRS). In the first three scenarios we fix the same maximal number of time points $M,$ and the same locations, and we then chose their actual number and locations according to a different scheme for each scenario. Scenario (IRS) is instead completely random. Precisely:
\begin{itemize}
	\item Scenario (R): for subject $i$, $Y_i(t)$ is fully observed at times $t_{ij}=\nicefrac{j}{M}$, $j=1,2,\ldots,M$.
	\item Scenario (RT): for subject $i$, $Y_i(t)$ is fully observed at times $t_{ij}=\nicefrac{j}{M} $, $j=1,2,\ldots,M_i$, with $M_i\sim\mathcal{Q}_c$.
	\item Scenario (RM): for subject $i$, $Y_i(t)$ is supposed to be observed at times $t_{ij}=\nicefrac{j}{M}$, $j=1,2,\ldots,M$; however, $Y_i(t_{ij})$ is missing at random according to an independent Bernoulli random variable with success probability $p_m$.
	\item Scenario (IRS): for subject $i$, $Y_i(t)$ is fully observed at times $t_{ij},~j=1,\ldots,M_{i}$, with $M_i\sim\mathcal{Q}_M$ and $t_{ij}\stackrel{iid}{\sim} \mathcal{T}_{ j}$.
\end{itemize}
The distributions $\mathcal{Q}_c$, $\mathcal{Q}_M$ and $\mathcal{T}_{j}$ can be chosen so that the sampling designs   are realistic as compared to what we observe in real data problems. For the scenario corresponding to the (RT) desing, we employ the distribution $\mathcal{Q}_c (a)=\kappa p_{c}^{(M-a)+1}$, $1\leq a\leq M$, where $\kappa$ is a normalizing constant and $0< p_{c}< 1$. For the (IRS) design, we fix $\mathcal{Q}_M$ to be the same as $\mathcal{Q}_c,$ and $\mathcal{T}_{j}$ is the uniform distribution over $[0,1]$. We also choose $r=1.5$, $p_c =0.5$ and $p_m=1-\nicefrac{\Bbb{E} \left[\mathcal{Q}_{c}\right]}{M}$. As a result, the expected size of the data generated in scenarios (RT), (RM) and (IRS) are equal, so that it is reasonable to compare the simulation results for inspecting the effect of the designs on the estimation of the model parameters.

For what concerns the other model parameters, the structure of the simulation is as follows:
\begin{itemize}
	\item each of the four scenarios (R), (RT), (RM), and (IRS) is used as sampling design.
	\item $N=50, 100$ is used respectively as small and large sample size.
	\item $M=12, 36$ is used respectively as small and large number of sampling points per curve.
	\item $\sigma^2 =0.2,  0.8$ is used respectively as small and large measurement error variance.
	\item $\rho=0.1,  0.4$ is used respectively as large and small correlation within the  binary variables in the sequence.
\end{itemize}
In the parameters estimation procedure, the minimal number of sampling points that is required to be measured per subject depends on the complexity of the functional parameters. Here we consider two cases for the functional regression coefficients: (1) a `simple' case in which $\beta_0(t)=-\cos(2\pi t)$ and $\beta_1(t)=1-2t$, and (2) a `complex' case in which $\beta_0(t)=-\cos(2\pi t)$ and $\beta_1(t)=-\sin(4\pi t)$.

The simulation study is run for all designs combinations, and for all combinations of the parameters $N$, $M$, $\sigma^2$ and $\rho$, by using an MCMC algorithm with 2,000 iterations. 
The accuracy of the AMCEM algorithm is evaluated by computing the mean square error (MSE) of estimations. The accuracy and efficiency are also compared to those of the alternative method implemented in the package \code{pffr}. Results are shown in Tables \ref{t.R} and \ref{t.RT}. 

\begin{figure}[tb!]
	
	\begin{minipage}{\dimexpr\linewidth-0cm\relax}%
		\begin{minipage}[b]{1\linewidth}
			\begin{minipage}[b]{0.25\linewidth}
				\centering
				\footnotesize{$\beta_{0}$}
			\end{minipage}
			\begin{minipage}[b]{0.25\linewidth}
				\centering
				\footnotesize{	$\beta_{1}$}
			\end{minipage}
			\begin{minipage}[b]{0.25\linewidth}
				\centering
				\footnotesize{PC 1 }
			\end{minipage}
			\begin{minipage}[b]{0.25\linewidth}
				\centering
				\footnotesize{PC 2 }
			\end{minipage}
		\end{minipage}
		
		\begin{minipage}[b]{0.25\linewidth}
			\centering
			\includegraphics[width=1\linewidth]{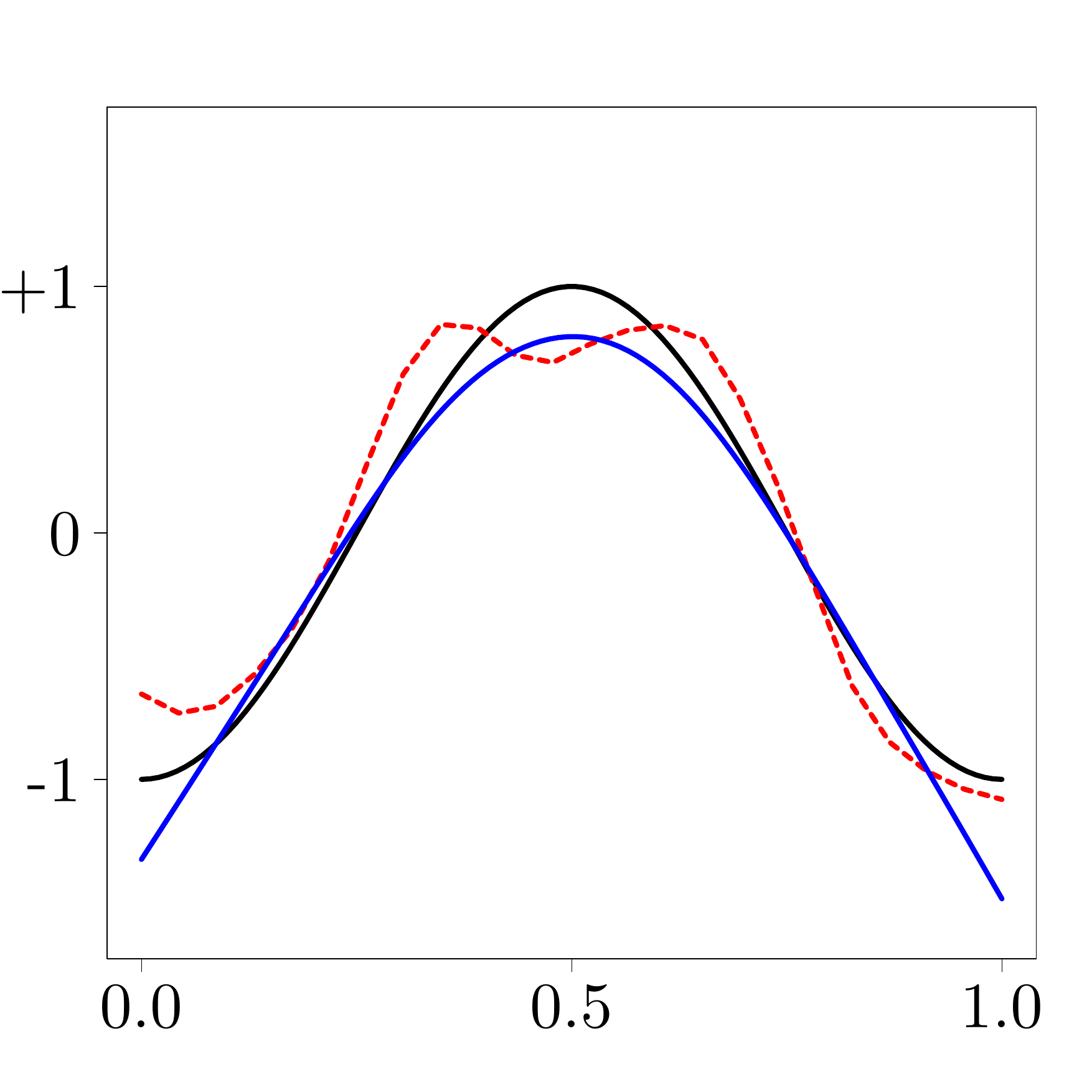} 
		\end{minipage}
		\begin{minipage}[b]{0.25\linewidth}
			\centering
			\includegraphics[width=1\linewidth]{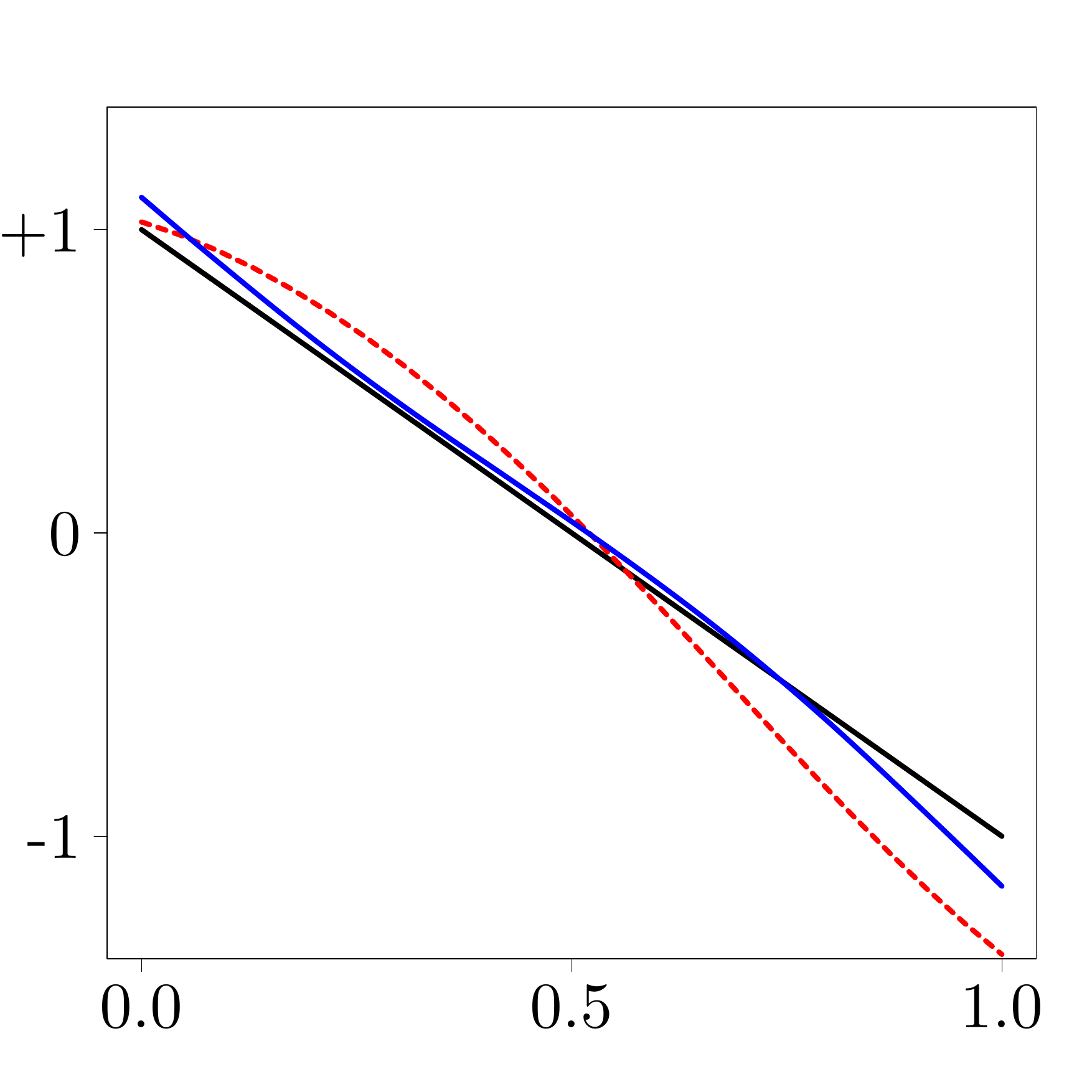} 
		\end{minipage}
		\begin{minipage}[b]{0.25\linewidth}
			\centering
			\includegraphics[width=1\linewidth]{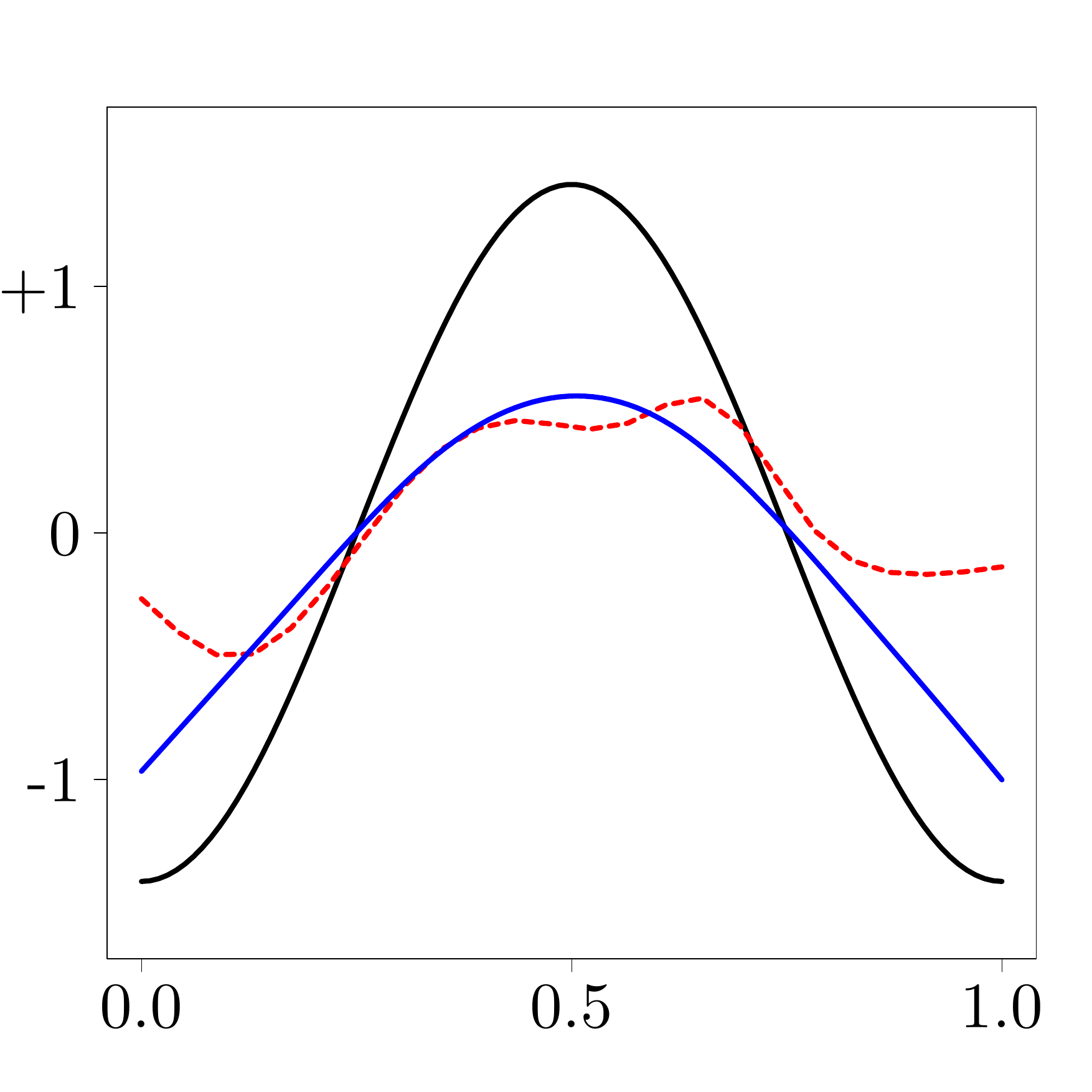} 
		\end{minipage}
		\begin{minipage}[b]{0.25\linewidth}
			\centering
			\includegraphics[width=1\linewidth]{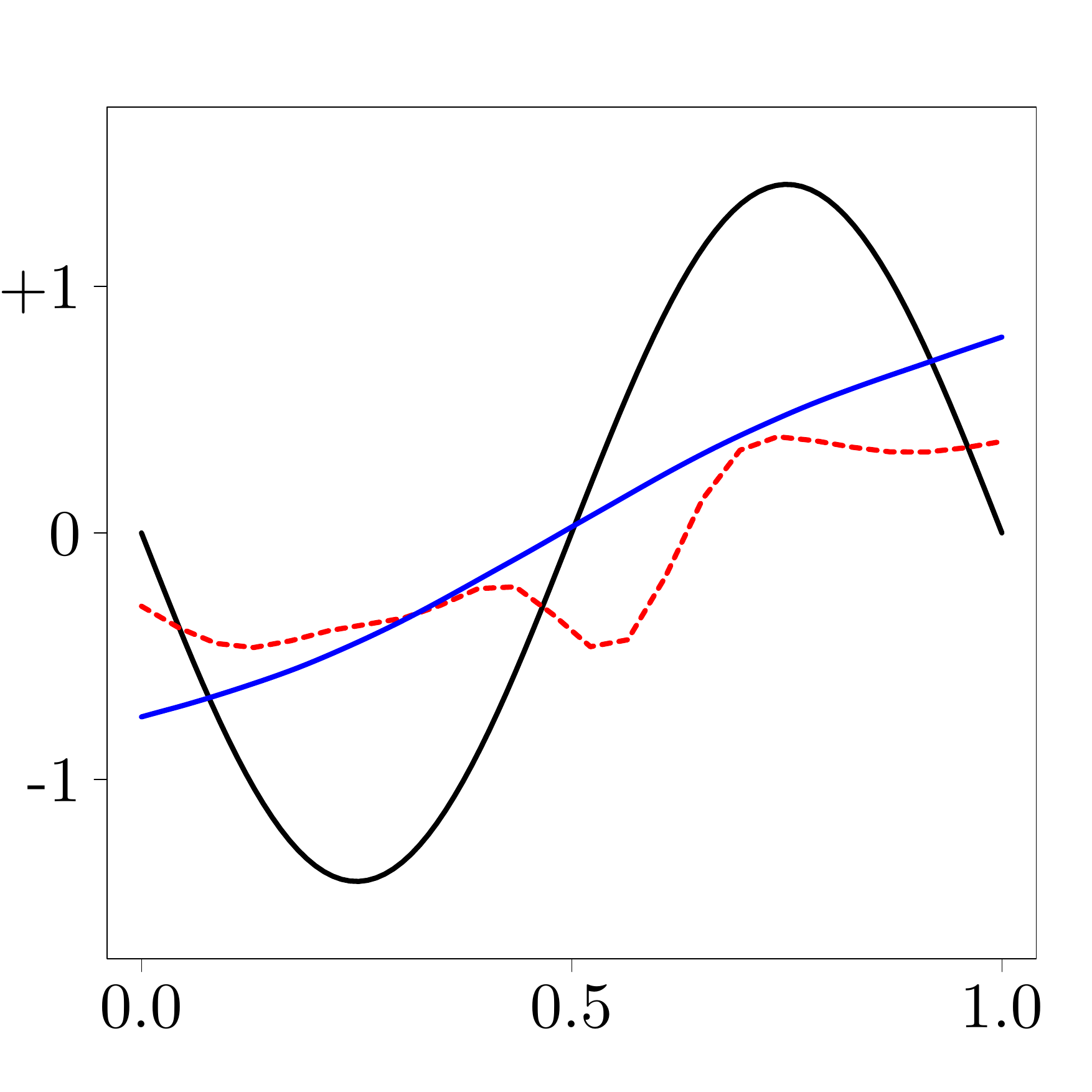} 
		\end{minipage}
		
		\begin{minipage}[b]{0.25\linewidth}
			\centering
			\includegraphics[width=1\linewidth]{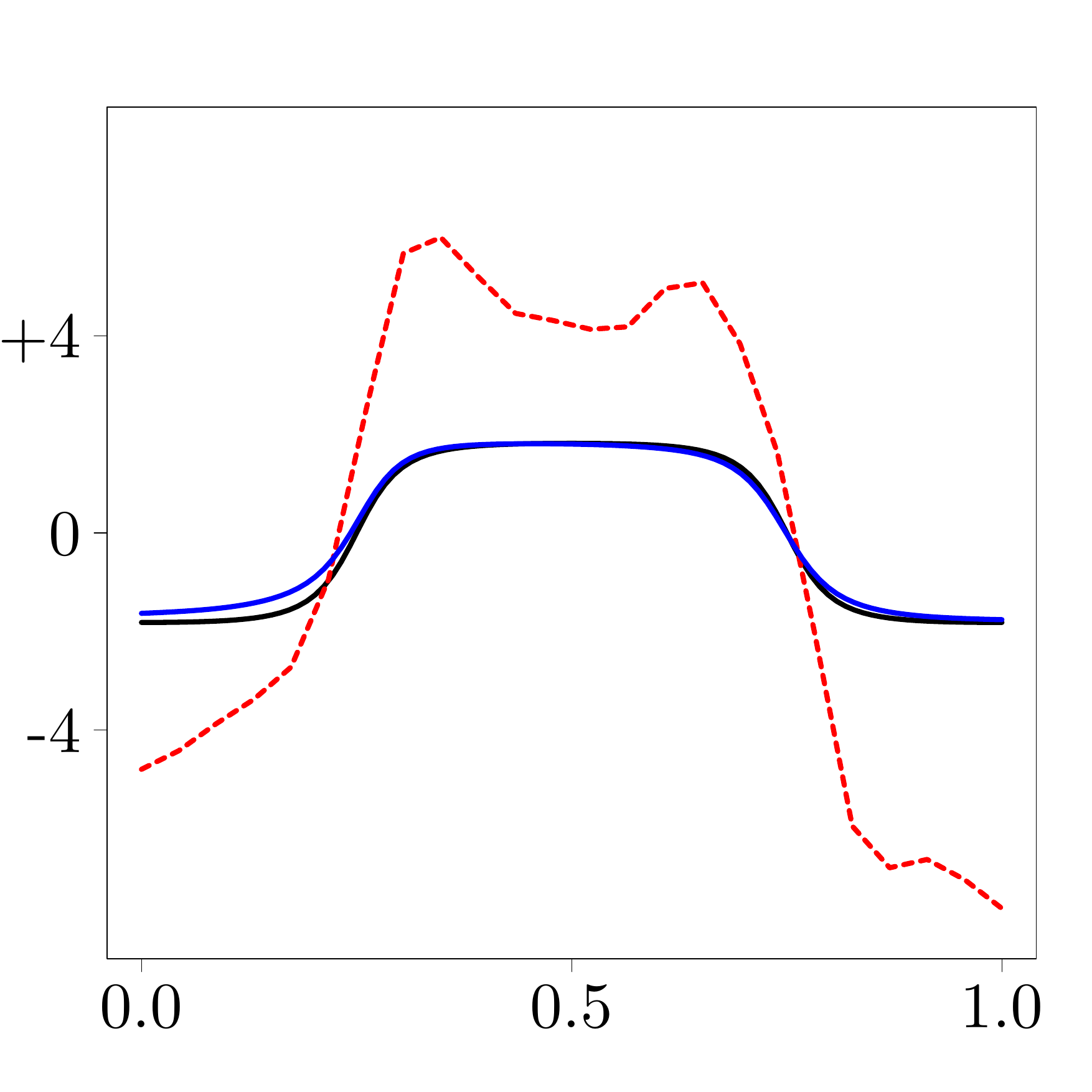} 
		\end{minipage}
		\begin{minipage}[b]{0.25\linewidth}
			\centering
			\includegraphics[width=1\linewidth]{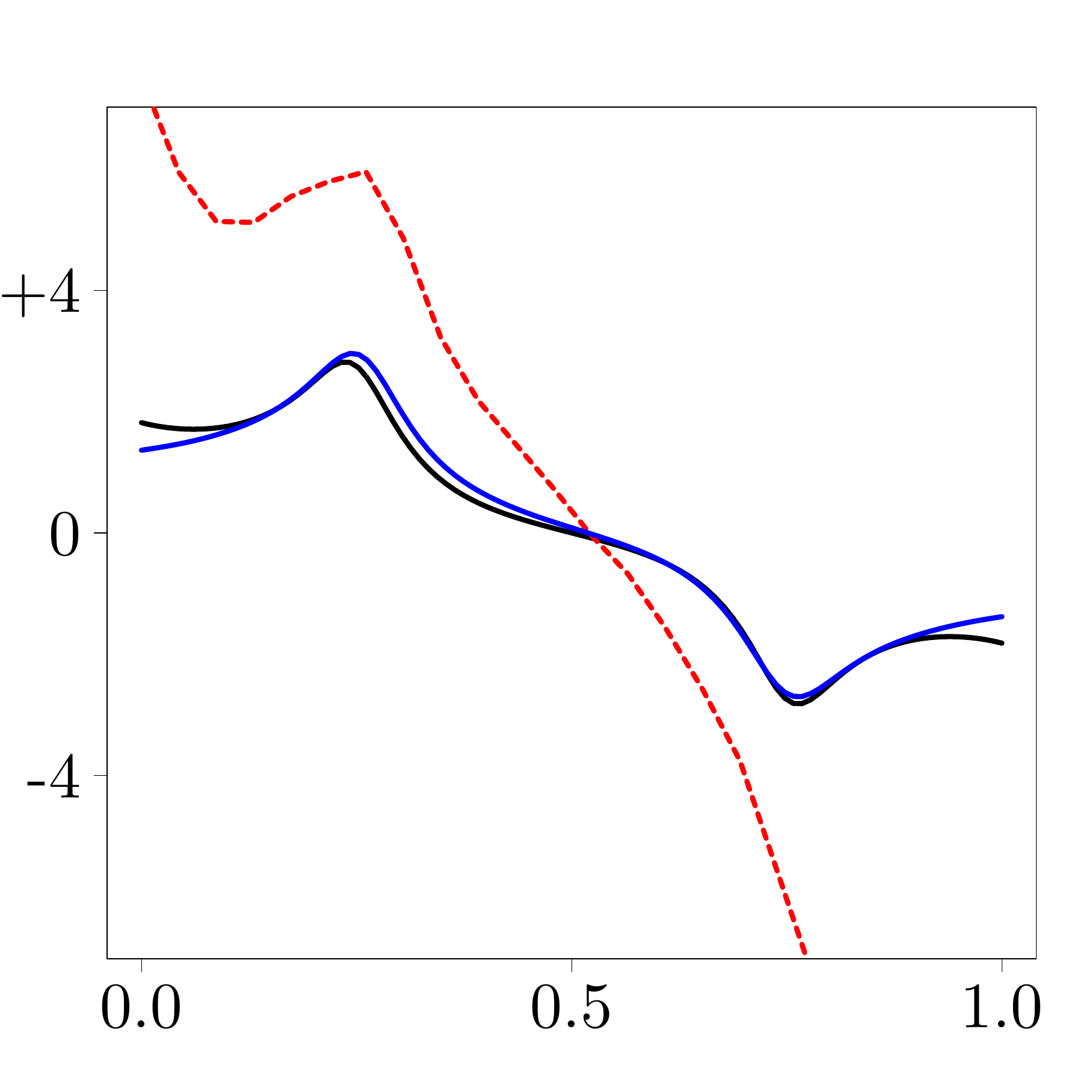} 
		\end{minipage}
		\begin{minipage}[b]{0.25\linewidth}
			\centering
			\includegraphics[width=1\linewidth]{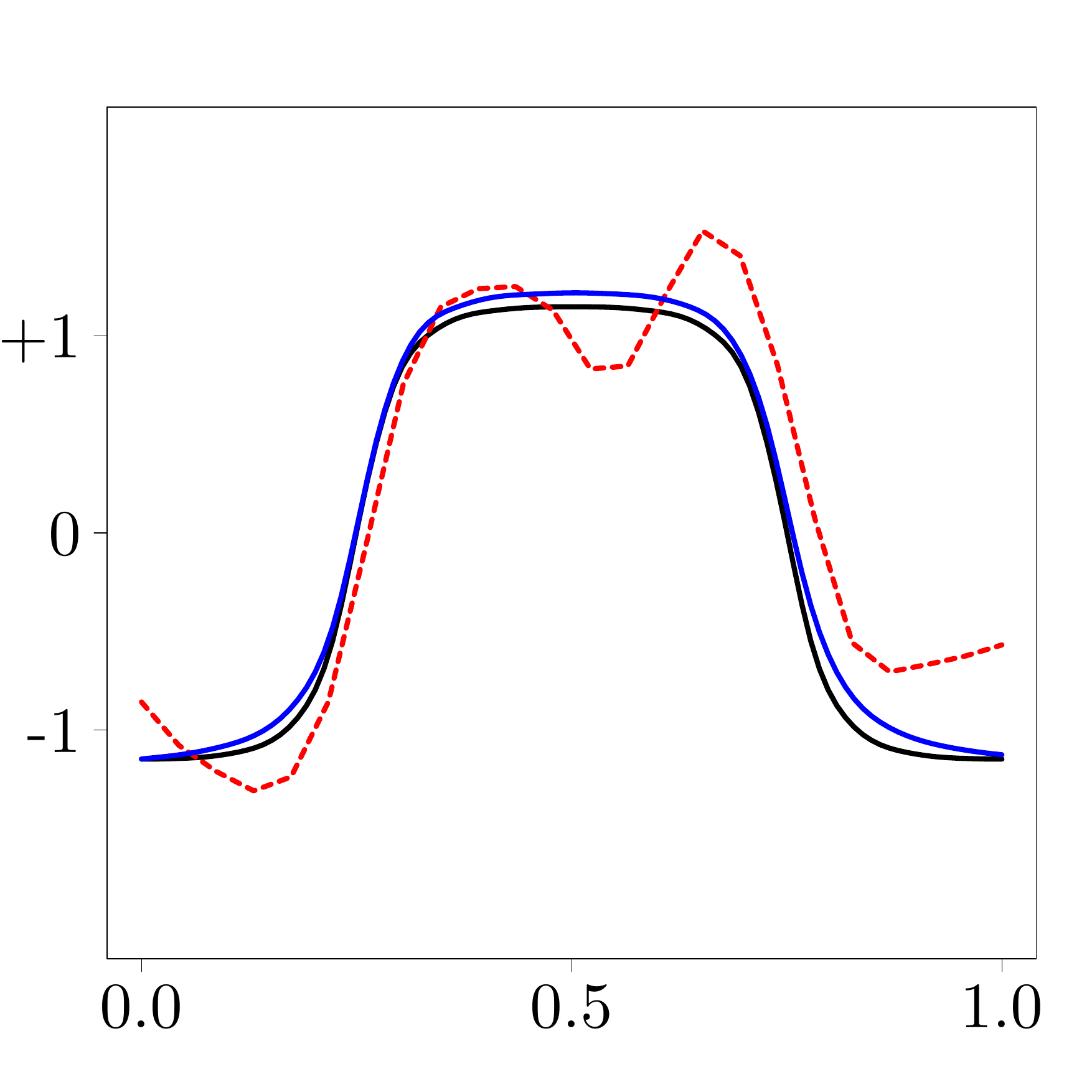} 
		\end{minipage}
		\begin{minipage}[b]{0.25\linewidth}
			\centering
			\includegraphics[width=1\linewidth]{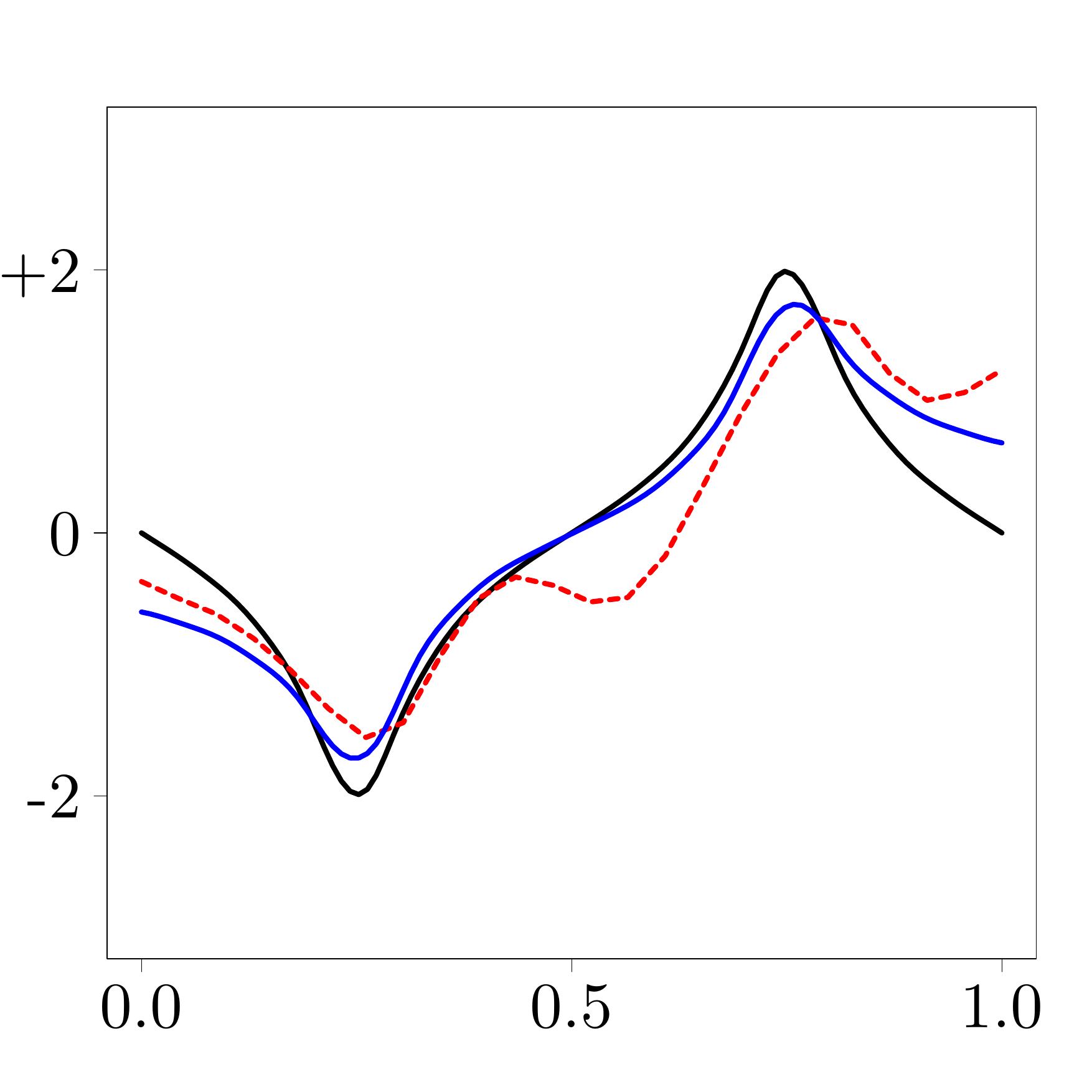} 
		\end{minipage}
		
		\vspace{-1em}
		\begin{minipage}[b]{1\linewidth}
			\centering
			\footnotesize{t}
		\end{minipage}
		
	\end{minipage}%
	\caption{A single run of simulation scenario R(s). Upper panels: unstandardized regression coefficients and first two principal components; lower panels: standardized regression coefficients and first two principal components. Black solid line: original parameter; blue solid line: AMCEM estimation; red dashed line: \code{pffr} estimation. }
	
	\label{sim:R1}
\end{figure}

\begin{figure}[tb!]
	\vspace*{.5cm}
	\begin{minipage}{\dimexpr\linewidth-0cm\relax}%
		\begin{minipage}[b]{1\linewidth}
			\begin{minipage}[b]{0.25\linewidth}
				\centering
				\footnotesize{$\beta_{0}$}
			\end{minipage}
			\begin{minipage}[b]{0.25\linewidth}
				\centering
				\footnotesize{$\beta_{1}$}
			\end{minipage}
			\begin{minipage}[b]{0.25\linewidth}
				\centering
				\footnotesize{PC 1 }
			\end{minipage}
			\begin{minipage}[b]{0.25\linewidth}
				\centering
				\footnotesize{PC 2 }
			\end{minipage}
		\end{minipage}
		
		\begin{minipage}[b]{0.25\linewidth}
			\centering
			\includegraphics[width=1\linewidth]{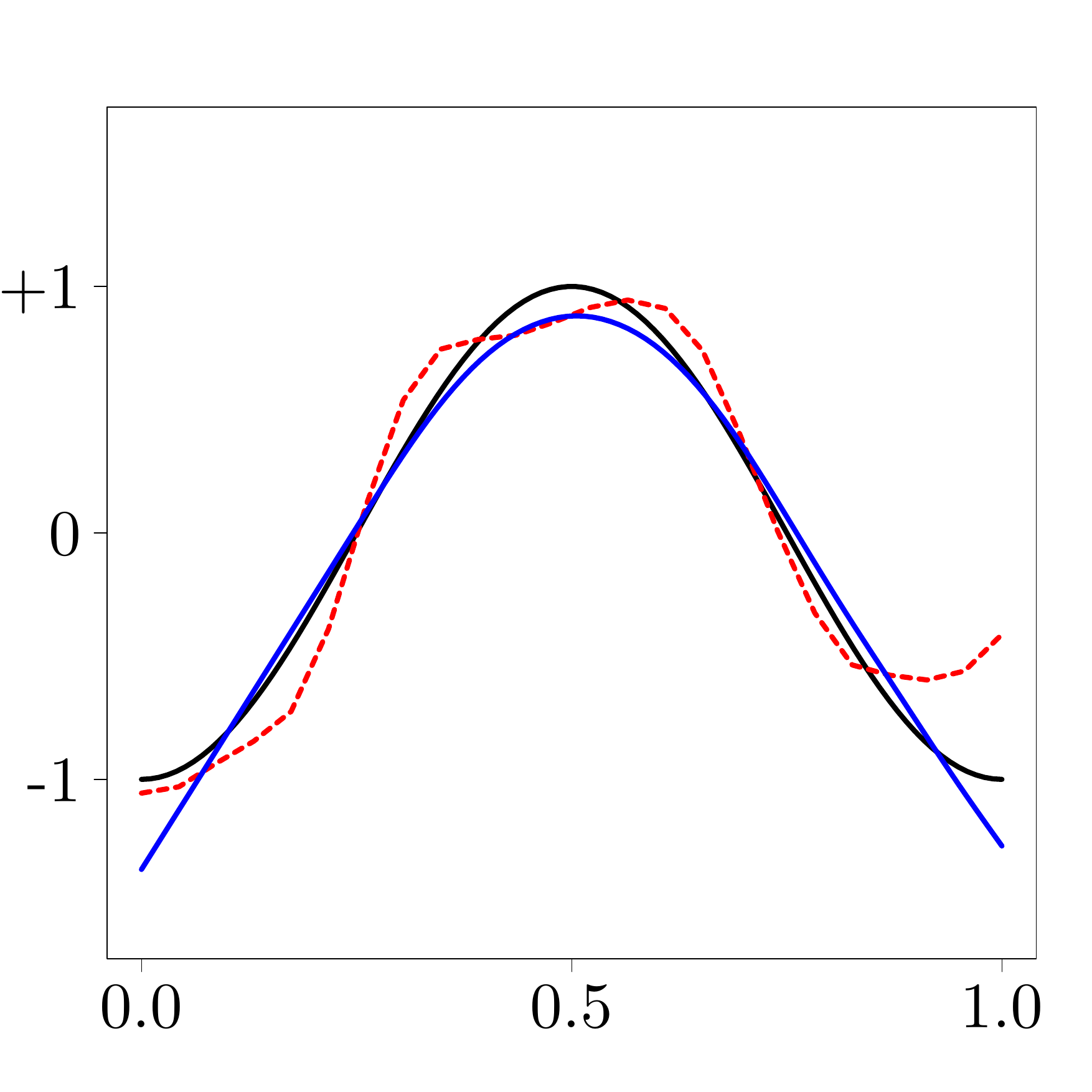} 
		\end{minipage}
		\begin{minipage}[b]{0.25\linewidth}
			\centering
			\includegraphics[width=1\linewidth]{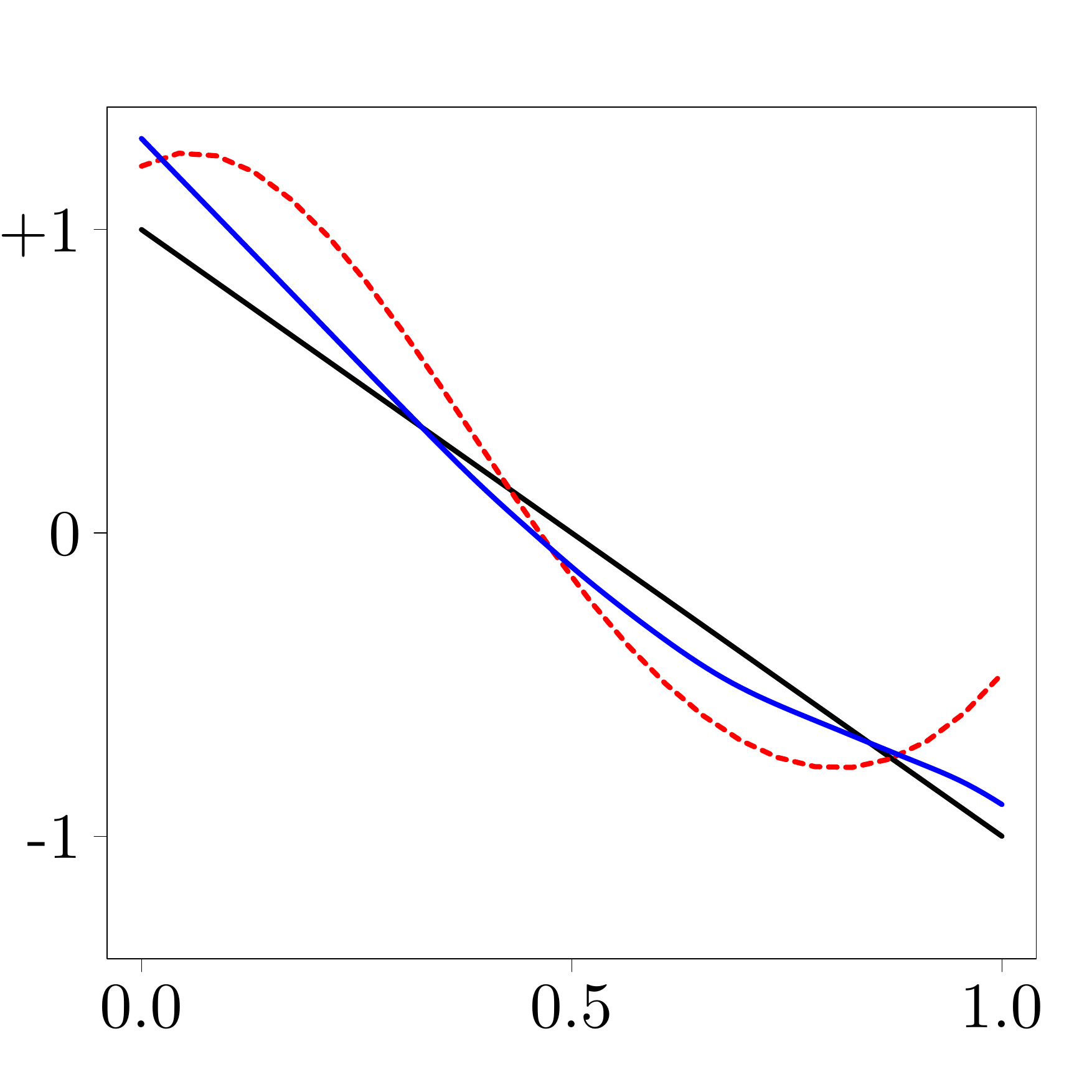} 
		\end{minipage}
		\begin{minipage}[b]{0.25\linewidth}
			\centering
			\includegraphics[width=1\linewidth]{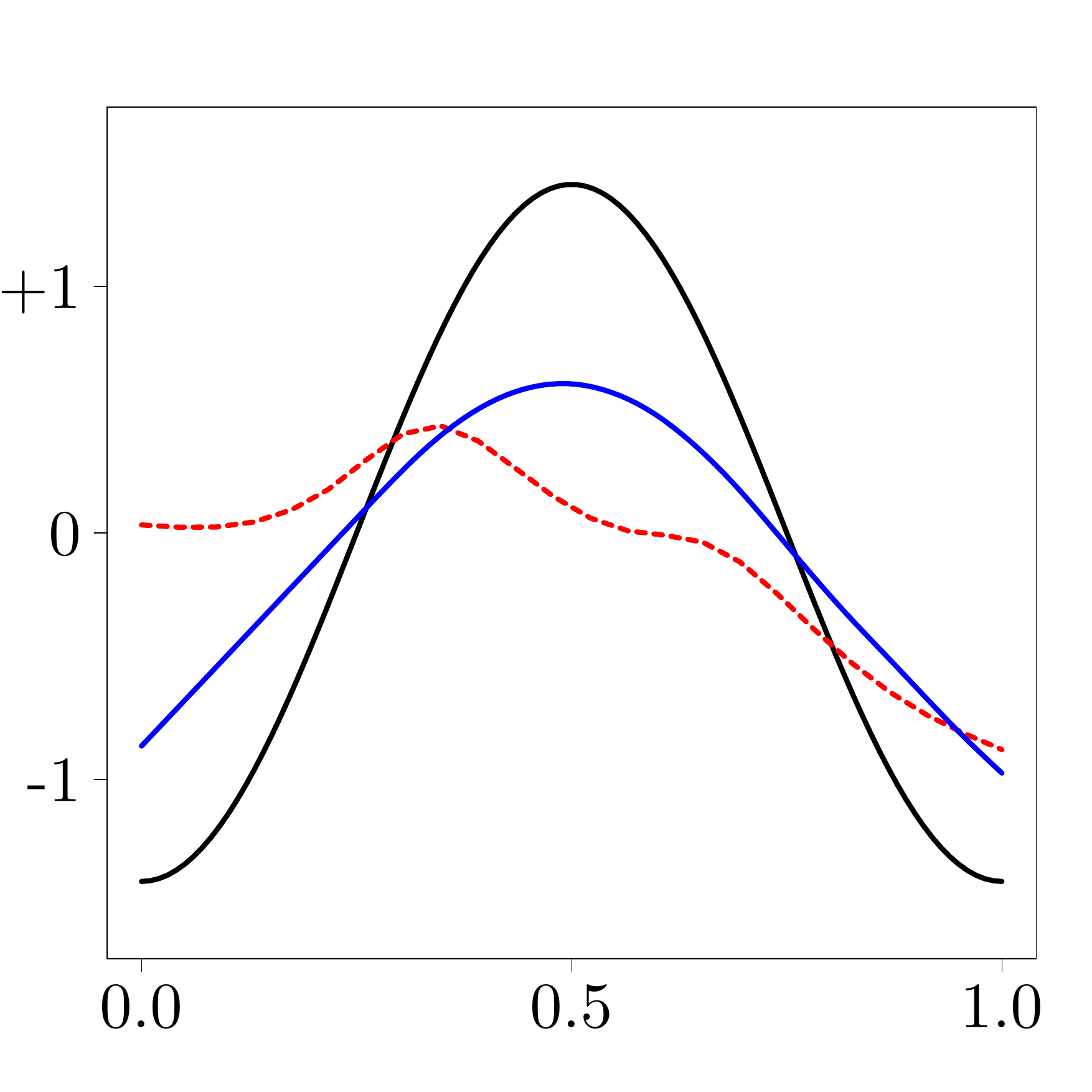} 
		\end{minipage}
		\begin{minipage}[b]{0.25\linewidth}
			\centering
			\includegraphics[width=1\linewidth]{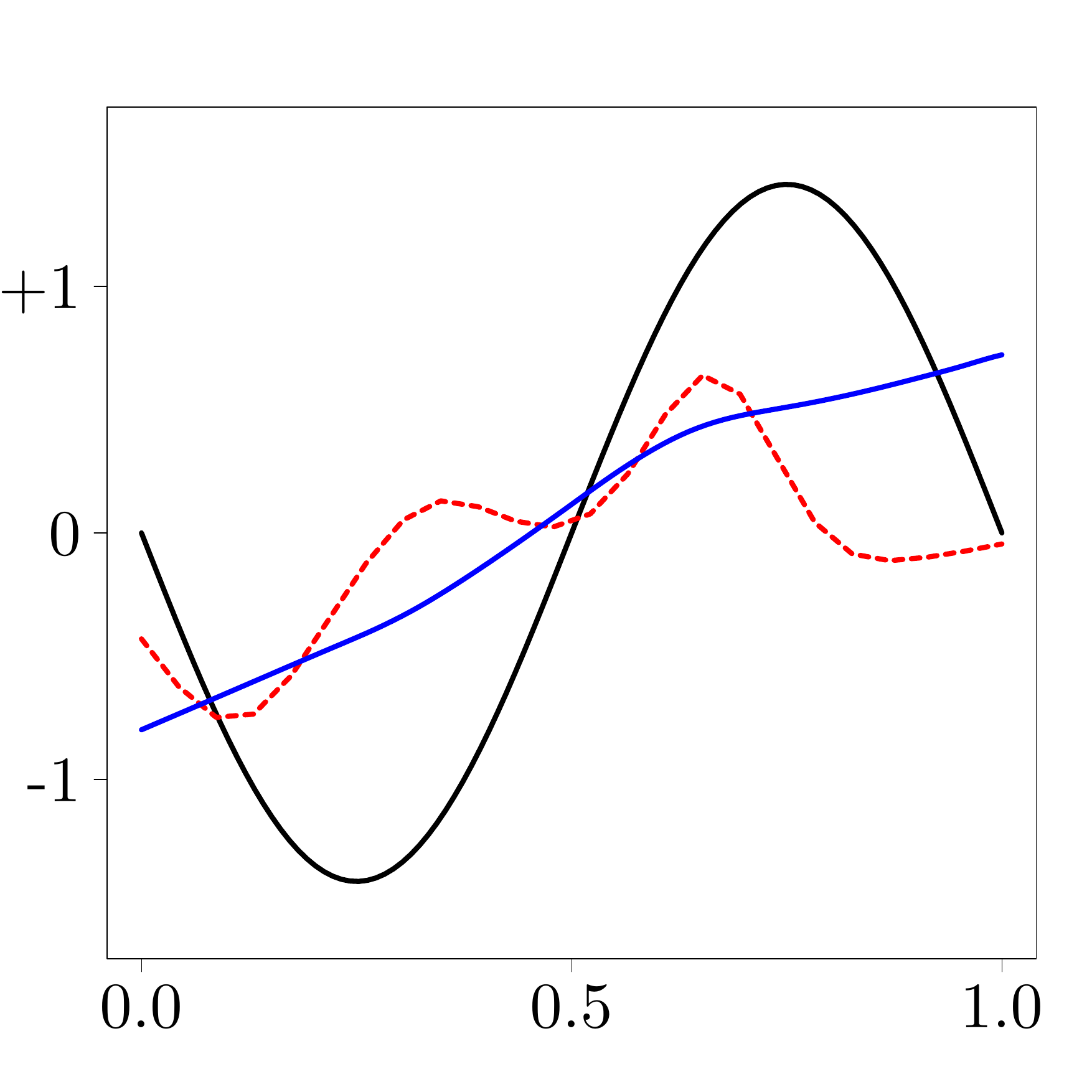} 
		\end{minipage}
		
		\begin{minipage}[b]{0.25\linewidth}
			\centering
			\includegraphics[width=1\linewidth]{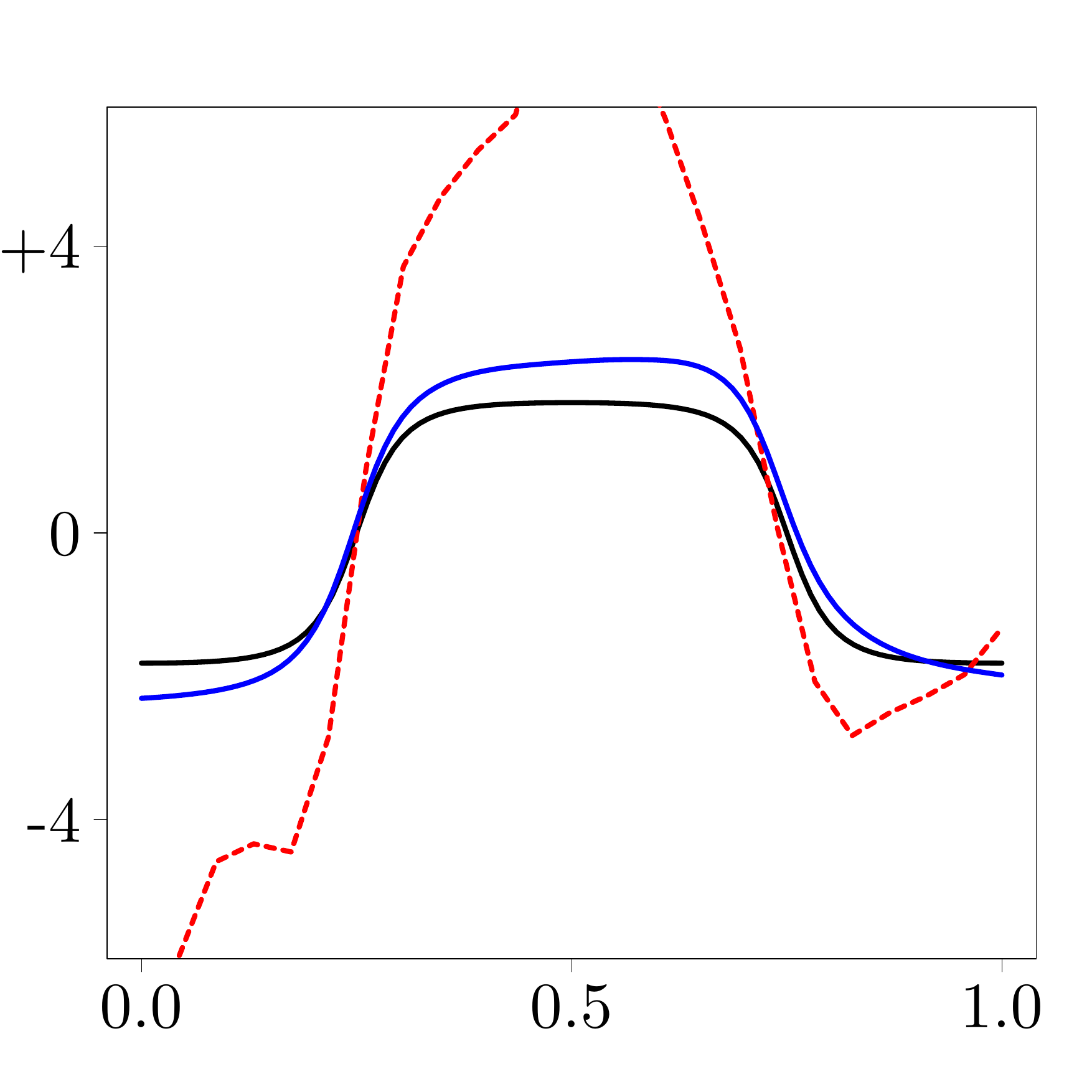} 
		\end{minipage}
		\begin{minipage}[b]{0.25\linewidth}
			\centering
			\includegraphics[width=1\linewidth]{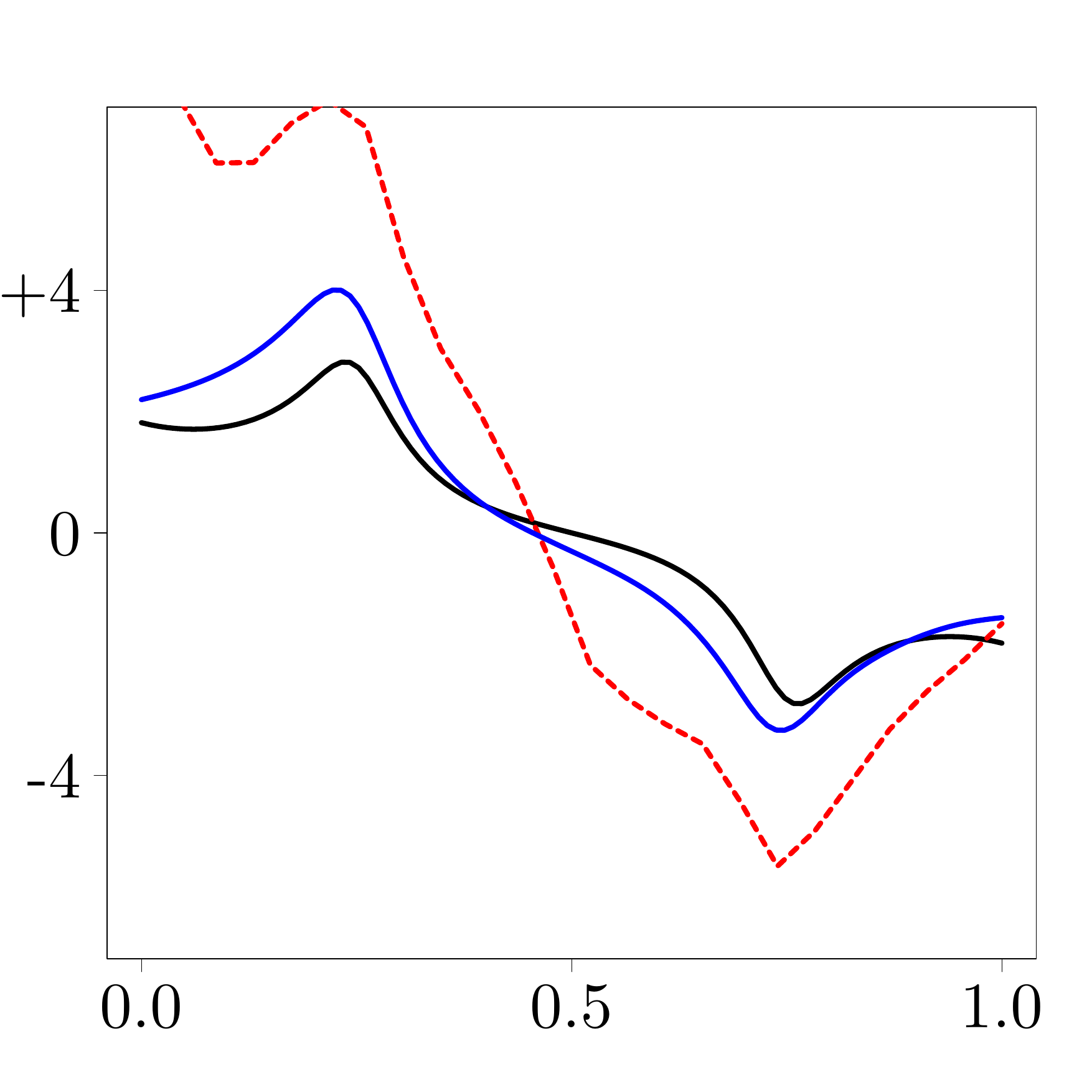} 
		\end{minipage}
		\begin{minipage}[b]{0.25\linewidth}
			\centering
			\includegraphics[width=1\linewidth]{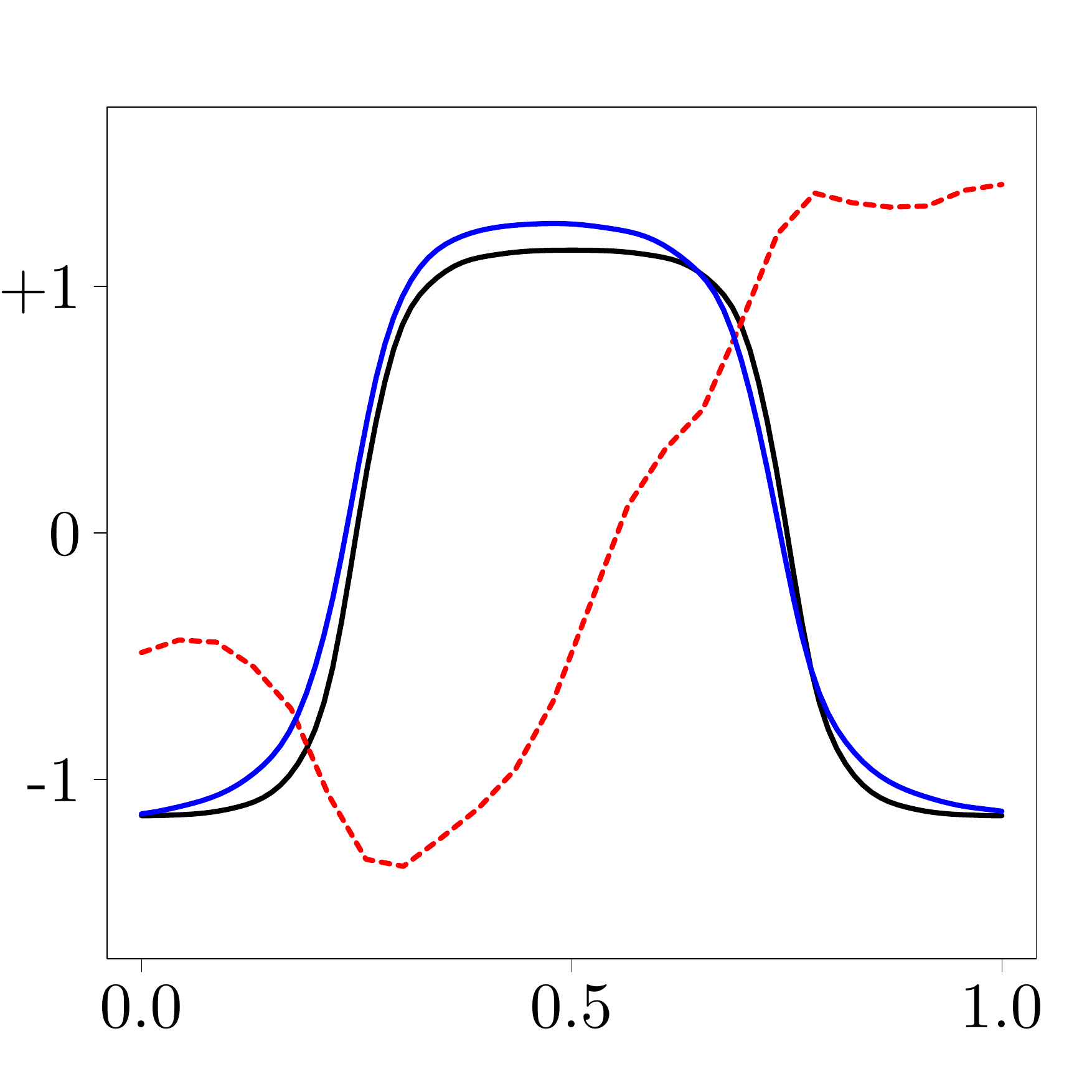} 
		\end{minipage}
		\begin{minipage}[b]{0.25\linewidth}
			\centering
			\includegraphics[width=1\linewidth]{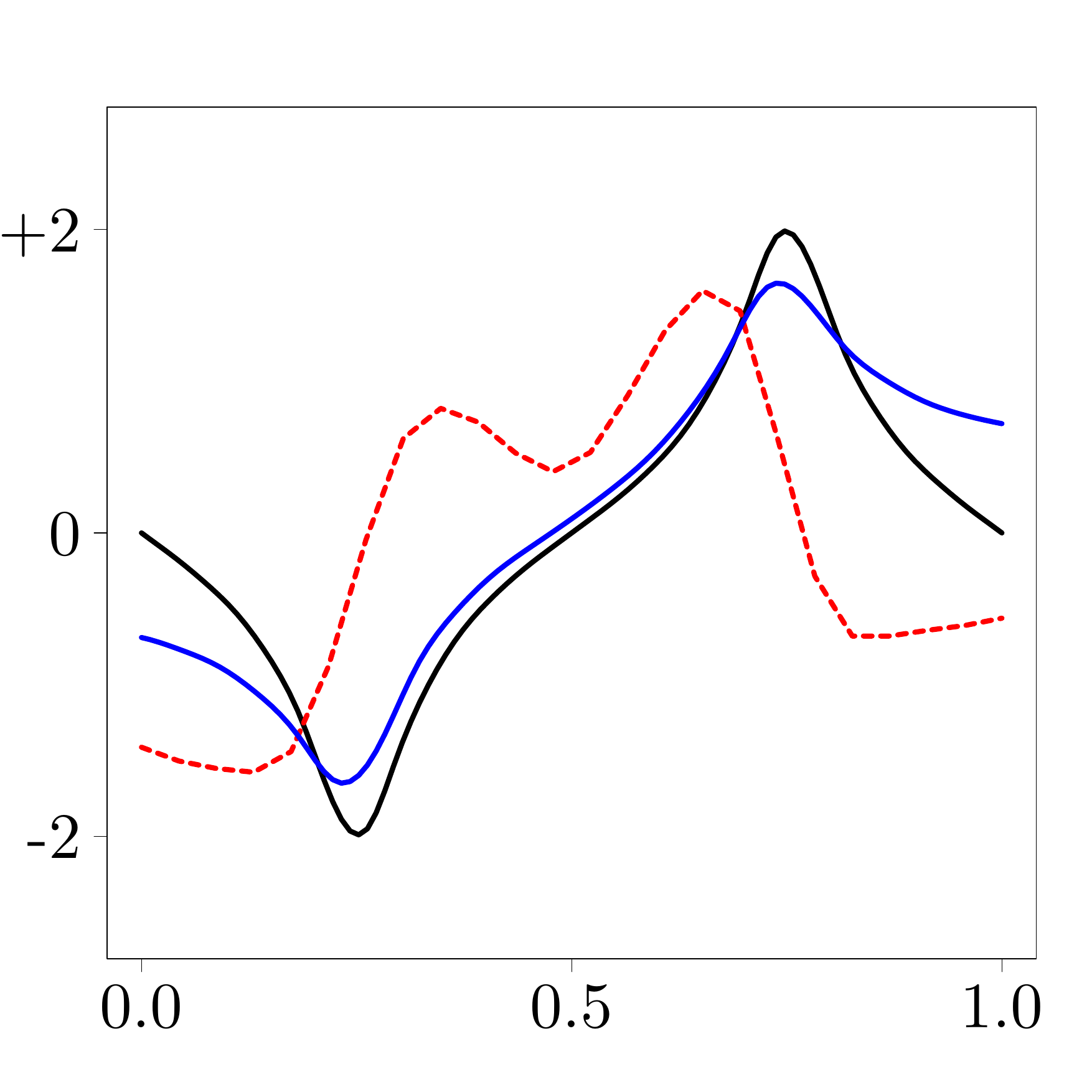} 
		\end{minipage}
		
		\vspace{-1em}
		\begin{minipage}[b]{1\linewidth}
			\centering
			\footnotesize{t}
		\end{minipage}
		
	\end{minipage}%
	\caption{A single run of simulation scenario RT(s). Upper panels: unstandardized regression coefficients and first two principal components; lower panels: standardized regression coefficients and first two principal components. Black solid line: original parameter; blue solid line: AMCEM estimation; red dashed line: \code{pffr} estimation. }
	
	\label{sim:RT1}
\end{figure}

\begin{figure}[tb!]
	
	\begin{minipage}{\dimexpr\linewidth-0cm\relax}%
		\begin{minipage}[b]{1\linewidth}
			\begin{minipage}[b]{0.25\linewidth}
				\centering
				\footnotesize{$\beta_{0}$}
			\end{minipage}
			\begin{minipage}[b]{0.25\linewidth}
				\centering
				\footnotesize{$\beta_{1}$}
			\end{minipage}
			\begin{minipage}[b]{0.25\linewidth}
				\centering
				\footnotesize{PC 1 }
			\end{minipage}
			\begin{minipage}[b]{0.25\linewidth}
				\centering
				\footnotesize{PC 2 }
			\end{minipage}
		\end{minipage}
		
		\begin{minipage}[b]{0.25\linewidth}
			\centering
			\includegraphics[width=1\linewidth]{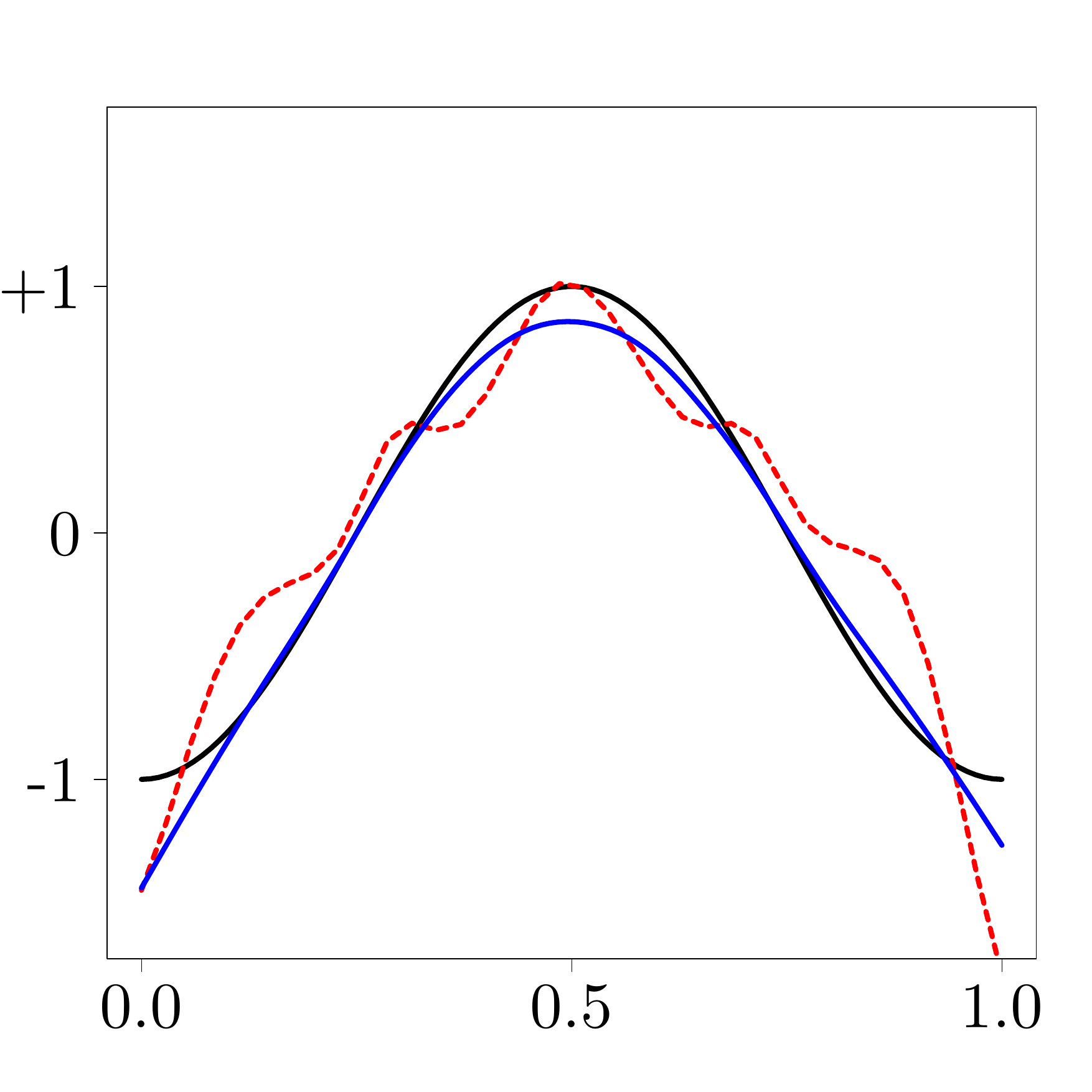} 
		\end{minipage}
		\begin{minipage}[b]{0.25\linewidth}
			\centering
			\includegraphics[width=1\linewidth]{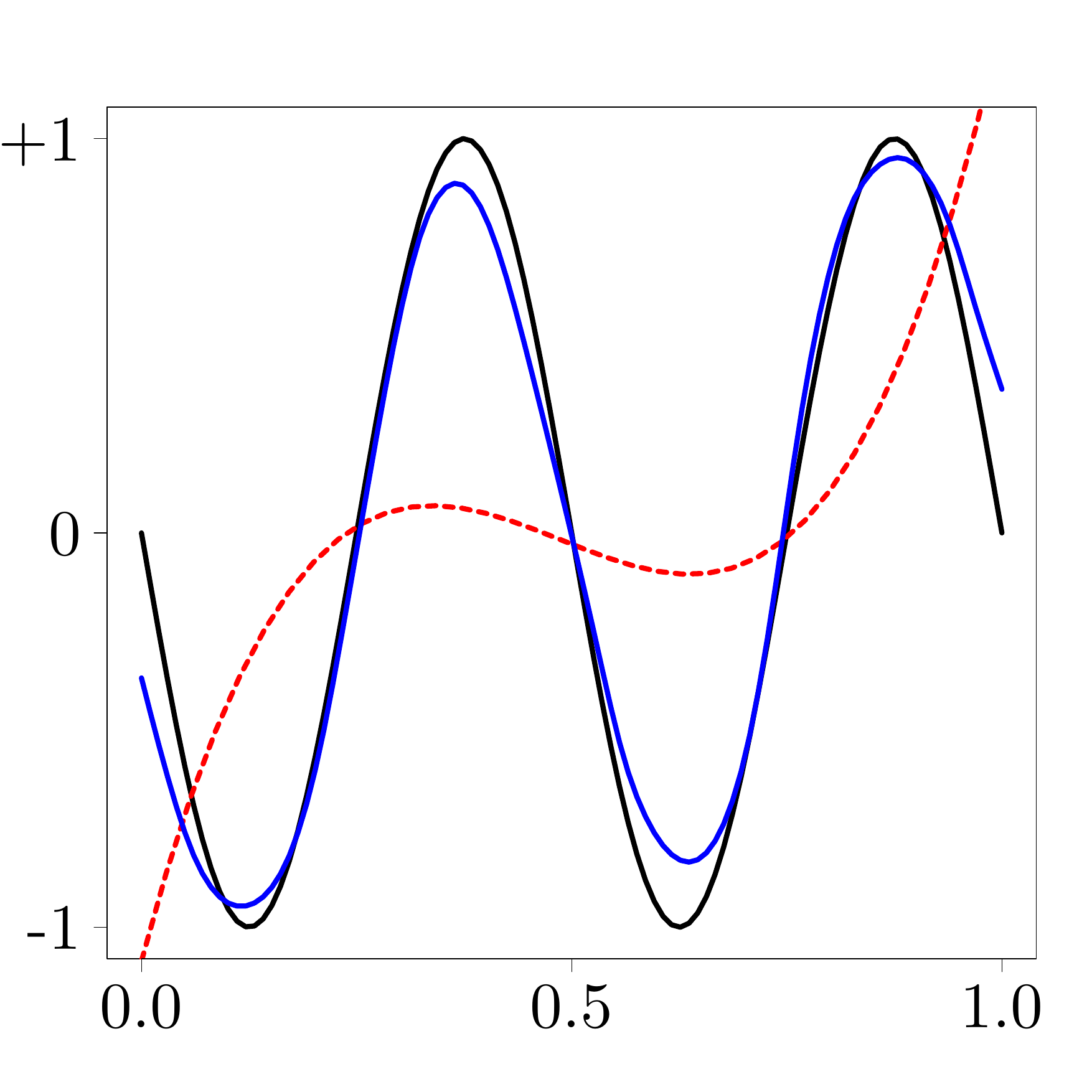} 
		\end{minipage}
		\begin{minipage}[b]{0.25\linewidth}
			\centering
			\includegraphics[width=1\linewidth]{images_simulation/plot_N100_M36_R_complex_reg1_nonstd.pdf} 
		\end{minipage}
		\begin{minipage}[b]{0.25\linewidth}
			\centering
			\includegraphics[width=1\linewidth]{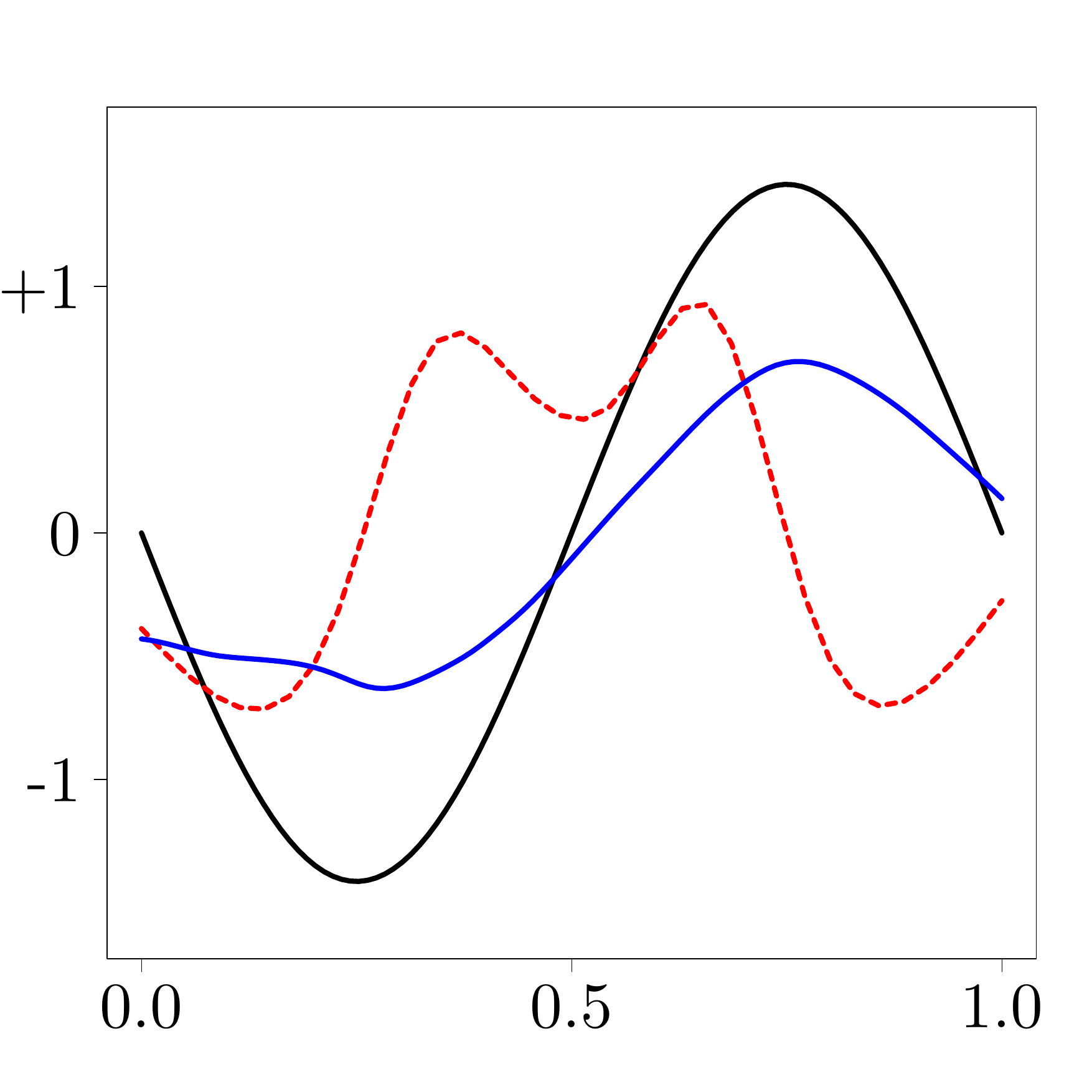} 
		\end{minipage}
		
		\begin{minipage}[b]{0.25\linewidth}
			\centering
			\includegraphics[width=1\linewidth]{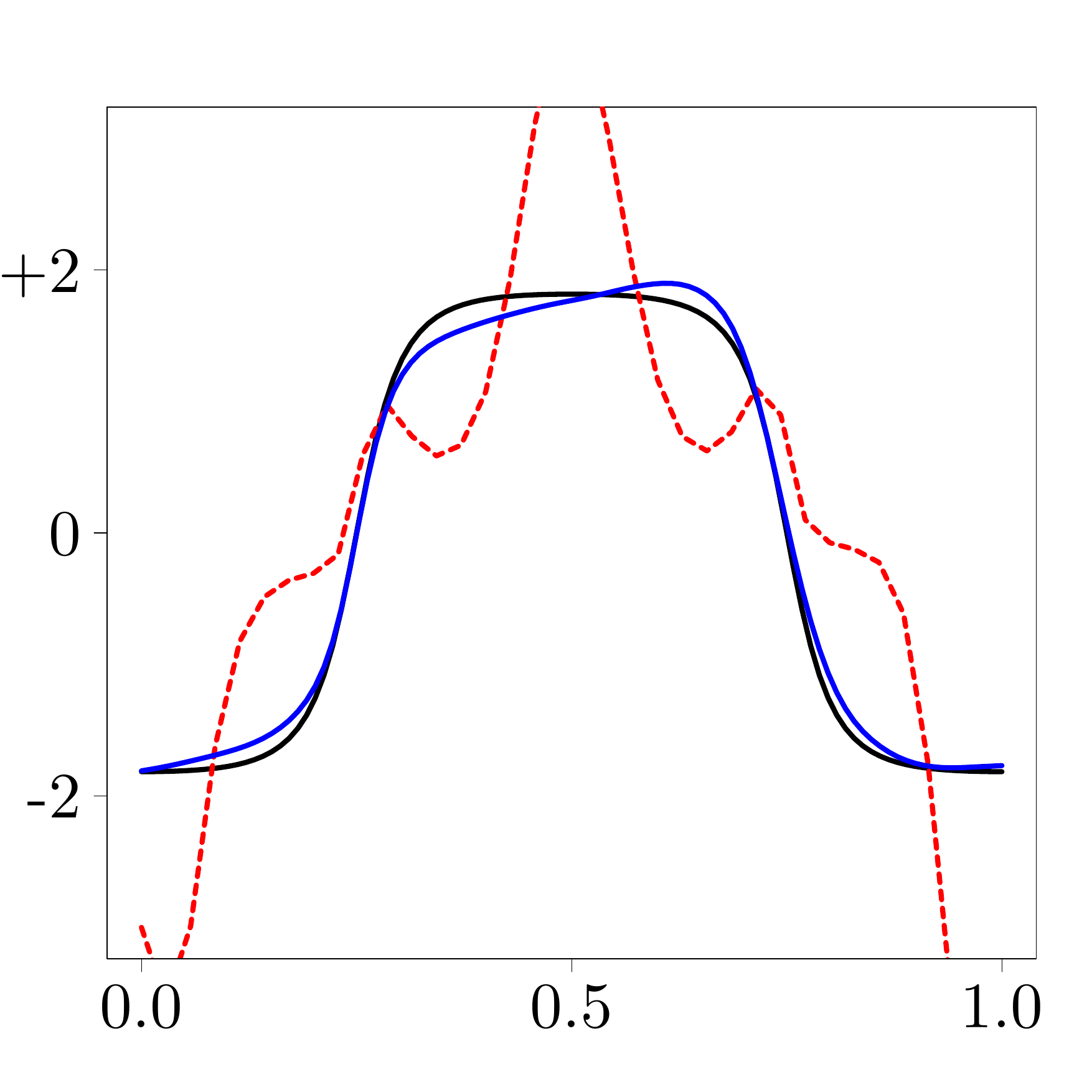} 
		\end{minipage}
		\begin{minipage}[b]{0.25\linewidth}
			\centering
			\includegraphics[width=1\linewidth]{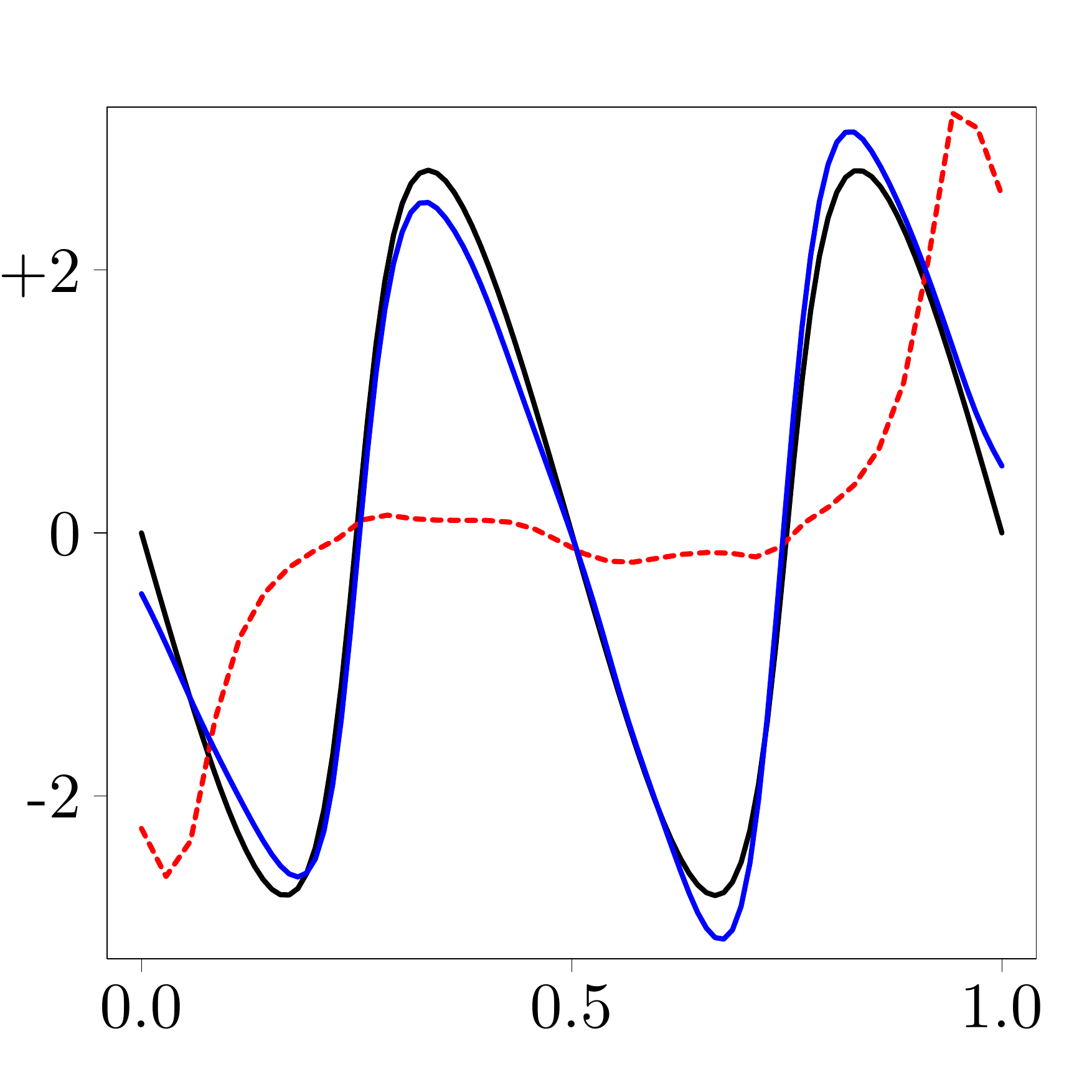} 
		\end{minipage}
		\begin{minipage}[b]{0.25\linewidth}
			\centering
			\includegraphics[width=1\linewidth]{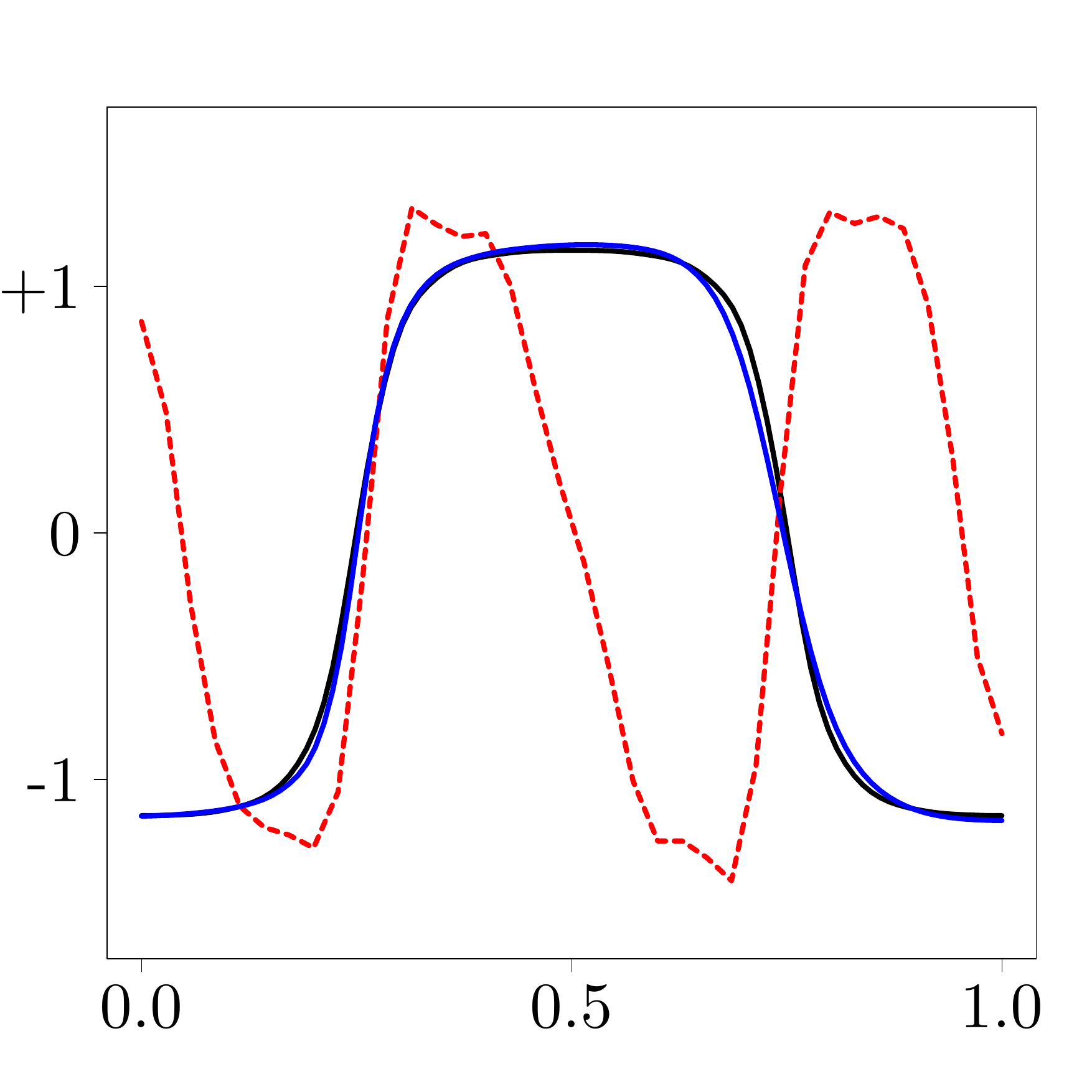} 
		\end{minipage}
		\begin{minipage}[b]{0.25\linewidth}
			\centering
			\includegraphics[width=1\linewidth]{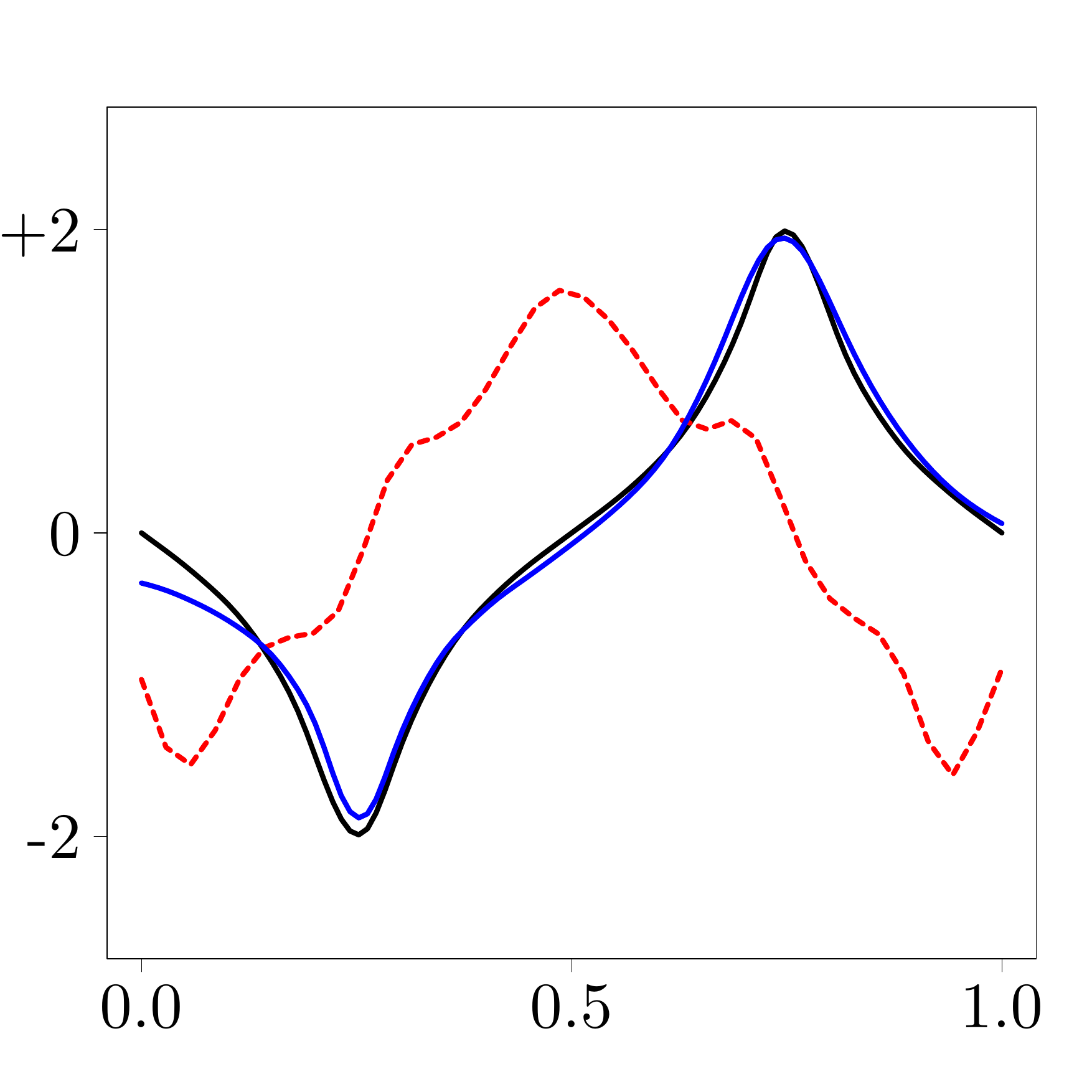} 
		\end{minipage}
		
		\vspace{-1em}
		\begin{minipage}[b]{1\linewidth}
			\centering
			\footnotesize{t}
		\end{minipage}
		
	\end{minipage}%
	\caption{A single run of simulation scenario R(c). Upper panels: unstandardized regression coefficients and first two principal components; lower panels: standardized regression coefficients and first two principal components. Black solid line: original parameter; blue solid line: the AMCEM estimation; red dashed line: \code{pffr} estimation.}
	
	\label{sim:R2}
\end{figure}

\begin{figure}[tb!]
	\vspace*{.5cm}
	\begin{minipage}{\dimexpr\linewidth-0cm\relax}%
		\begin{minipage}[b]{1\linewidth}
			\begin{minipage}[b]{0.25\linewidth}
				\centering
				\footnotesize{$\beta_{0}$}
			\end{minipage}
			\begin{minipage}[b]{0.25\linewidth}
				\centering
				\footnotesize{$\beta_{1}$}
			\end{minipage}
			\begin{minipage}[b]{0.25\linewidth}
				\centering
				\footnotesize{PC 1 }
			\end{minipage}
			\begin{minipage}[b]{0.25\linewidth}
				\centering
				\footnotesize{PC 2 }
			\end{minipage}
		\end{minipage}
		
		\begin{minipage}[b]{0.25\linewidth}
			\centering
			\includegraphics[width=1\linewidth]{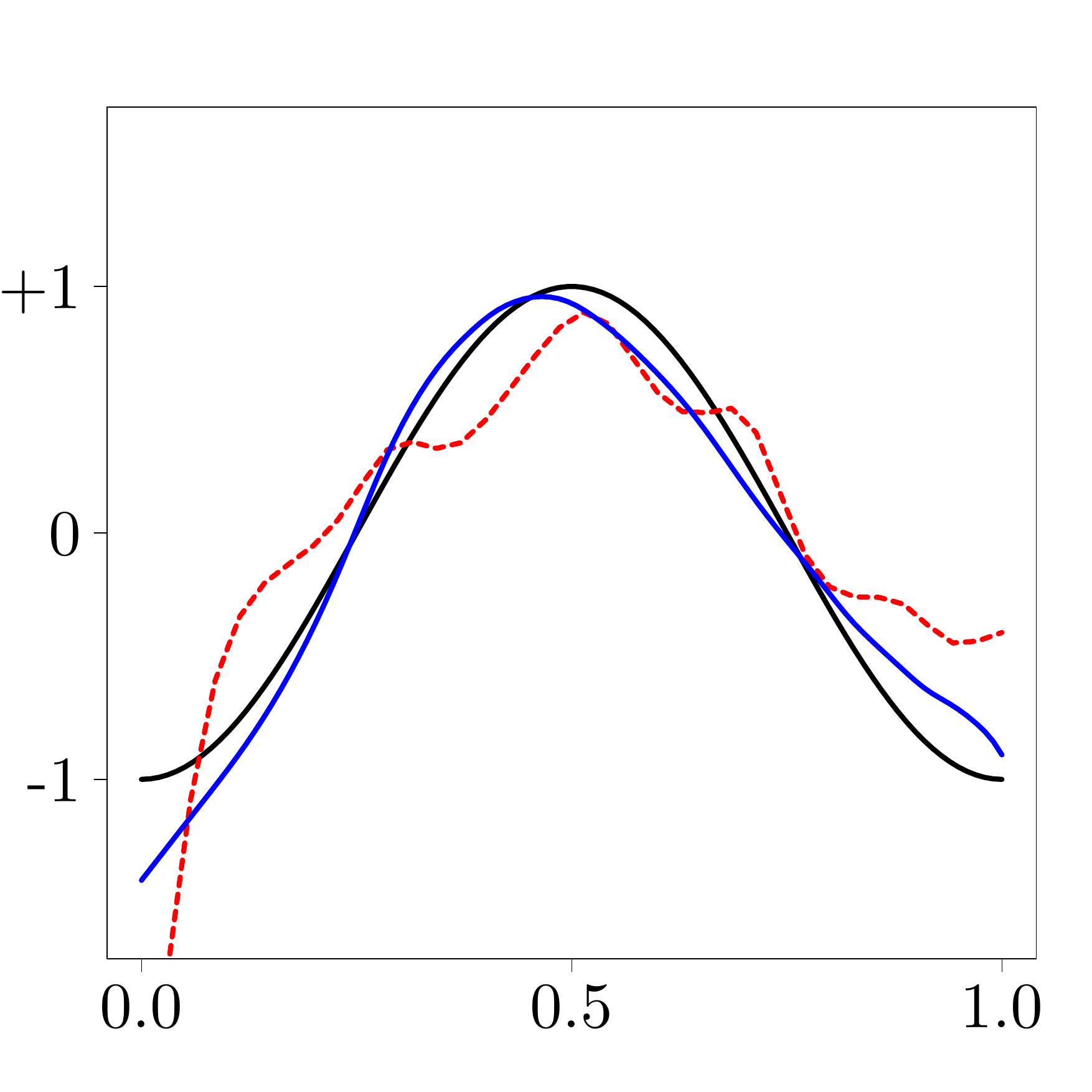} 
		\end{minipage}
		\begin{minipage}[b]{0.25\linewidth}
			\centering
			\includegraphics[width=1\linewidth]{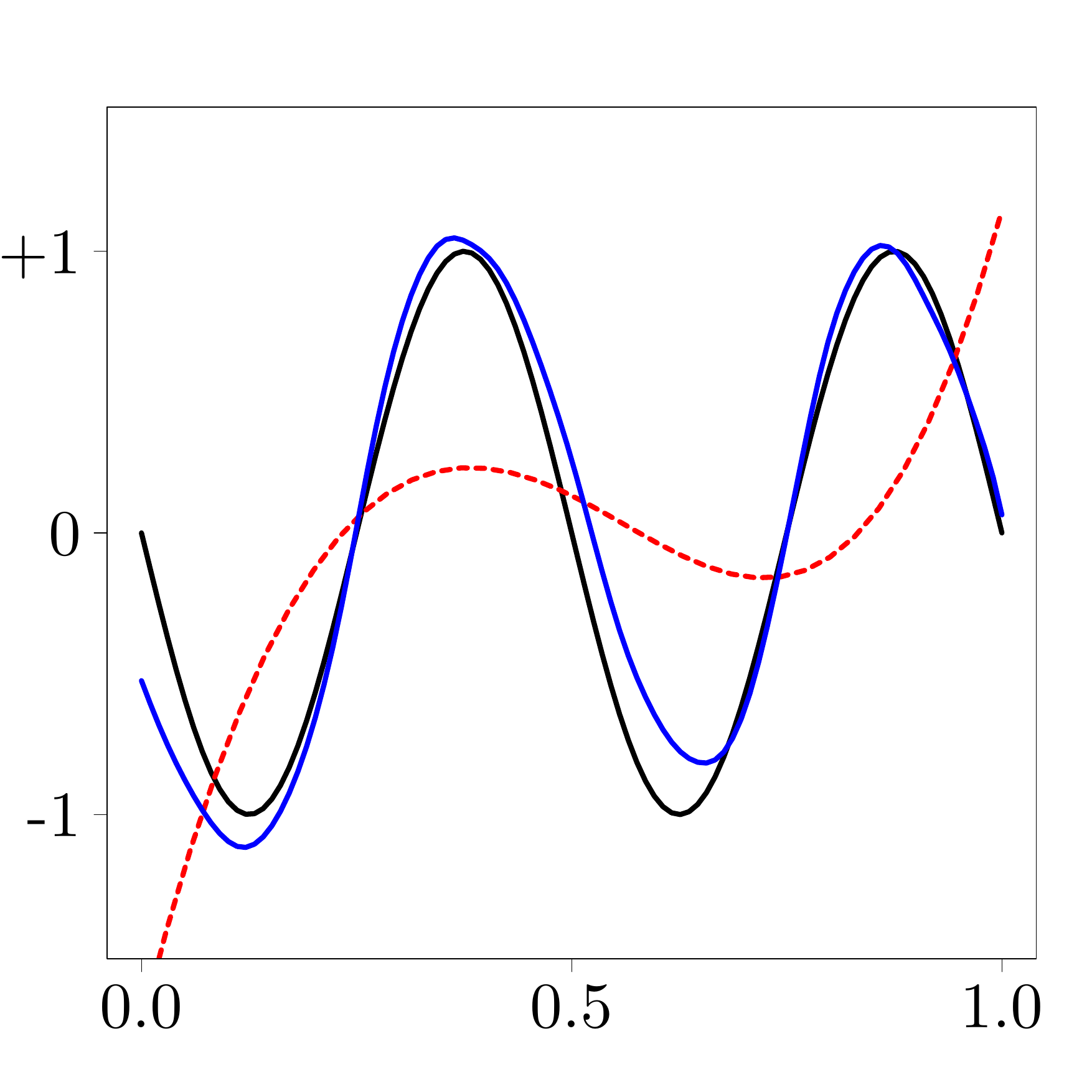} 
		\end{minipage}
		\begin{minipage}[b]{0.25\linewidth}
			\centering
			\includegraphics[width=1\linewidth]{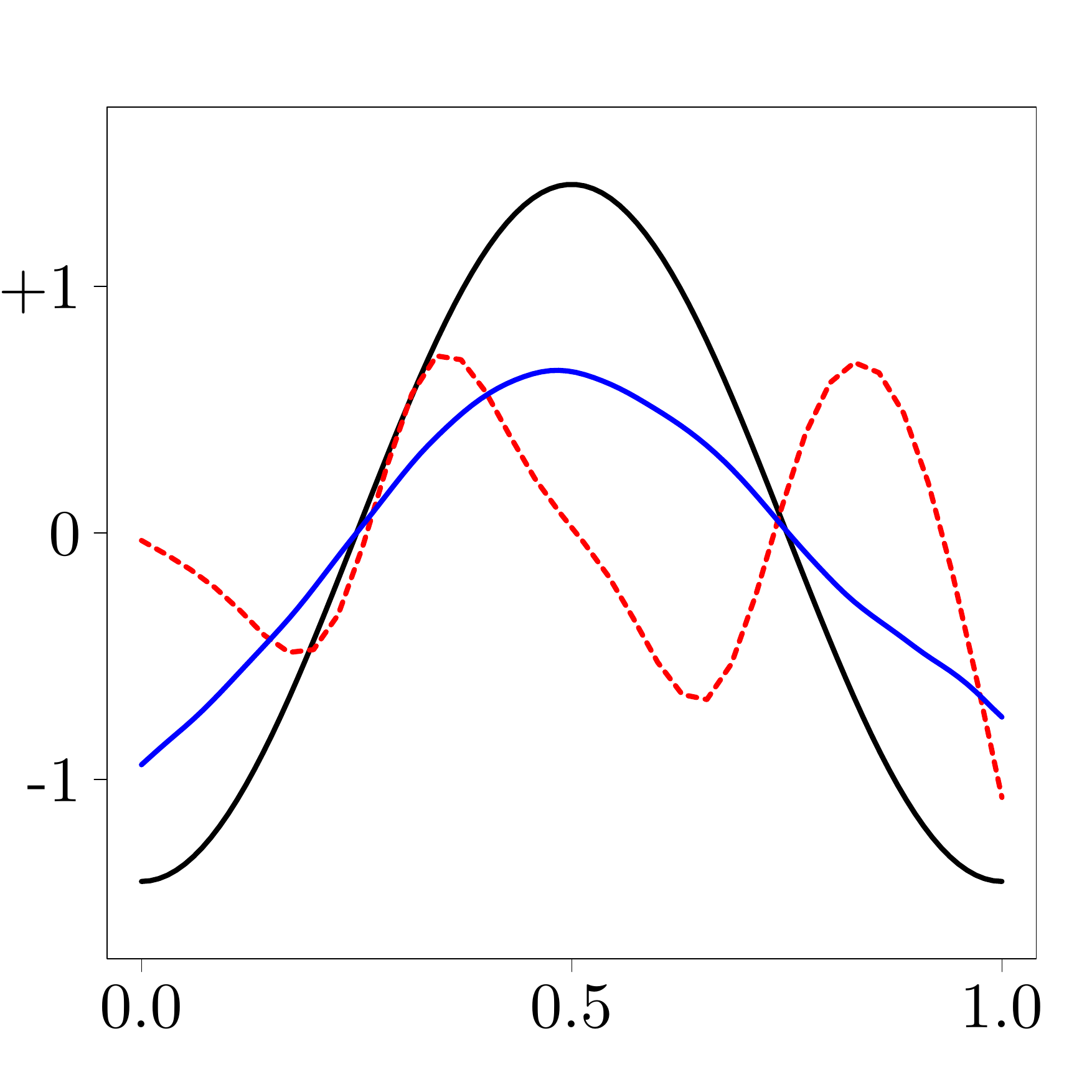} 
		\end{minipage}
		\begin{minipage}[b]{0.25\linewidth}
			\centering
			\includegraphics[width=1\linewidth]{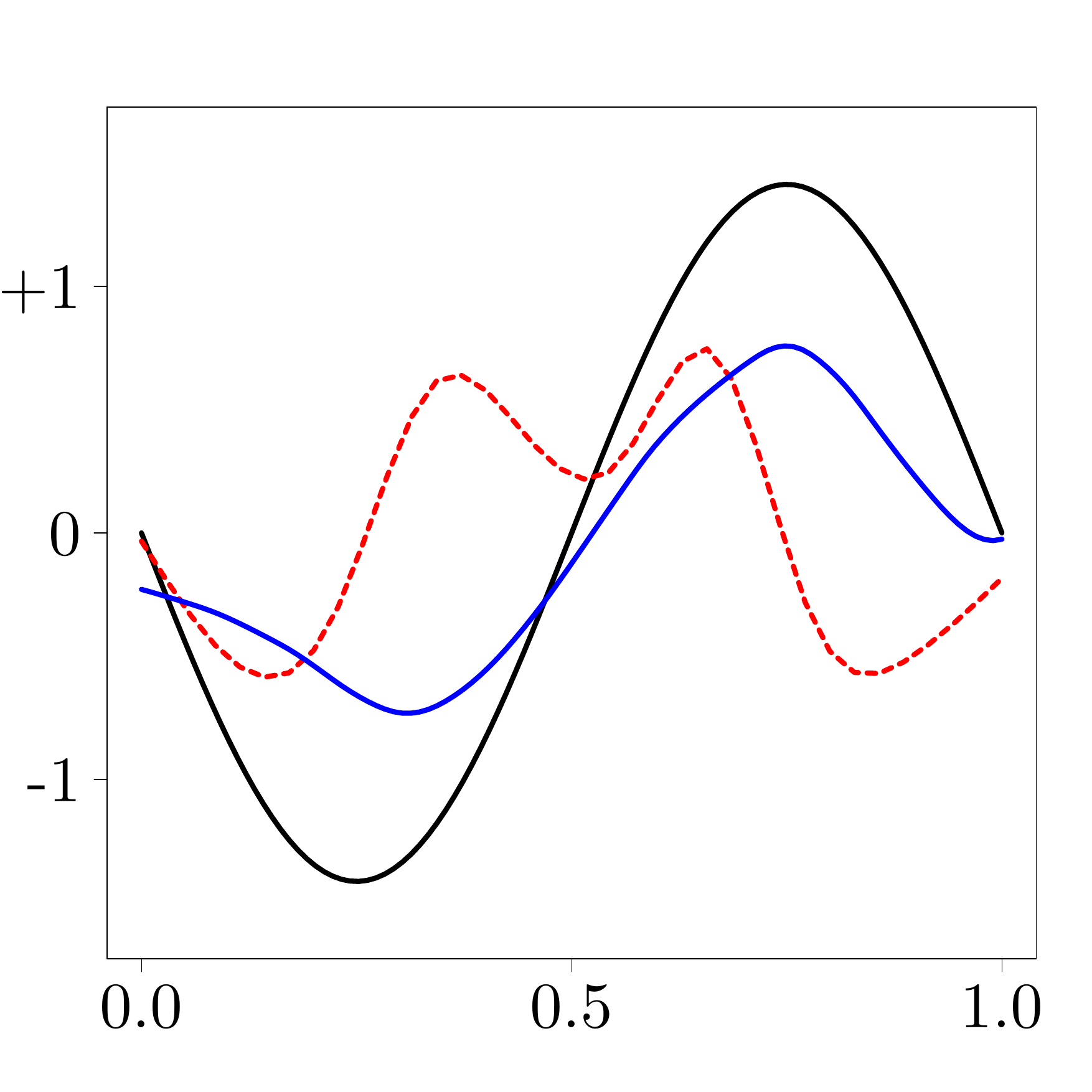} 
		\end{minipage}
		
		\begin{minipage}[b]{0.25\linewidth}
			\centering
			\includegraphics[width=1\linewidth]{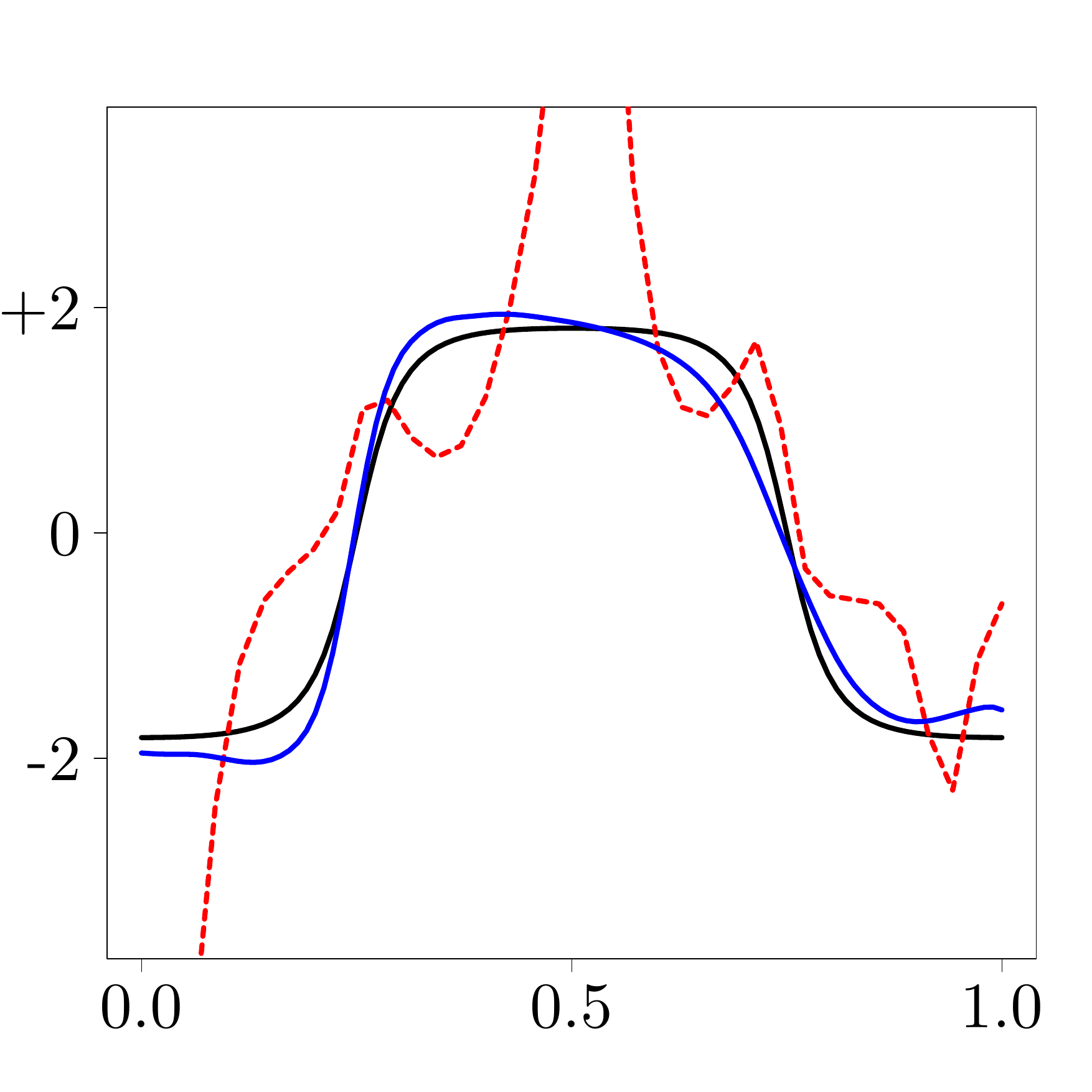} 
		\end{minipage}
		\begin{minipage}[b]{0.25\linewidth}
			\centering
			\includegraphics[width=1\linewidth]{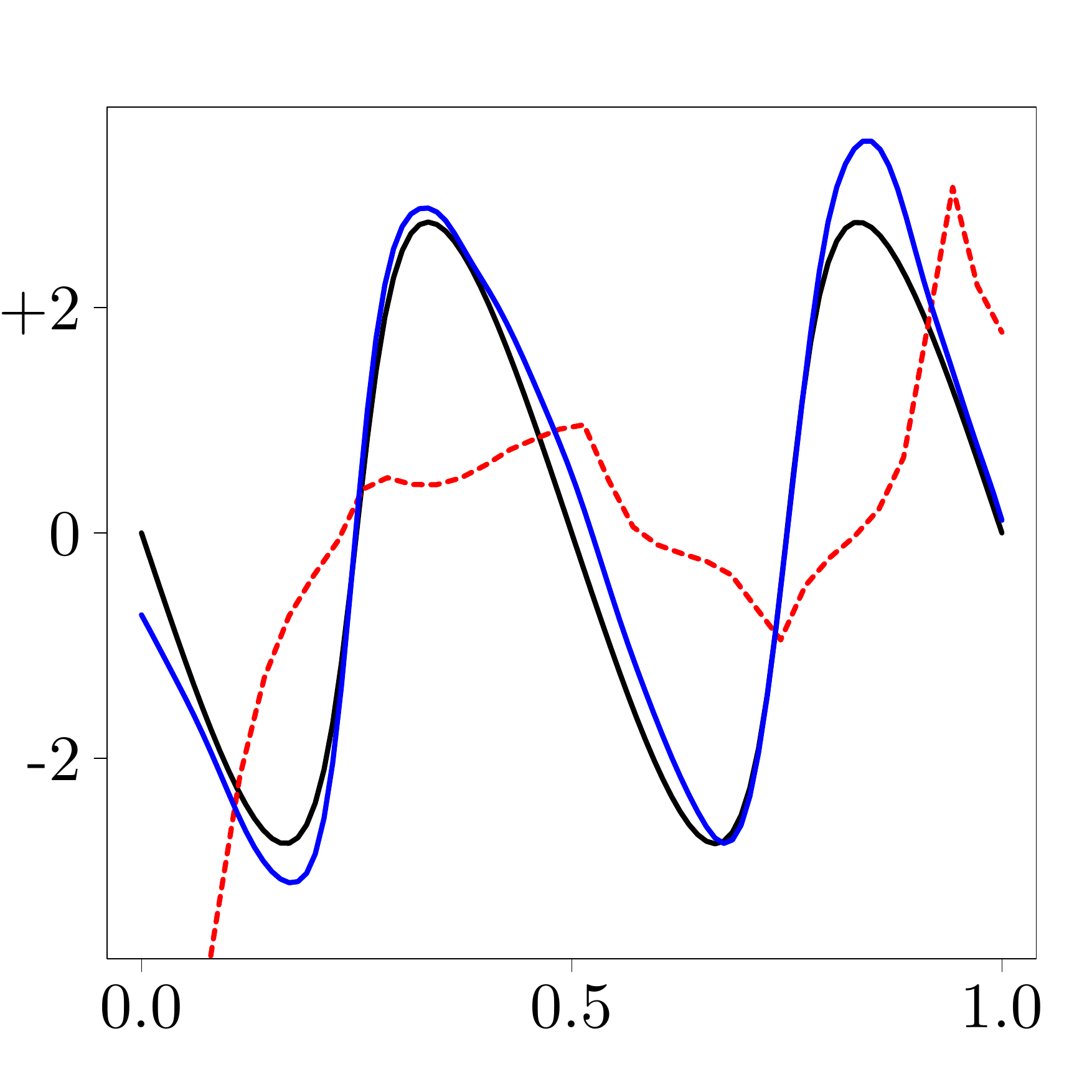} 
		\end{minipage}
		\begin{minipage}[b]{0.25\linewidth}
			\centering
			\includegraphics[width=1\linewidth]{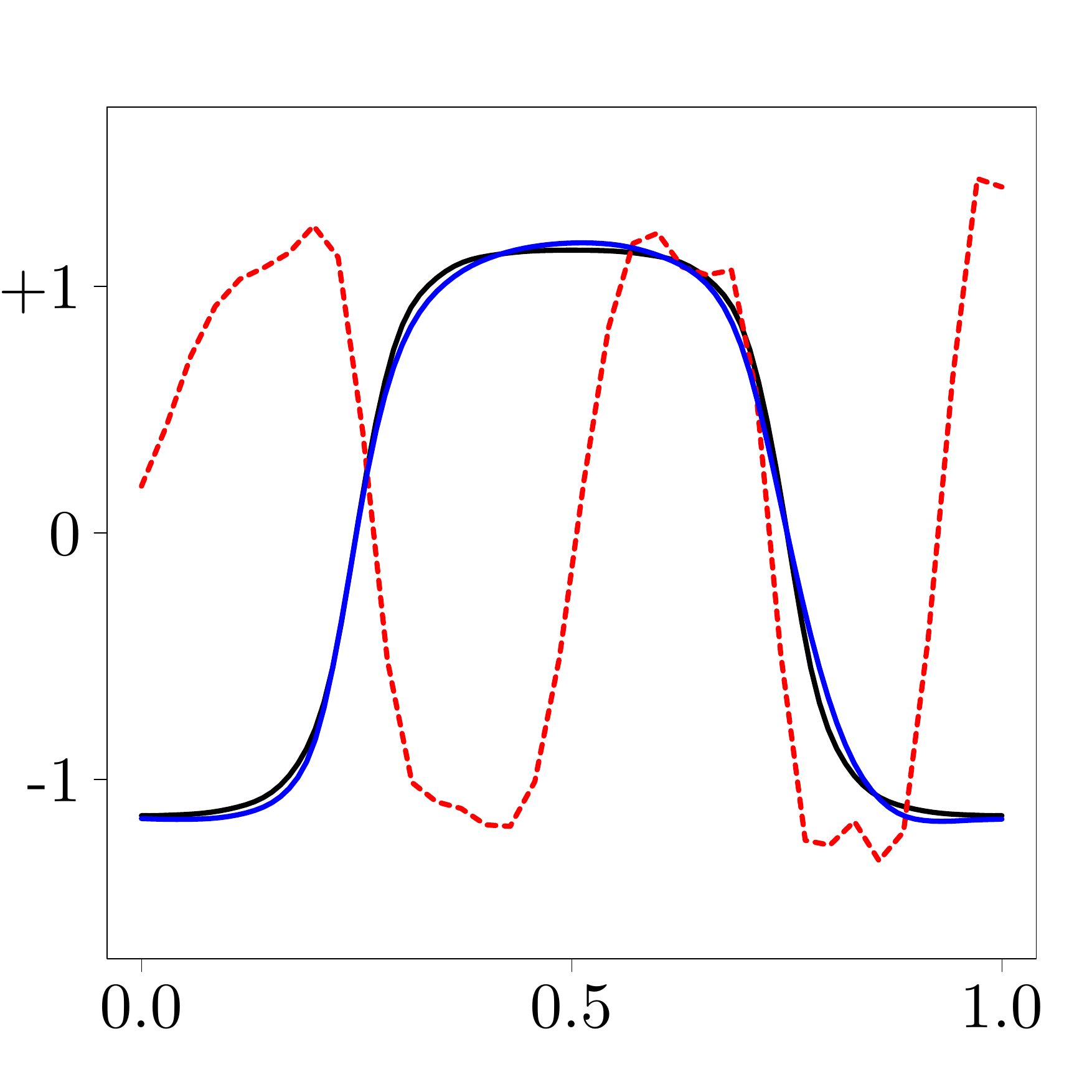} 
		\end{minipage}
		\begin{minipage}[b]{0.25\linewidth}
			\centering
			\includegraphics[width=1\linewidth]{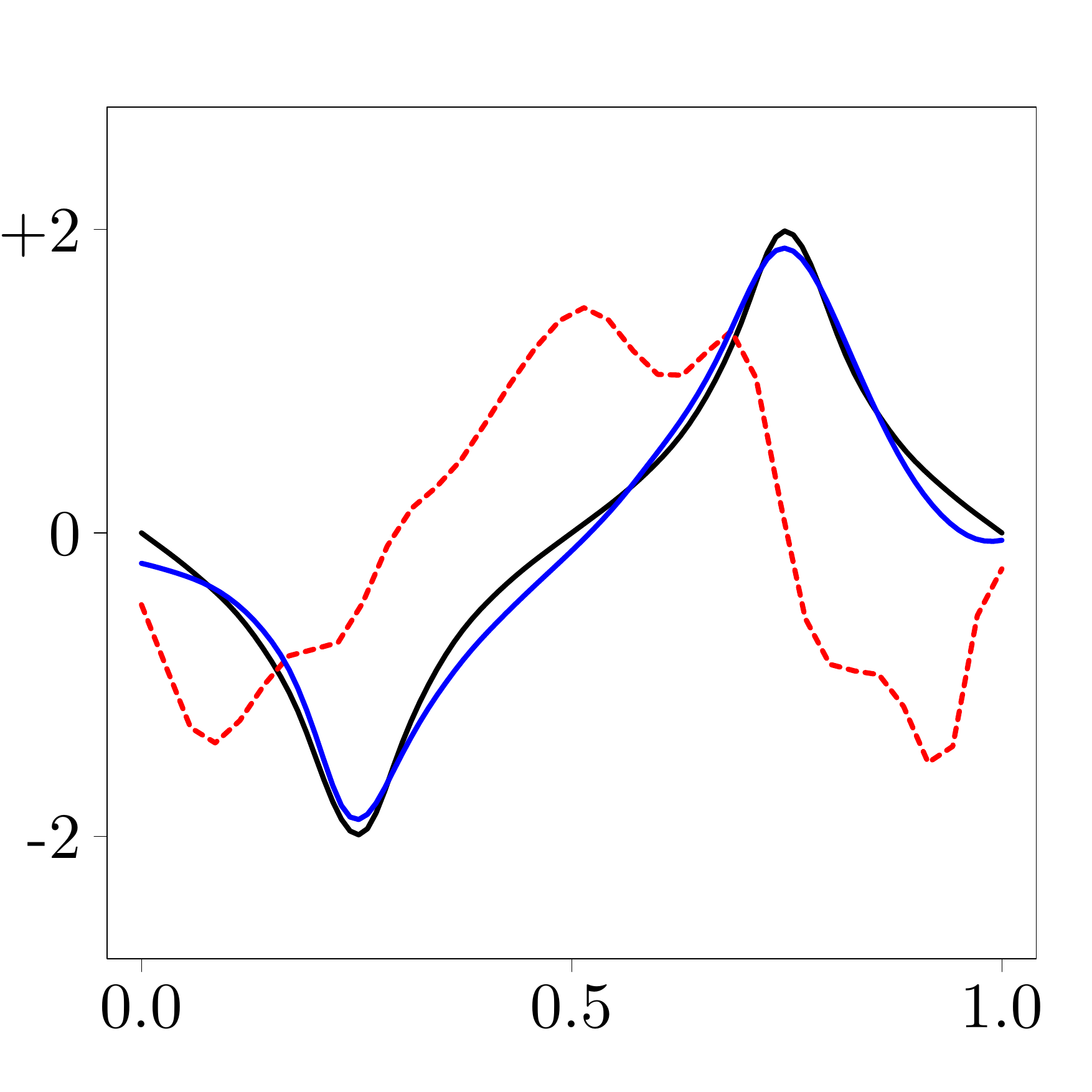} 
		\end{minipage}
		
		\vspace{-1em}
		\begin{minipage}[b]{1\linewidth}
			\centering
			\footnotesize{t}
		\end{minipage}
		
	\end{minipage}%
	\caption{A single run of simulation scenario RT(c). Upper panels: unstandardized regression coefficients and first two principal components; lower panels: standardized regression coefficients and first two principal components. Black solid line: original parameter; blue solid line: the AMCEM estimation; red dashed line: \code{pffr} estimation.}
	
	\label{sim:RT2}
\end{figure}

To provide a better comparison of AMCEM and \code{pffr}, the unstandardized and standardized estimations of parameters  are illustrated in Figures \ref{sim:R1} and \ref{sim:RT2} for a single run of the simulation study.
In the caption of each figure, a short combination of letters  is given to indicate the combination of sampling design and complexity of regression coefficients the figure refers to. For example, R(s) denotes the case of design (R) and `simple' regression coefficients, while RT(c) stands for the case of design (RT) and `complex' regression coefficients. The results of a single run of the simulations R(s) and  RT(s)  with $N=100$, $M=24$, $\sigma^2=0.1$ and $\rho=0.1$ are shown in Figure \ref{sim:R1} and Figure \ref{sim:RT1}, respectively. Similarly, a single run of the simulations R(c) and  RT(c)  with $N=100$, $M=36$, $\sigma^2=0.1$ and $\rho=0.1$ are shown in Figure \ref{sim:R2} and Figure \ref{sim:RT2}, respectively.

The first row of panels in each figure shows the unstandardized functional parameters, while the second row shows the standardized parameters. 
Since the unstandardized regression coefficients and the covariance operator are not identifiable, the functions illustrated in the first row are multiplied by a constant scalar value. The constant scalar is selected so that the unstandardized $\beta_0$ are similar in their $L^2$-norm. 
For comparing the performance of AMCEM and \code{pffr},  we need to consider the standardized functional parameters, which are uniquely identifiable. To obtain a standardized version of \code{pffr}  estimations, we need to first obtain an estimation of the covariance function, which is not provided directly by \code{pffr}. To fix this drawback, we used the residuals provided by \code{pffr} and the function \code{fpca.sc} in the package \code{refund}.

It can be noticed that in all simulation scenarios shown in the figures, e.g., scenario R(s), RT(s), R(c), and RT(c) with the prespecified aforementioned simulation parameters, the unstandardized estimations provided by AMCEM are smoother than the ones obtained via \code{pffr}. Furthermore, the standardized estimations provided by AMCEM are more accurate. 

The results given in Table \ref{t.R} and Table \ref{t.RT} show that the MSE of the estimations obtained both via AMCEM and via \code{pffr} decrease as the parameters $N$ and $M$ increase, as reasonably expected. Moreover, the precision of the estimation of the covariance operator (including the eigen-values and eigen-functions) decreases when the value of $\sigma^2$ increases. This implies that the accuracy of AMCEM in estimating the covariance operator decreases (increases) as the total variation of the outcome is less (more) explained by the functional part of the model.


\begin{table}[b!]
	\centering
	\caption{Mean square error of estimation of parameters ($\times 10^{4}$) with $\rho=0.1$. }
	\label{t.R}
	\resizebox{\textwidth}{!}{
		\begin{tabular}{cccclccccccclcc}
			\multicolumn{4}{c}{}                                                                       &   & \multicolumn{7}{c}{AMCEM}                                                                  &   & \multicolumn{2}{c}{PFFR}  \\
			Design\vspace*{0.3em}                & N                    & M                    & $\sigma^{2}$         & ~ & $\beta_{0}$ & $\beta_{1}$ & $\psi_{1}$ & $\psi_{2}$ & $\nu_{1}$ & $\nu_{2}$ & $\sigma^{2}$ & ~ & $\beta_{0}$ & $\beta_{1}$ \\
			R                     & 50                   & 12                   & 0.2                  &   & 3.256       & 5.540       & 1.039      & 2.626      & 0.085     & 0.080     & 0.144        &   & 213.466     & 543.760     \\
			&                      &                      & 0.8                  &   & 3.141       & 5.340       & 1.327      & 3.838      & 0.184     & 0.062     & 2.322        &   & 118.373     & 399.279     \\
			&                      & 36                   & 0.2                  &   & 2.915       & 4.434       & 0.773      & 4.106      & 0.212     & 0.048     & 0.297        &   & 105.625     & 403.354     \\
			&                      &                      & 0.8                  &   & 3.462       & 5.130       & 1.337      & 6.155      & 1.367     & 0.042     & 8.624        &   & 64.458      & 317.890     \\
			& 100                  & 12                   & 0.2                  &   & 1.382       & 2.399       & 0.589      & 1.172      & 0.058     & 0.032     & 0.052        &   & 132.097     & 428.435     \\
			&                      &                      & 0.8                  &   & 1.437       & 2.455       & 0.643      & 1.534      & 0.101     & 0.033     & 0.668        &   & 79.790      & 343.974     \\
			&                      & 36                   & 0.2                  &   & 1.139       & 1.823       & 0.245      & 1.465      & 0.068     & 0.021     & 0.092        &   & 80.196      & 353.965     \\
			&                      &                      & 0.8                  &   & 1.192       & 1.843       & 0.315      & 1.752      & 0.217     & 0.025     & 1.645        &   & 47.118      & 292.963     \\
			\multicolumn{1}{l}{~} & \multicolumn{1}{l}{} & \multicolumn{1}{l}{} & \multicolumn{1}{l}{} &   &             &             &            &            &           &           &              &   &             &             \\
			RT                    & 50                   & 12                   & 0.2                  &   & 3.552       & 6.678       & 1.145      & 2.928      & 0.074     & 0.070     & 0.148        &   & 195.507     & 522.159     \\
			&                      &                      & 0.8                  &   & 3.597       & 6.342       & 1.260      & 4.139      & 0.125     & 0.056     & 2.043        &   & 123.039     & 400.171     \\
			&                      & 36                   & 0.2                  &   & 3.227       & 4.804       & 0.892      & 4.252      & 0.229     & 0.054     & 0.290        &   & 133.905     & 435.993     \\
			&                      &                      & 0.8                  &   & 3.446       & 4.985       & 1.243      & 6.806      & 1.130     & 0.041     & 8.008        &   & 65.659      & 322.279     \\
			& 100                  & 12                   & 0.2                  &   & 1.406       & 2.703       & 0.670      & 1.472      & 0.070     & 0.041     & 0.060        &   & 134.882     & 439.225     \\
			&                      &                      & 0.8                  &   & 1.709       & 2.856       & 0.757      & 1.681      & 0.121     & 0.044     & 0.809        &   & 78.686      & 337.273     \\
			&                      & 36                   & 0.2                  &   & 1.023       & 1.795       & 0.256      & 1.345      & 0.056     & 0.020     & 0.087        &   & 76.791      & 347.971     \\
			&                      &                      & 0.8                  &   & 1.257       & 1.940       & 0.347      & 1.836      & 0.208     & 0.017     & 1.403        &   & 50.760      & 296.436     \\
			\multicolumn{1}{l}{~} & \multicolumn{1}{l}{} & \multicolumn{1}{l}{} & \multicolumn{1}{l}{} &   &             &             &            &            &           &           &              &   &             &             \\
			RM                    & 50                   & 12                   & 0.2                  &   & 3.489       & 6.289       & 1.200      & 3.025      & 0.090     & 0.088     & 0.134        &   & 212.224     & 560.007     \\
			&                      &                      & 0.8                  &   & 3.535       & 5.946       & 1.433      & 3.948      & 0.172     & 0.051     & 2.433        &   & 116.894     & 398.856     \\
			&                      & 36                   & 0.2                  &   & 3.168       & 4.630       & 0.894      & 4.399      & 0.252     & 0.062     & 0.308        &   & 128.676     & 421.158     \\
			&                      &                      & 0.8                  &   & 3.262       & 5.263       & 1.280      & 7.158      & 1.302     & 0.042     & 8.770        &   & 69.891      & 328.868     \\
			& 100                  & 12                   & 0.2                  &   & 1.408       & 2.511       & 0.646      & 1.270      & 0.073     & 0.038     & 0.059        &   & 132.515     & 431.424     \\
			&                      &                      & 0.8                  &   & 1.547       & 2.827       & 0.729      & 1.826      & 0.109     & 0.039     & 0.814        &   & 77.409      & 337.420     \\
			&                      & 36                   & 0.2                  &   & 1.152       & 1.838       & 0.283      & 1.623      & 0.062     & 0.023     & 0.092        &   & 84.342      & 356.910     \\
			&                      &                      & 0.8                  &   & 1.204       & 1.891       & 0.350      & 1.763      & 0.214     & 0.018     & 1.461        &   & 51.087      & 294.211     \\
			\multicolumn{1}{l}{~} & \multicolumn{1}{l}{} & \multicolumn{1}{l}{} & \multicolumn{1}{l}{} &   &             &             &            &            &           &           &              &   &             &             \\
			IRS                   & 50                   & 12                   & 0.2                  &   & 4.105       & 7.954       & 1.410      & 3.700      & 0.080     & 0.095     & 0.097        &   & 308.945     & 677.738     \\
			&                      &                      & 0.8                  &   & 4.029       & 8.114       & 1.473      & 4.862      & 0.087     & 0.081     & 1.709        &   & 132.110     & 414.834     \\
			&                      & 36                   & 0.2                  &   & 3.425       & 5.798       & 0.885      & 4.362      & 0.106     & 0.046     & 0.099        &   & 164.808     & 468.643     \\
			&                      &                      & 0.8                  &   & 3.061       & 5.110       & 1.111      & 5.525      & 0.436     & 0.038     & 2.805        &   & 87.554      & 348.130     \\
			& 100                  & 12                   & 0.2                  &   & 1.742       & 3.444       & 0.763      & 1.695      & 0.069     & 0.055     & 0.052        &   & 140.797     & 435.026     \\
			&                      &                      & 0.8                  &   & 1.815       & 3.349       & 0.826      & 2.155      & 0.089     & 0.042     & 0.704        &   & 87.428      & 347.862     \\
			&                      & 36                   & 0.2                  &   & 1.272       & 2.073       & 0.278      & 1.452      & 0.048     & 0.023     & 0.036        &   & 93.853      & 374.392     \\
			&                      &                      & 0.8                  &   & 1.302       & 2.051       & 0.341      & 1.711      & 0.112     & 0.021     & 0.806        &   & 54.912      & 299.900    
		\end{tabular}
	}
\end{table}

\begin{table}[ht!]
	\centering
	\caption{Mean square error of estimation of parameters ($\times 10^{4}$) with $\rho=0.4$. }\label{t.RT}
	
	\resizebox{\textwidth}{!}{
		\begin{tabular}{cccclccccccclcc}
			\multicolumn{4}{c}{}                                                                       &   & \multicolumn{7}{c}{AMCEM}                                                                                                                                      &   & \multicolumn{2}{c}{PFFR}                    \\
			Design\vspace*{0.3em}                & N                    & M                    & $\sigma^{2}$         & ~ & $\beta_{0}$          & $\beta_{1}$          & $\psi_{1}$           & $\psi_{2}$           & $\nu_{1}$            & $\nu_{2}$            & $\sigma^{2}$         & ~ & $\beta_{0}$          & $\beta_{1}$          \\
			R                     & 50                   & 12                   & 0.2                  &   & 0.936                & 1.112                & 1.391                & 2.329                & 0.052                & 0.042                & 0.202                &   & 2.210                & 27.342               \\
			&                      &                      & 0.8                  &   & 1.138                & 1.433                & 1.946                & 2.824                & 0.111                & 0.070                & 2.103                &   & 2.664                & 27.393               \\
			&                      & 36                   & 0.2                  &   & 0.809                & 0.982                & 1.201                & 2.278                & 0.049                & 0.026                & 0.433                &   & 2.055                & 29.099               \\
			&                      &                      & 0.8                  &   & 1.066                & 1.200                & 1.775                & 3.529                & 0.325                & 0.113                & 7.369                &   & 1.940                & 27.729               \\
			& 100                  & 12                   & 0.2                  &   & 0.490                & 0.532                & 0.493                & 0.878                & 0.059                & 0.033                & 0.177                &   & 1.708                & 27.495               \\
			&                      &                      & 0.8                  &   & 0.713                & 0.842                & 0.885                & 1.407                & 0.165                & 0.055                & 4.857                &   & 1.746                & 26.554               \\
			&                      & 36                   & 0.2                  &   & 0.382                & 0.421                & 0.426                & 0.768                & 0.016                & 0.013                & 0.066                &   & 1.121                & 26.231               \\
			&                      &                      & 0.8                  &   & 0.475                & 0.521                & 0.690                & 1.235                & 0.035                & 0.022                & 0.847                &   & 1.107                & 25.852               \\
			\multicolumn{1}{l}{~} & \multicolumn{1}{l}{} & \multicolumn{1}{l}{} & \multicolumn{1}{l}{} &   & \multicolumn{1}{l}{} & \multicolumn{1}{l}{} & \multicolumn{1}{l}{} & \multicolumn{1}{l}{} & \multicolumn{1}{l}{} & \multicolumn{1}{l}{} & \multicolumn{1}{l}{} &   & \multicolumn{1}{l}{} & \multicolumn{1}{l}{} \\
			RT                    & 50                   & 12                   & 0.2                  &   & 1.008                & 1.186                & 1.287                & 2.284                & 0.055                & 0.043                & 0.221                &   & 2.660                & 29.274               \\
			&                      &                      & 0.8                  &   & 1.306                & 1.556                & 2.374                & 3.589                & 0.131                & 0.059                & 2.445                &   & 2.908                & 27.994               \\
			&                      & 36                   & 0.2                  &   & 0.841                & 1.018                & 1.218                & 2.405                & 0.053                & 0.026                & 0.439                &   & 2.153                & 27.121               \\
			&                      &                      & 0.8                  &   & 1.035                & 1.197                & 2.007                & 4.045                & 0.330                & 0.099                & 6.747                &   & 2.016                & 27.904               \\
			& 100                  & 12                   & 0.2                  &   & 0.529                & 0.540                & 0.569                & 0.937                & 0.061                & 0.039                & 0.231                &   & 1.568                & 26.505               \\
			&                      &                      & 0.8                  &   & 0.821                & 0.952                & 1.068                & 1.669                & 0.119                & 0.077                & 5.596                &   & 1.947                & 26.332               \\
			&                      & 36                   & 0.2                  &   & 0.406                & 0.452                & 0.477                & 0.926                & 0.016                & 0.014                & 0.061                &   & 1.451                & 27.269               \\
			&                      &                      & 0.8                  &   & 0.489                & 0.536                & 0.674                & 1.250                & 0.026                & 0.015                & 0.723                &   & 1.190                & 26.359               \\
			\multicolumn{1}{l}{~} & \multicolumn{1}{l}{} & \multicolumn{1}{l}{} & \multicolumn{1}{l}{} &   & \multicolumn{1}{l}{} & \multicolumn{1}{l}{} & \multicolumn{1}{l}{} & \multicolumn{1}{l}{} & \multicolumn{1}{l}{} & \multicolumn{1}{l}{} & \multicolumn{1}{l}{} &   & \multicolumn{1}{l}{} & \multicolumn{1}{l}{} \\
			RM                    & 50                   & 12                   & 0.2                  &   & 0.957                & 1.114                & 1.220                & 2.143                & 0.065                & 0.046                & 0.185                &   & 2.847                & 28.986               \\
			&                      &                      & 0.8                  &   & 1.319                & 1.393                & 1.740                & 2.862                & 0.148                & 0.059                & 1.686                &   & 2.539                & 26.266               \\
			&                      & 36                   & 0.2                  &   & 0.853                & 0.944                & 0.938                & 2.014                & 0.053                & 0.025                & 0.450                &   & 1.912                & 27.176               \\
			&                      &                      & 0.8                  &   & 1.067                & 1.173                & 1.971                & 3.967                & 0.365                & 0.120                & 7.594                &   & 1.893                & 27.975               \\
			& 100                  & 12                   & 0.2                  &   & 0.497                & 0.609                & 0.509                & 0.862                & 0.091                & 0.027                & 0.314                &   & 1.723                & 27.606               \\
			&                      &                      & 0.8                  &   & 0.767                & 0.975                & 1.077                & 1.641                & 0.136                & 0.063                & 4.641                &   & 1.743                & 26.416               \\
			&                      & 36                   & 0.2                  &   & 0.401                & 0.433                & 0.473                & 0.909                & 0.015                & 0.015                & 0.064                &   & 1.493                & 27.474               \\
			&                      &                      & 0.8                  &   & 0.458                & 0.547                & 0.706                & 1.298                & 0.027                & 0.019                & 0.907                &   & 1.019                & 25.576               \\
			\multicolumn{1}{l}{~} & \multicolumn{1}{l}{} & \multicolumn{1}{l}{} & \multicolumn{1}{l}{} &   & \multicolumn{1}{l}{} & \multicolumn{1}{l}{} & \multicolumn{1}{l}{} & \multicolumn{1}{l}{} & \multicolumn{1}{l}{} & \multicolumn{1}{l}{} & \multicolumn{1}{l}{} &   & \multicolumn{1}{l}{} & \multicolumn{1}{l}{} \\
			IRS                   & 50                   & 12                   & 0.2                  &   & 1.028                & 1.395                & 1.566                & 2.951                & 0.075                & 0.054                & 0.133                &   & 3.396                & 29.038               \\
			&                      &                      & 0.8                  &   & 1.424                & 1.722                & 3.273                & 4.794                & 0.137                & 0.080                & 2.191                &   & 3.422                & 28.029               \\
			&                      & 36                   & 0.2                  &   & 0.891                & 1.039                & 1.242                & 2.431                & 0.036                & 0.026                & 0.146                &   & 2.863                & 30.101               \\
			&                      &                      & 0.8                  &   & 1.061                & 1.285                & 1.990                & 3.674                & 0.086                & 0.041                & 1.903                &   & 2.134                & 27.351               \\
			& 100                  & 12                   & 0.2                  &   & 0.558                & 0.641                & 0.789                & 1.430                & 0.077                & 0.042                & 0.204                &   & 2.380                & 28.359               \\
			&                      &                      & 0.8                  &   & 0.733                & 0.900                & 1.626                & 2.600                & 0.079                & 0.087                & 3.603                &   & 1.998                & 26.388               \\
			&                      & 36                   & 0.2                  &   & 0.405                & 0.485                & 0.506                & 0.986                & 0.016                & 0.015                & 0.027                &   & 1.865                & 28.474               \\
			&                      &                      & 0.8                  &   & 0.539                & 0.598                & 0.713                & 1.366                & 0.023                & 0.016                & 0.305                &   & 1.314                & 26.556              
		\end{tabular}
	}
\end{table}

By comparing  Table  \ref{t.R} and Table \ref{t.RT}, it can be noticed that an increase in $\rho$ leads to a reduction in the MSE of the estimations of the functional regression coefficients in the AMCEM method. As expected, if the value of the parameter $\rho$ increases,  the accuracy of \code{pffr} in estimating the functional regression coefficients increases dramatically. This can be explained by the fact that, as $\rho$ approaches 1, the within-subject correlation of the sampling points decreases, an assumption on which the likelihood function in \code{pffr} model is based.  In summary, according to the results presented in Table \ref{t.R} and Table \ref{t.RT}, AMCEM outperforms \code{pffr} in estimating the functional regression coefficients in a general situation.

In connection to the results mentioned above, an increase in $\rho$ causes a decrease in the accuracy of the estimation of the first principal component in AMCEM, and an increase in the accuracy when estimating the second. This can be explained by noting that an increase in $\rho$ causes a decrease in the proportion of variation explained by the first eigen-function, and an increase in the proportion explained by the second.

As a final note, we emphasize that the accuracy in the estimation of functional regression coefficients and functional principal components for incomplete  designs is the highest in design (RT), followed by designs (RM) and (IRS), as expected from the pattern of missingness in the data.

\subsection{Analysis of the Madras dataset}

\begin{figure}[tb!]
	
	\begin{minipage}{\dimexpr\linewidth-0cm\relax}%
		\begin{minipage}[b]{0.25\linewidth}
			\centering
			\includegraphics[width=1\linewidth]{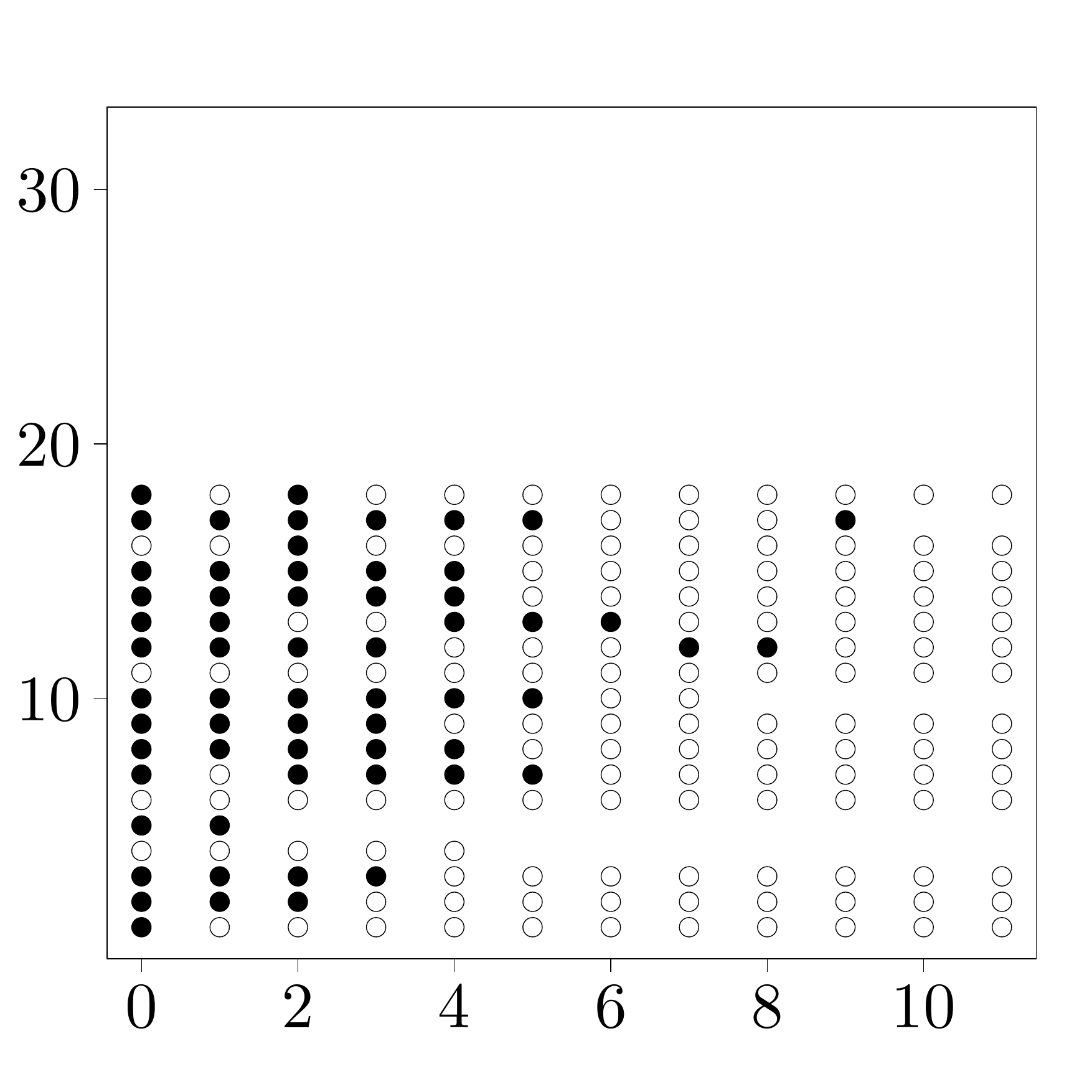} 
		\end{minipage}
		\begin{minipage}[b]{0.25\linewidth}
			\centering
			\includegraphics[width=1\linewidth]{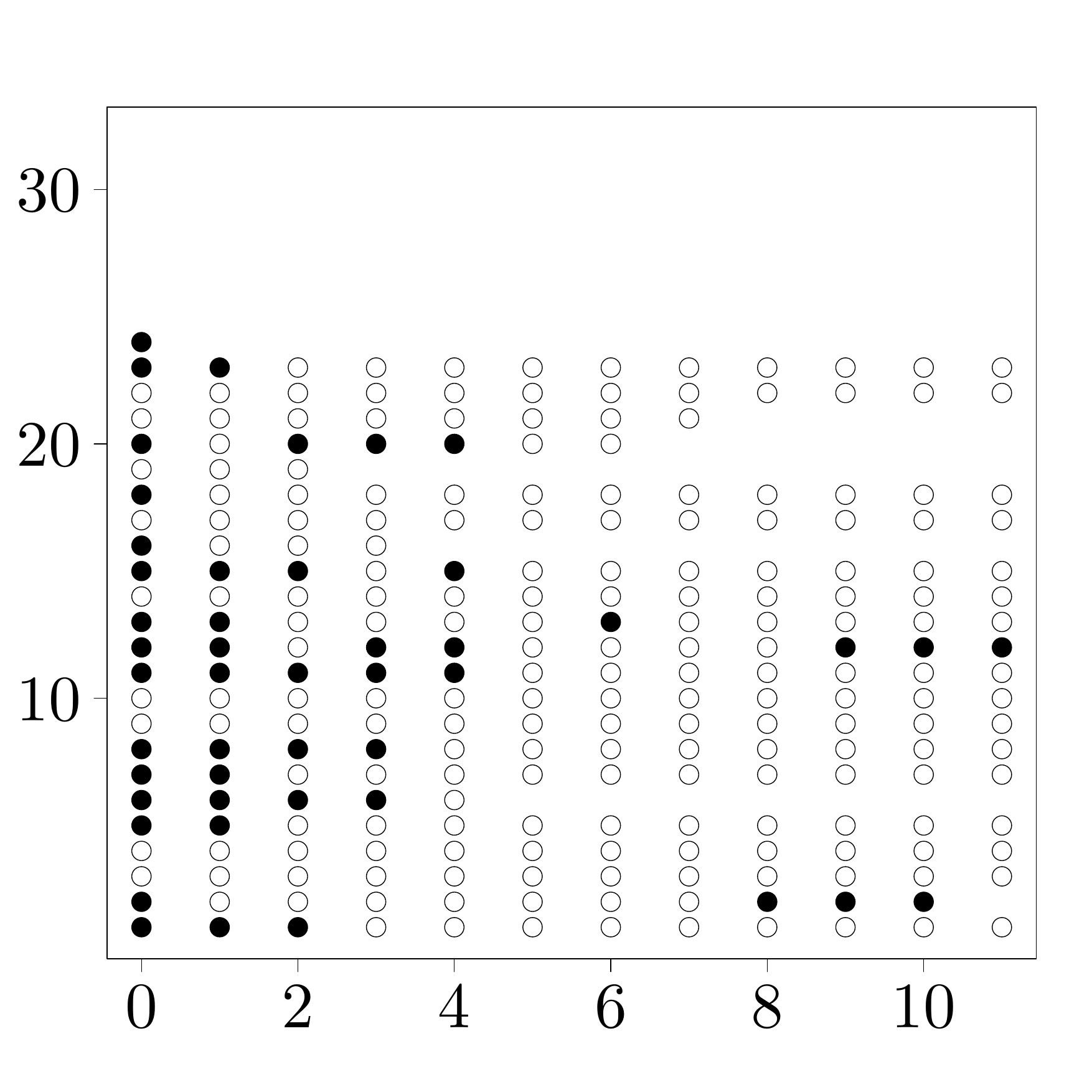} 
		\end{minipage}
		\begin{minipage}[b]{0.25\linewidth}
			\centering
			\includegraphics[width=1\linewidth]{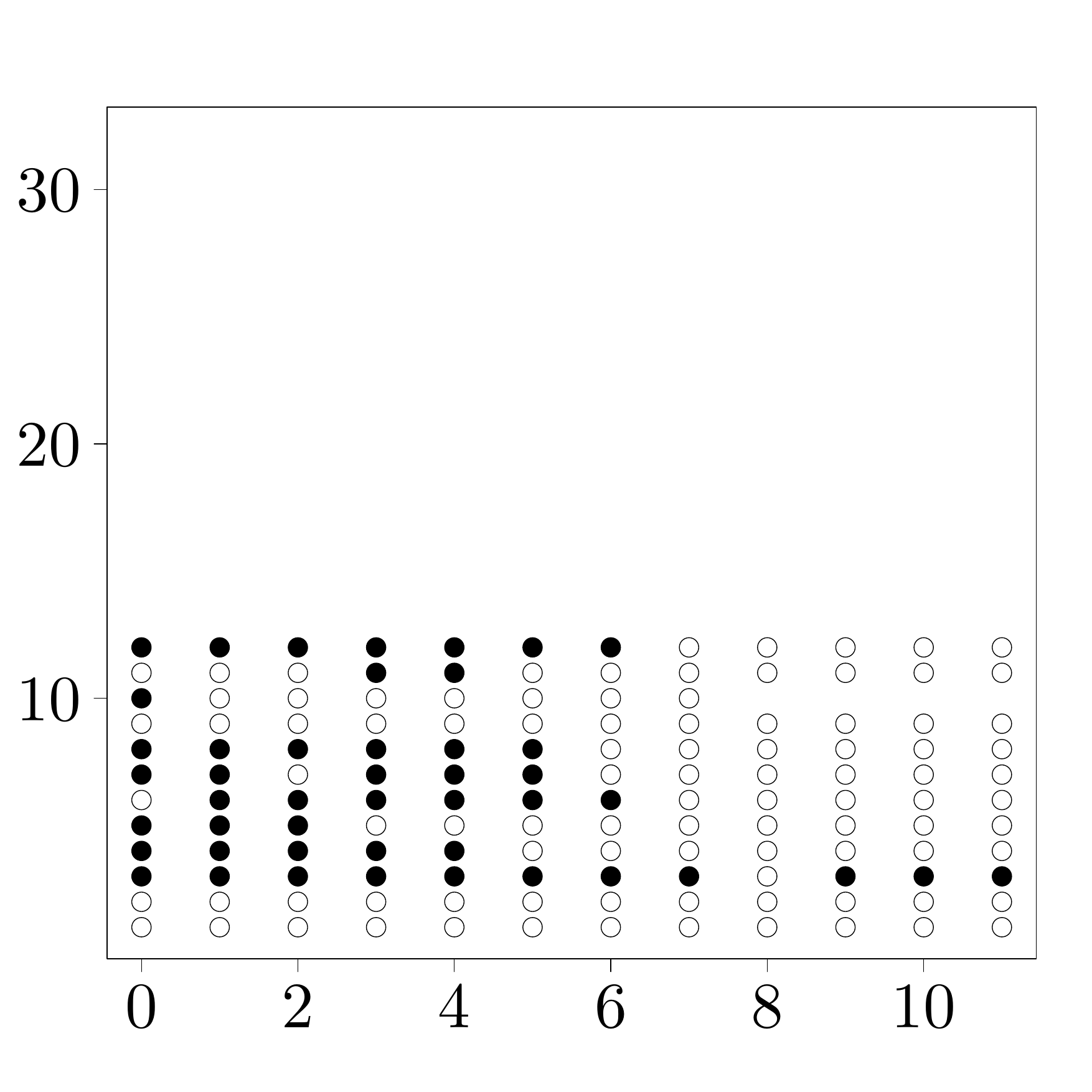} 
		\end{minipage}
		\begin{minipage}[b]{0.25\linewidth}
			\centering
			\includegraphics[width=1\linewidth]{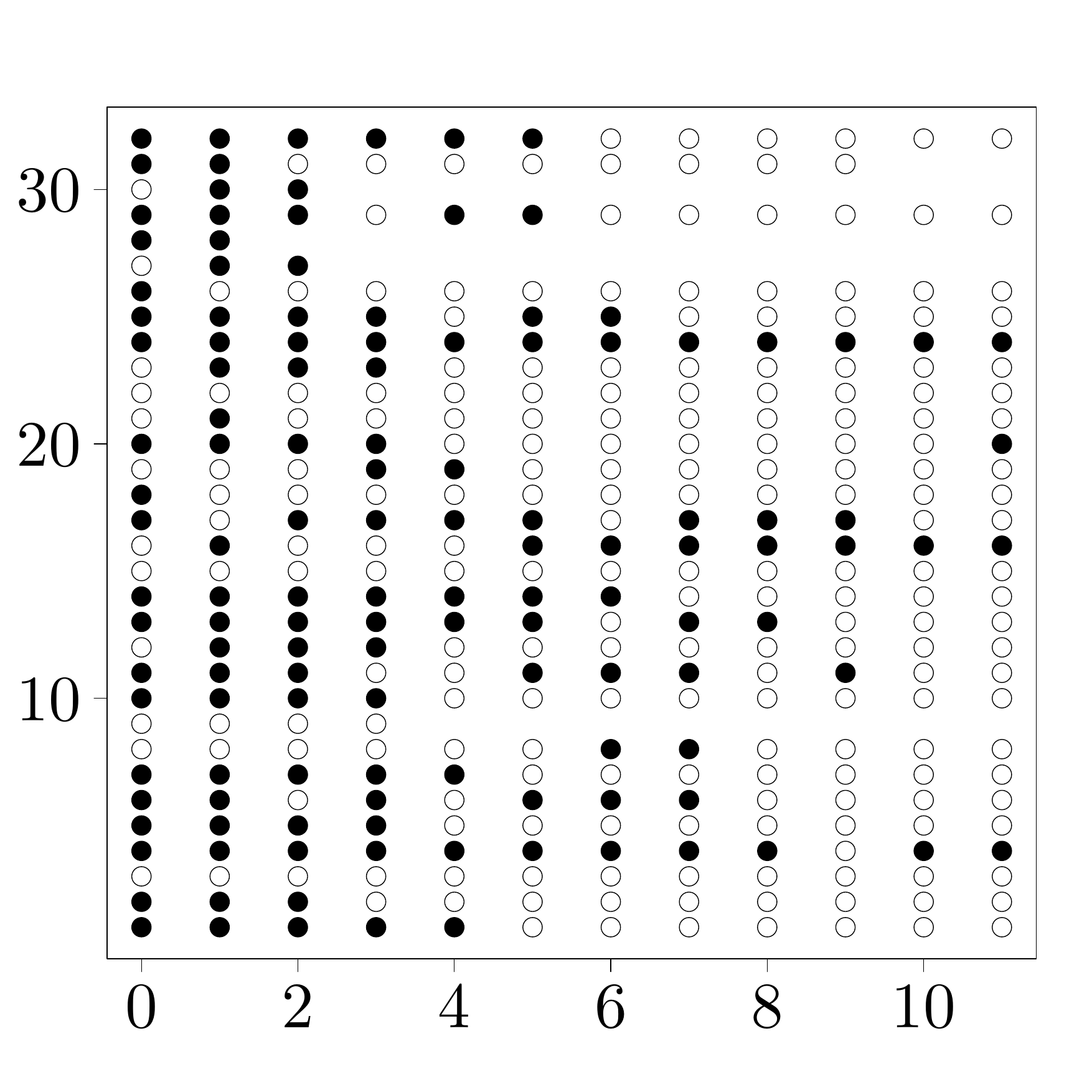} 
		\end{minipage}
		\vspace{-1em}
		\begin{minipage}[b]{1\linewidth}
			\centering
			\footnotesize{Month}
		\end{minipage}
		\vspace{-1em}
		
		\begin{minipage}[b]{1\linewidth}
			\begin{minipage}[b]{0.25\linewidth}
				\centering
				(a)
			\end{minipage}
			\begin{minipage}[b]{0.25\linewidth}
				\centering (b)
			\end{minipage}
			\begin{minipage}[b]{0.25\linewidth}
				\centering
				(c)
			\end{minipage}
			\begin{minipage}[b]{0.25\linewidth}
				\centering
				(d)
			\end{minipage}
		\end{minipage}
	\end{minipage}%
	\caption{binary sequence of observations of presence/absence of `thought disorder' in 86 patients along 12 months after initial hospitalization. Circle: absence of `thought disorder'; filled circle: presence of `thought disorder'. Panel (a): females under 20 years old; panel (b): females aged 20 and over; panel (c): males under 20 years old; panel (d): males aged 20 and over. The $y$-axis simply represents the subject indexing.}
	
	\label{mad:observation}
\end{figure}

The Madras dataset (shortly, Madras-Data) is a collection of monthly records that indicates the presence or absence of a set of six positive (hallucinations, delusions, thought disorders) and negative  (flat affect, apathy, withdrawal) psychiatric symptoms \citep[pp. 234-243]{diggle2002analysis}. In this section, we apply our method to a dataset that consists of a sequence of binary observations indicating the presence or absence of ‘thought disorder’ in 86 patients over one year after initial hospitalization. The presence or absence of the symptom is recorded for each patient regularly every month, but for a few patients data is missing for some months. Each patient's record includes two features: age-at-onset and gender at the initial stage, which are considered as covariates in this study. The aim is to evaluate the effect of age and gender on the course of hospitalization. 
Concerning the response variable $y_{ij}$, values of zero or one  indicate presence or absence of the `thought disorder' symptom respectively for the $i$\textsuperscript{th} subject at month $j$, $j=0,1,\ldots,11$. The response $y_{ij}$ is missing for some random $i$ and $j$, thus this dataset can be considered as data showing an irregular design, as described in Sec. \ref{sec.1}. 
The collection of data is illustrated in Figure \ref{mad:observation}. It can be noticed that the sample size is not balanced in the four different groups defined by gender and age above/below 20 years old.

There are two scalar covariates in the dataset. The variable age is coded as zero and one, with $0=\text{age-at-onset}\geq 20$ and $1=\text{age-at-onset}<20$. The scalar variable gender is also coded as zero and one, with $0 =\text{male}$, $1 = \text{female}$. 
The major question regarding this study is to assess the effect of covariates on the course of illness. We apply a FoSR model to Madras-data, to investigate the relationship of age and gender to the `thought disorder' symptom. In this model, the binary response $Y$ indicates the presence or absence of the symptom, while the latent function $\Ztipa$ is a proxy for the thought disorder intensity (TDI). Our analysis also reveals the smooth principal components, which can provide valuable information about the latent pattern of variability of TDI. 

\begin{figure}[tb!]
	\begin{minipage}{\dimexpr 0.2\linewidth-1.25cm\relax}
		\rotatebox{90}{$~$}
		
	\end{minipage}%
	\begin{minipage}{0.5cm}
		\rotatebox{90}{$\log \xi$}
		
	\end{minipage}%
	\begin{minipage}{\dimexpr\linewidth-0.5cm\relax}%
		\begin{minipage}{0.3\linewidth}
			\centering
			\includegraphics[width=.95\linewidth]{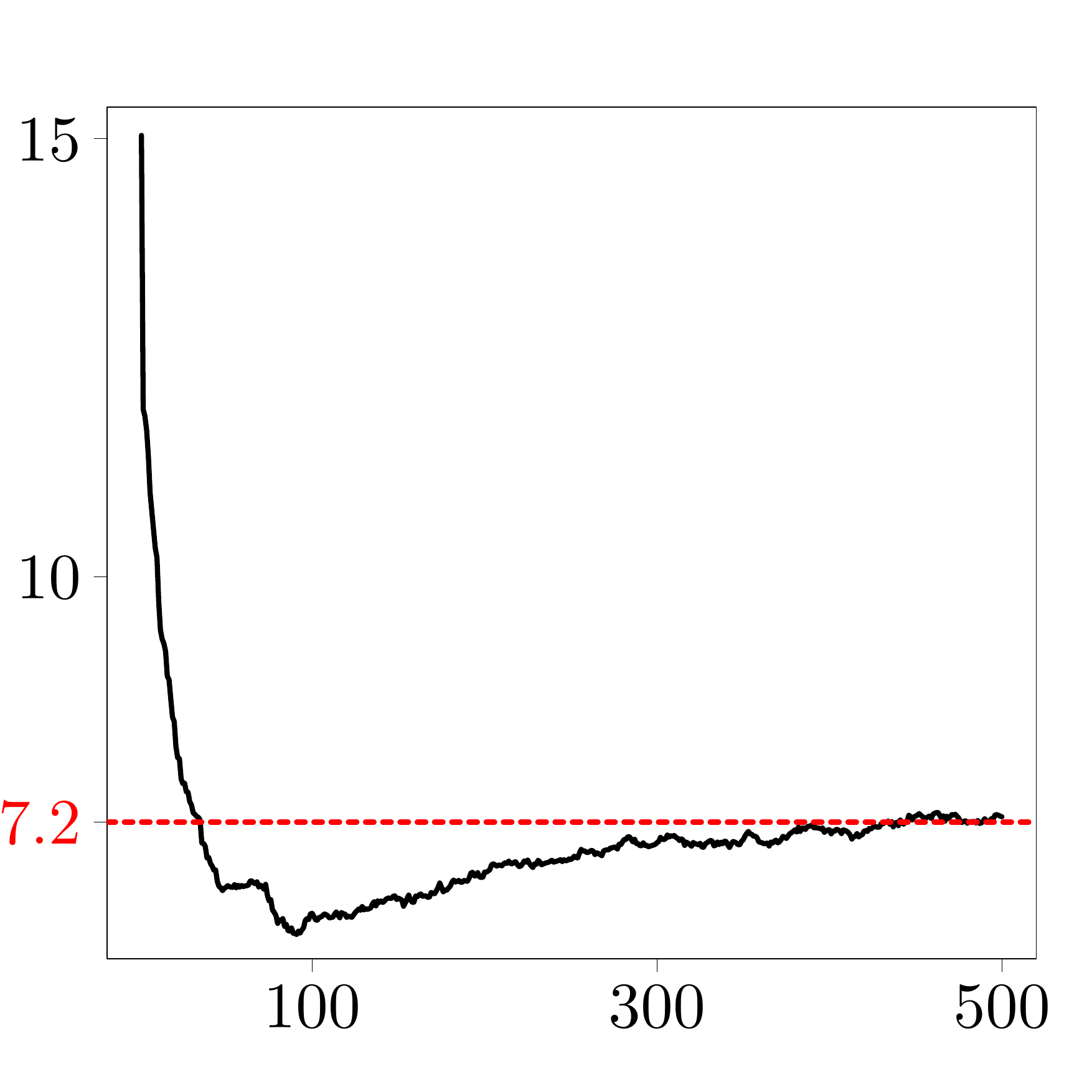}		
		\end{minipage}
		\begin{minipage}{0.5cm}
			\rotatebox{90}{$~$}
			
		\end{minipage}%
		\begin{minipage}{\dimexpr 0.5cm\relax}
			\rotatebox{90}{$\sigma^2$}
		\end{minipage}%
		\begin{minipage}{0.3\linewidth}
			\centering
			\includegraphics[width=.95\linewidth]{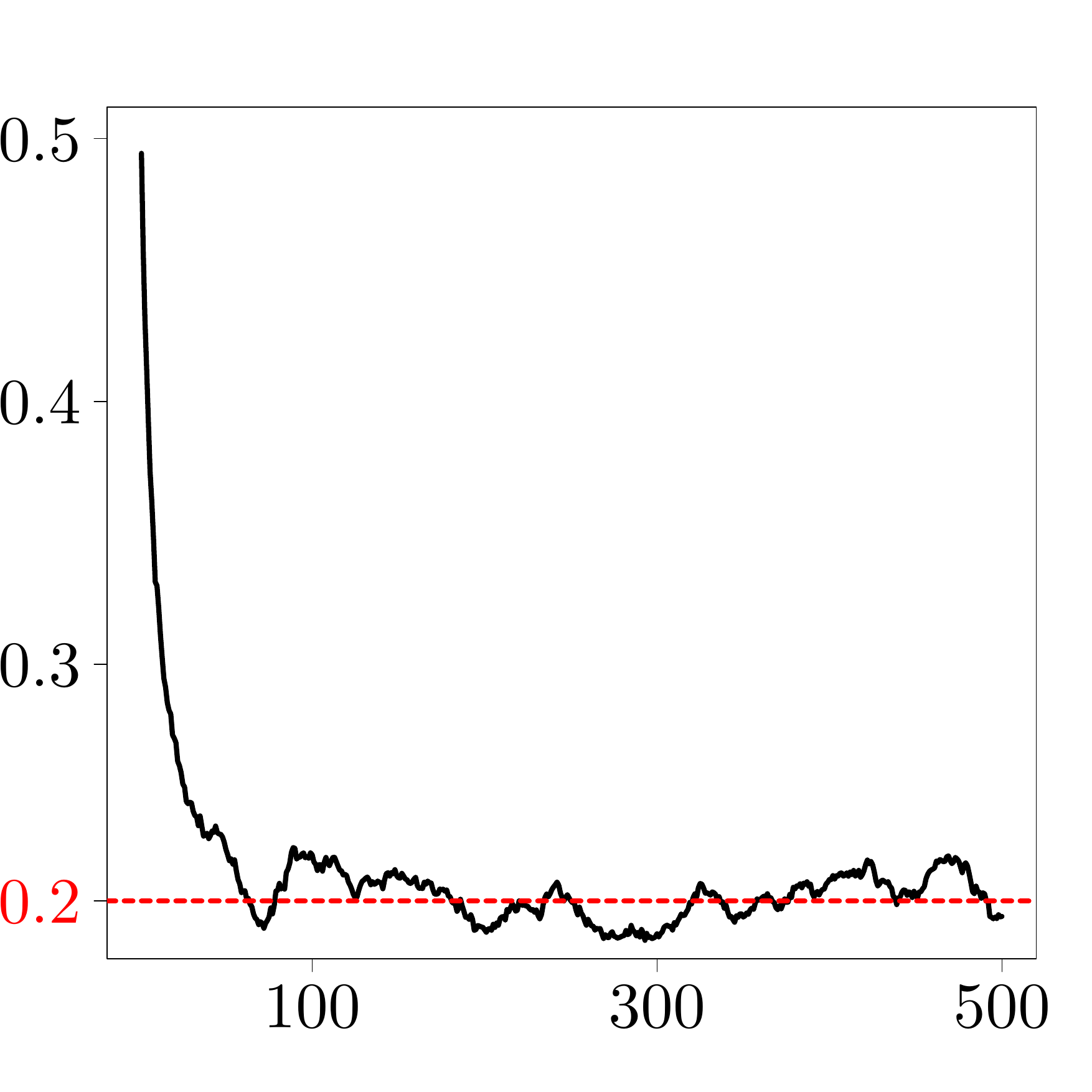} 	
		\end{minipage}
		\begin{minipage}{\dimexpr 0.2\linewidth-1.25cm\relax}
			\rotatebox{90}{$~$}
			
		\end{minipage}%
	\end{minipage}%
	
	\begin{minipage}{1\linewidth}
		\centering
		\footnotesize{AMCEM iteration}
	\end{minipage}
	\caption{convergence of the sequence of scalar-valued parameters along iterations of the AMCEM algorithm. Left panel: log of the estimated average degree of smoothness of the latent functions, as defined in (\ref{eq:avSmooth}); right panel: estimated measurement error.}
	
	\label{mad:lambda}
\end{figure}

To investigate the effect of Age and Gender on TDI, the following FoSR model is considered:
\begin{align*}
\text{TDI}_{i}\left(t\right)&=\beta_0\left(t\right)+\beta_1\left(t\right)\text{Age}_{i}+
\beta_2\left(t\right)\text{Gender}_{i}+\beta_3\left(t\right)\left(\text{Age}_{i}\times\text{Gender}_{i}\right)+\varepsilon_{i}\left(t\right),\\
\text{Y}_{i}\left(t\right)&=\text{I}\left(\text{TDI}_{i}\left(t\right)+\epsilon_{it}>0\right).
\end{align*}
This model is fitted to Madras-data by employing the AMCEM algorithm. Figure \ref{mad:lambda} shows the convergence of the estimation of the scalar-valued parameters of the model, including the measurement error $\sigma^2$ and the average degree of smoothness in the sample, defined as
\begin{equation}\label{eq:avSmooth}
\xi=\frac{1}{N}\sum_{i=1}^{N}\Bbb{E}_{\mathbf{W}\mid\mathbf{Y}=\mathbf{y}}\left[\Bbb{E}_{\mathbf{Z}\mid \mathbf{W}, V_{\delta^{(m)}}}\left(\mathbf{Z}^{\top}_{i}\mathbf{P}^{(n)}\mathbf{Z}_{i}\right)\right],
\end{equation}
The left panel shows the estimated average degree of smoothness of the latent functions in the log scale along the AMCEM algorithm iterations, and the right panel shows the estimated measurement error. It can be noticed from this figure that the AMCEM algorithm converges. Furthermore, the measurement error is estimated as $\hat{\sigma}^{2}=0.2,$ which demonstrates that a major portion of the variation in the observed binary sequences is described by the functional part of the model.

\begin{figure}[tb!]
	
	\begin{minipage}{\dimexpr\linewidth\relax}%
		\begin{minipage}[b]{1\linewidth}
			\begin{minipage}[b]{0.25\linewidth}
				\centering
				\footnotesize{Sample Id=1}
			\end{minipage}
			\begin{minipage}[b]{0.25\linewidth}
				\centering
				\footnotesize{Sample Id=43}
			\end{minipage}
			\begin{minipage}[b]{0.25\linewidth}
				\centering
				\footnotesize{Sample Id=50}
			\end{minipage}
			\begin{minipage}[b]{0.25\linewidth}
				\centering
				\footnotesize{Sample Id=80}
			\end{minipage}
		\end{minipage}
		
		\begin{minipage}[b]{0.25\linewidth}
			\centering
			\includegraphics[width=.95\linewidth]{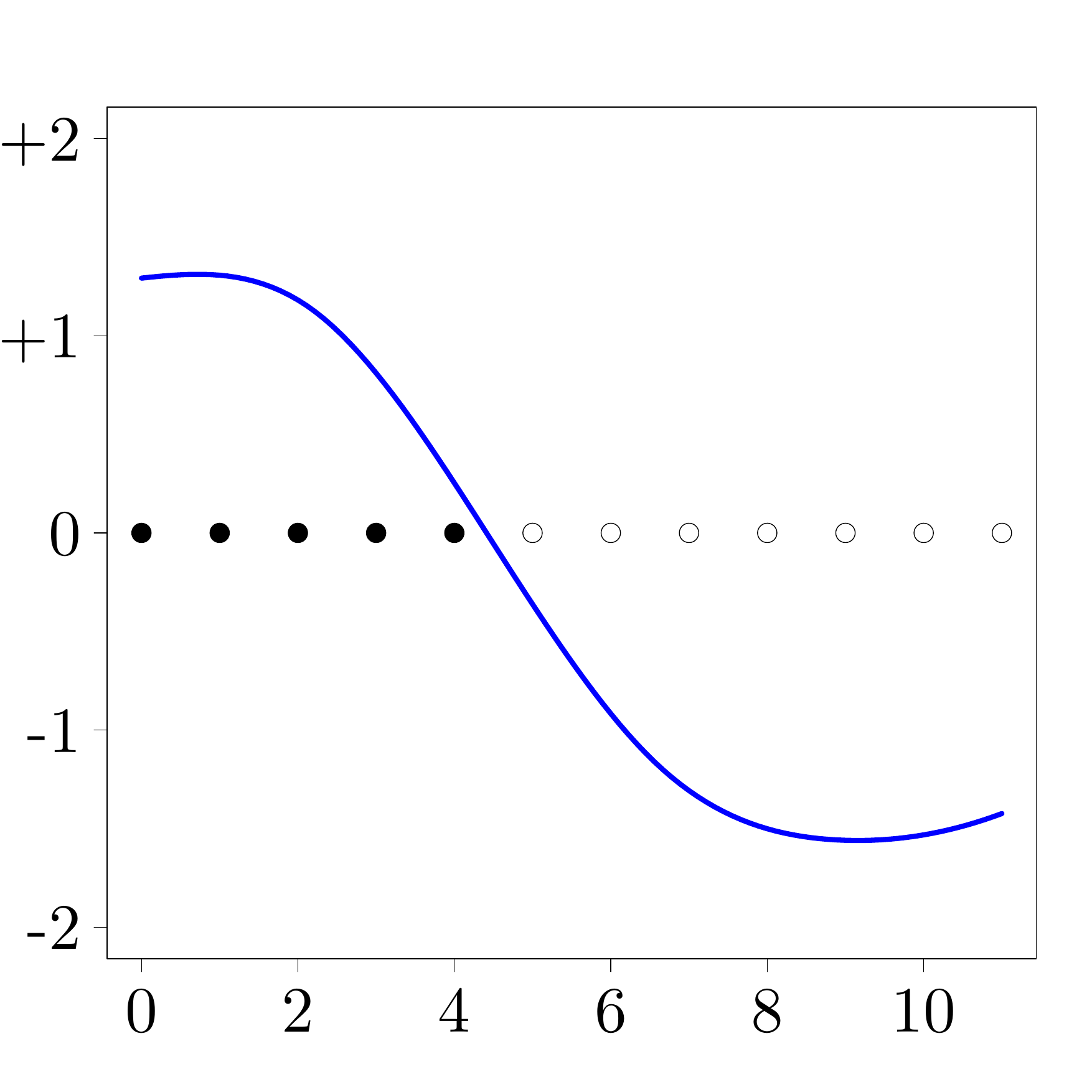} 
		\end{minipage}
		\begin{minipage}[b]{0.25\linewidth}
			\centering
			\includegraphics[width=.95\linewidth]{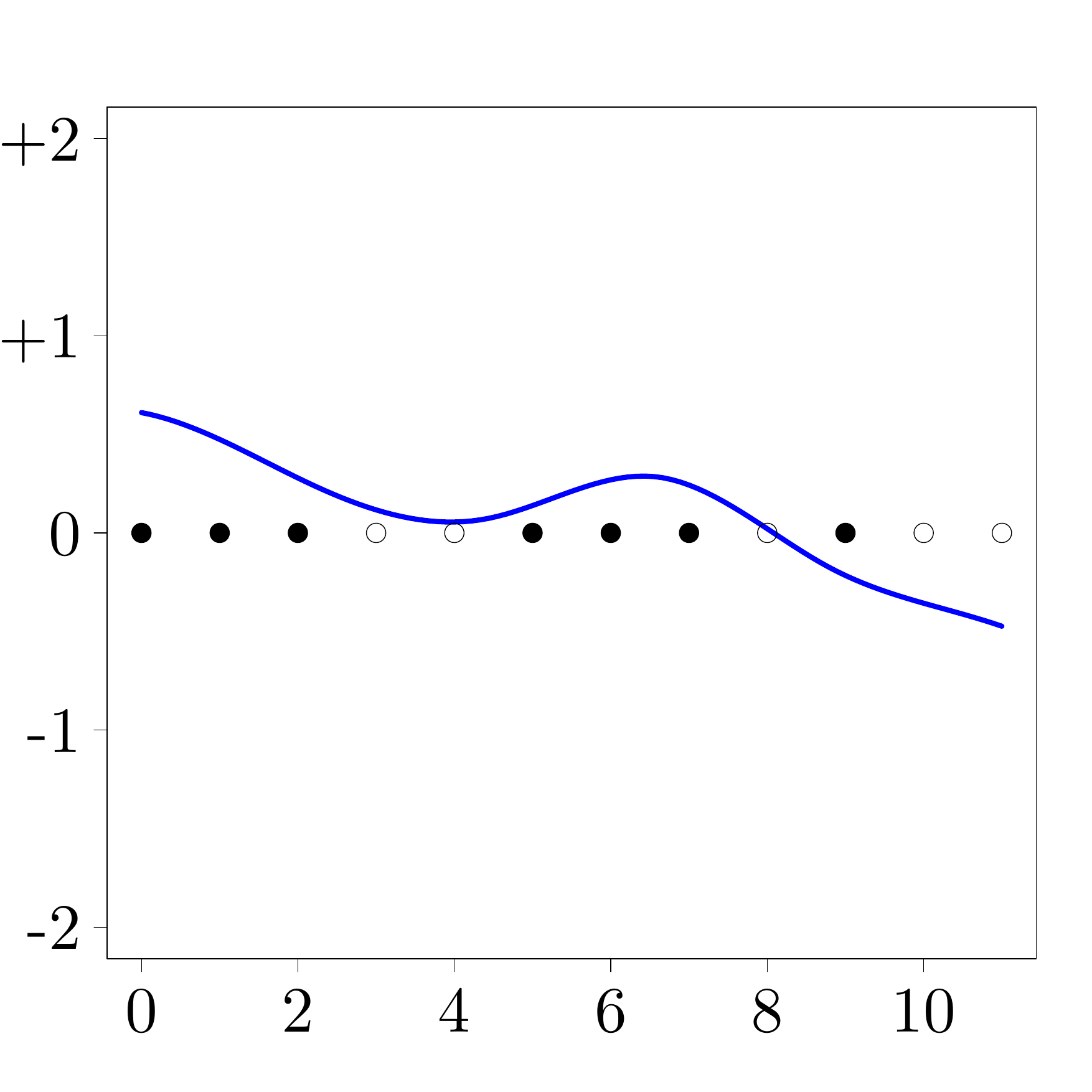} 
		\end{minipage}
		\begin{minipage}[b]{0.25\linewidth}
			\centering
			\includegraphics[width=.95\linewidth]{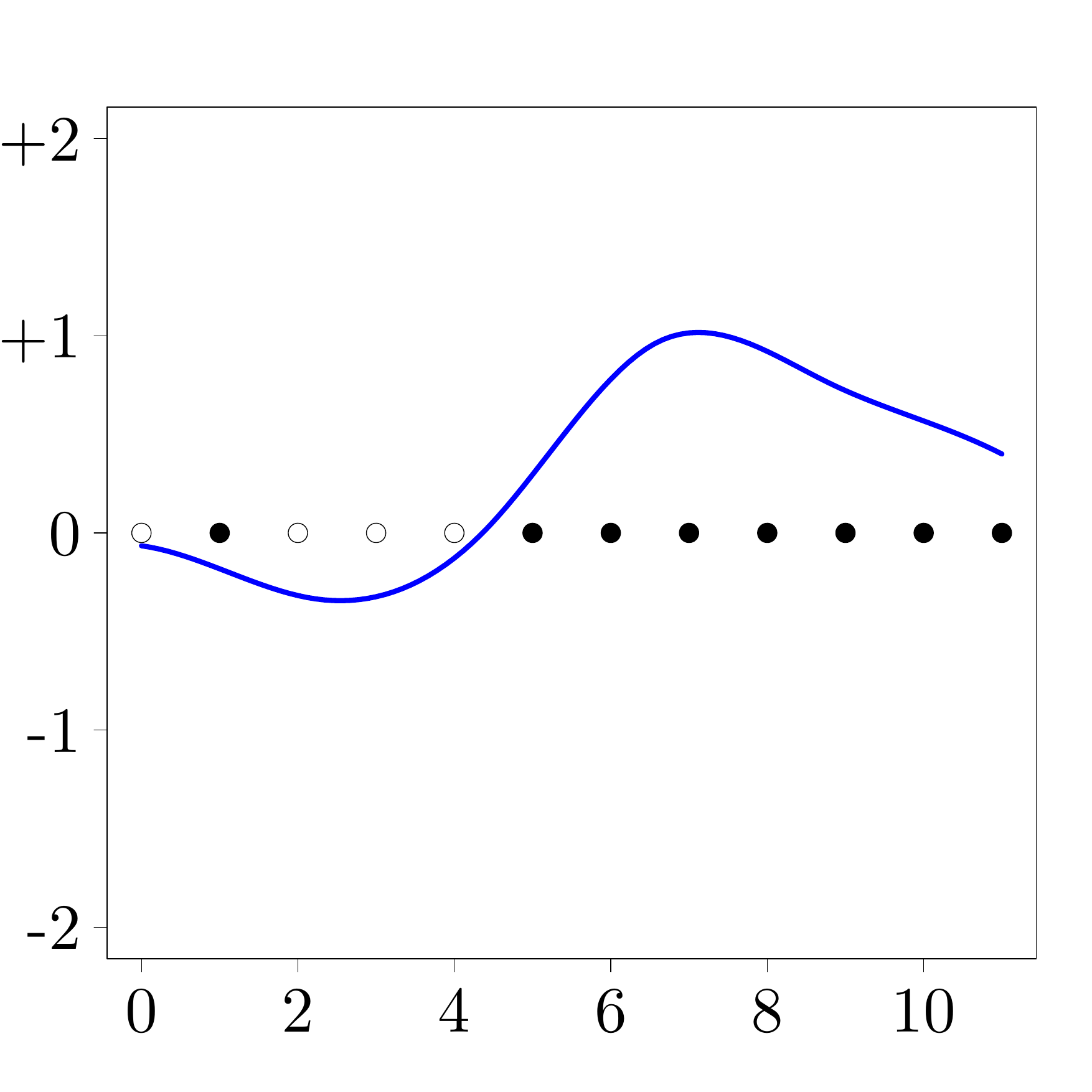} 
		\end{minipage}
		\begin{minipage}[b]{0.25\linewidth}
			\centering
			\includegraphics[width=.95\linewidth]{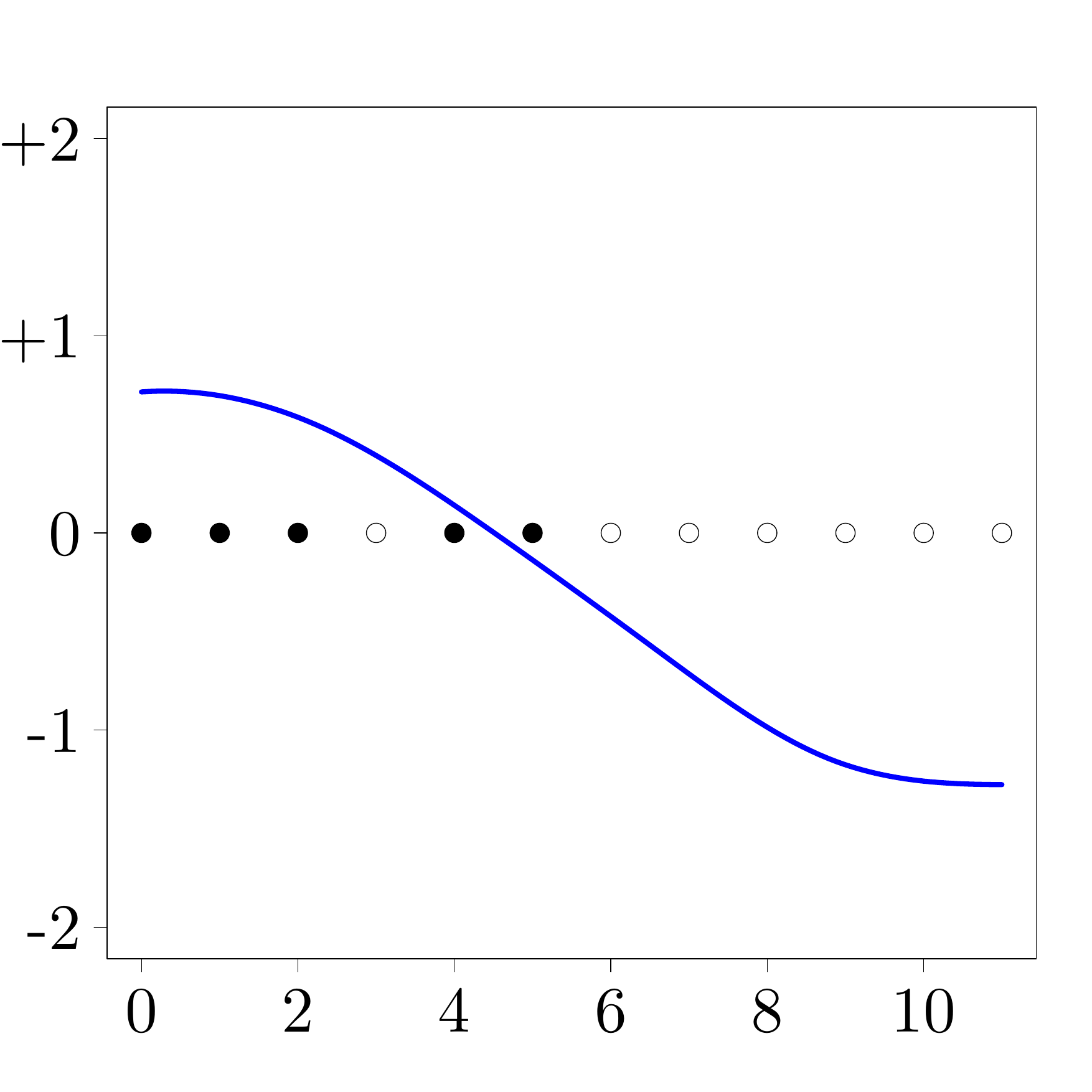} 
		\end{minipage}
		
		\vspace{-1em}
		\begin{minipage}[b]{1\linewidth}
			\centering
			{\footnotesize{Month}}
		\end{minipage}
		
	\end{minipage}%
	\caption{estimated latent TDI functions for four different subjects in the sample. Dots in the plot indicate the sequence of observations of the `thought disorder' symptom for the subject: empty circles for absence, filled circles for presence. The blue solid line is the estimated subject-specific latent TDI function, given the subject covariates and the sequence of observed binary responses.}
	
	\label{mad:zhat}
\end{figure}

The strategy employed in the AMCEM algorithm for providing smooth estimations of functional parameters is based on fitting smooth functions to the sequence of binary data, consequently leading to smooth functional regression coefficients and covariance function estimates. The smooth estimations of the latent TDI functions fitted to four selected binary sequences is shown in Figure \ref{mad:zhat}. 

\begin{figure}[tb!]
	
	\begin{minipage}{\dimexpr\linewidth-0cm\relax}%
		\begin{minipage}[b]{0.25\linewidth}
			\centering
			\includegraphics[width=1\linewidth]{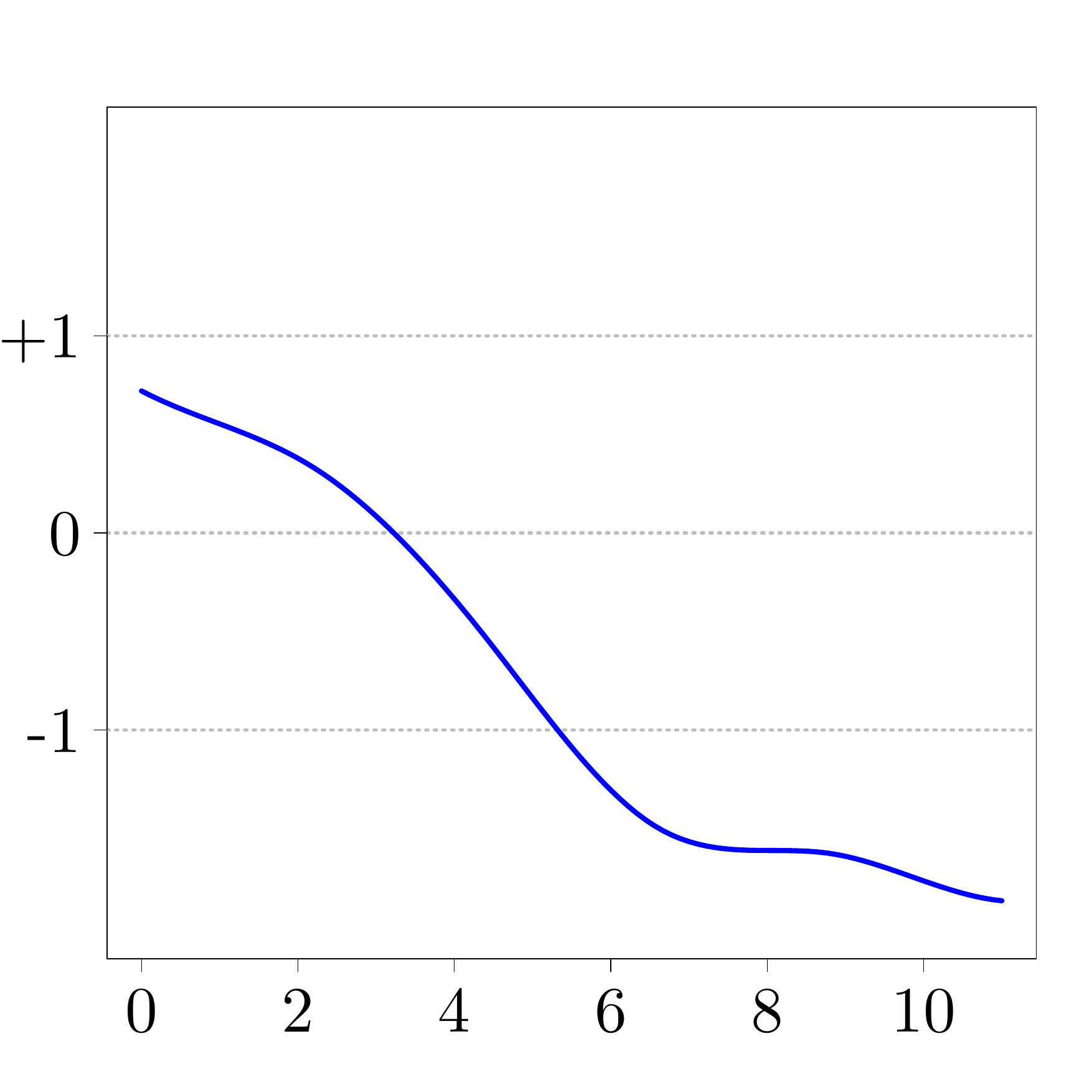} 
		\end{minipage}
		\begin{minipage}[b]{0.25\linewidth}
			\centering
			\includegraphics[width=1\linewidth]{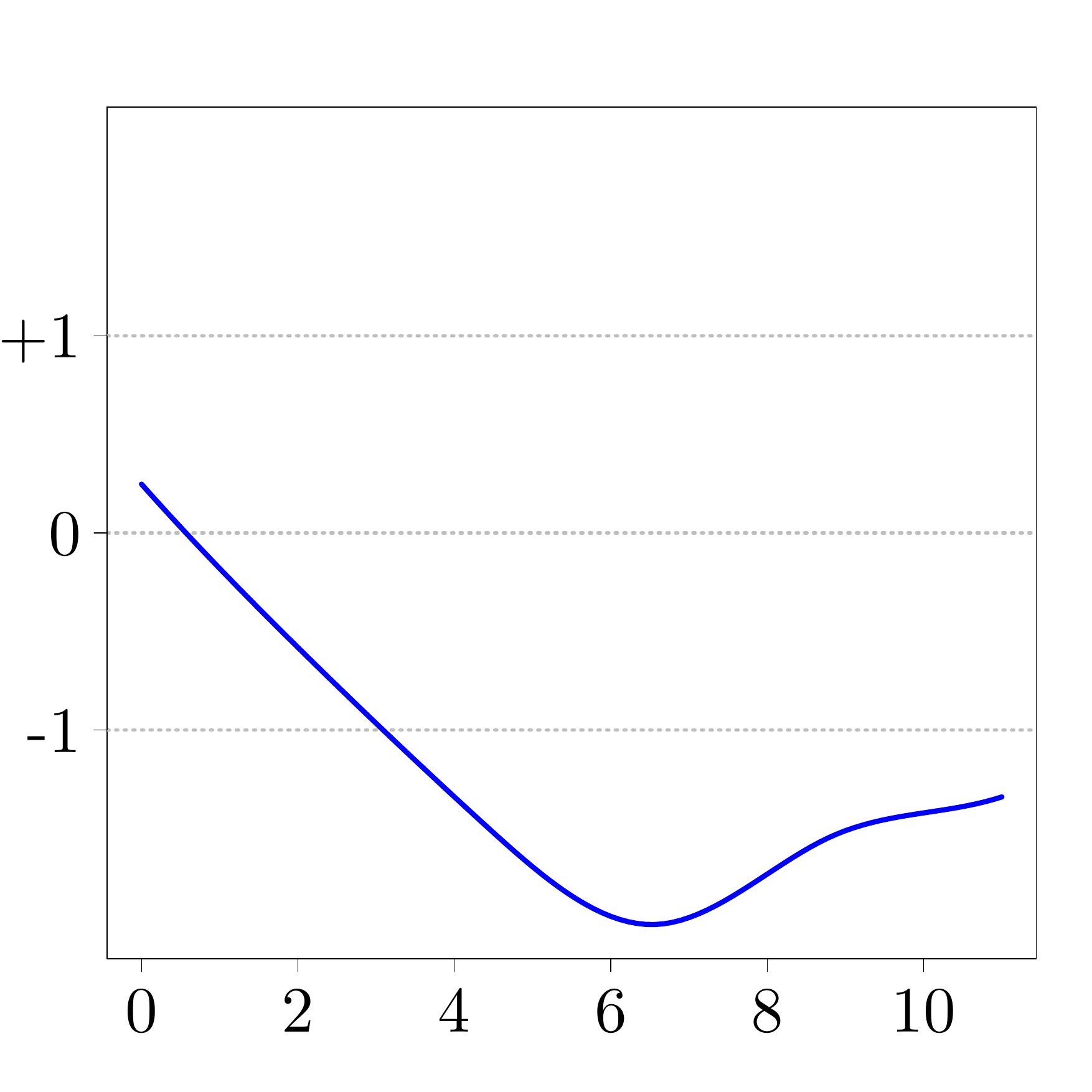} 
		\end{minipage}
		\begin{minipage}[b]{0.25\linewidth}
			\centering
			\includegraphics[width=1\linewidth]{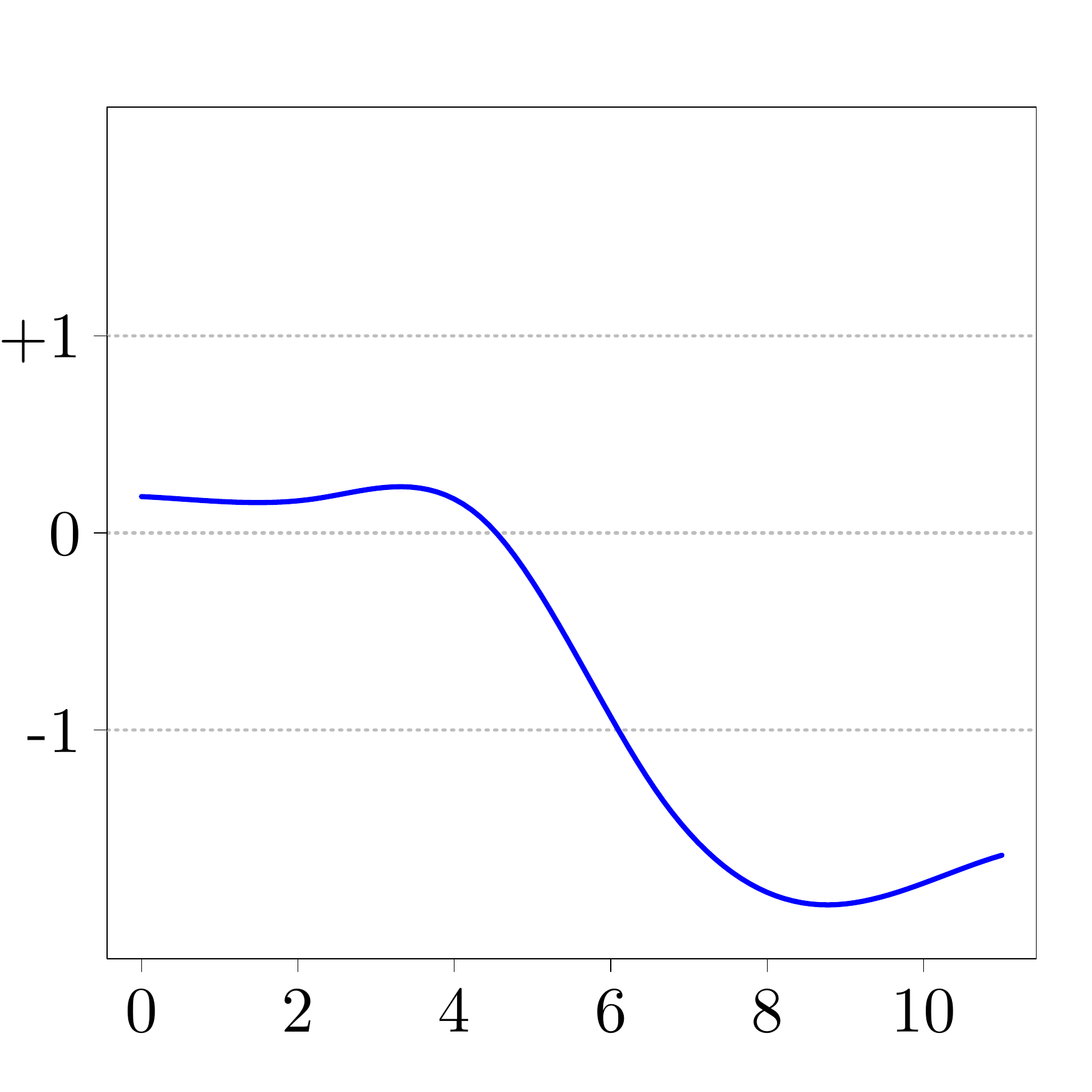} 
		\end{minipage}
		\begin{minipage}[b]{0.25\linewidth}
			\centering
			\includegraphics[width=1\linewidth]{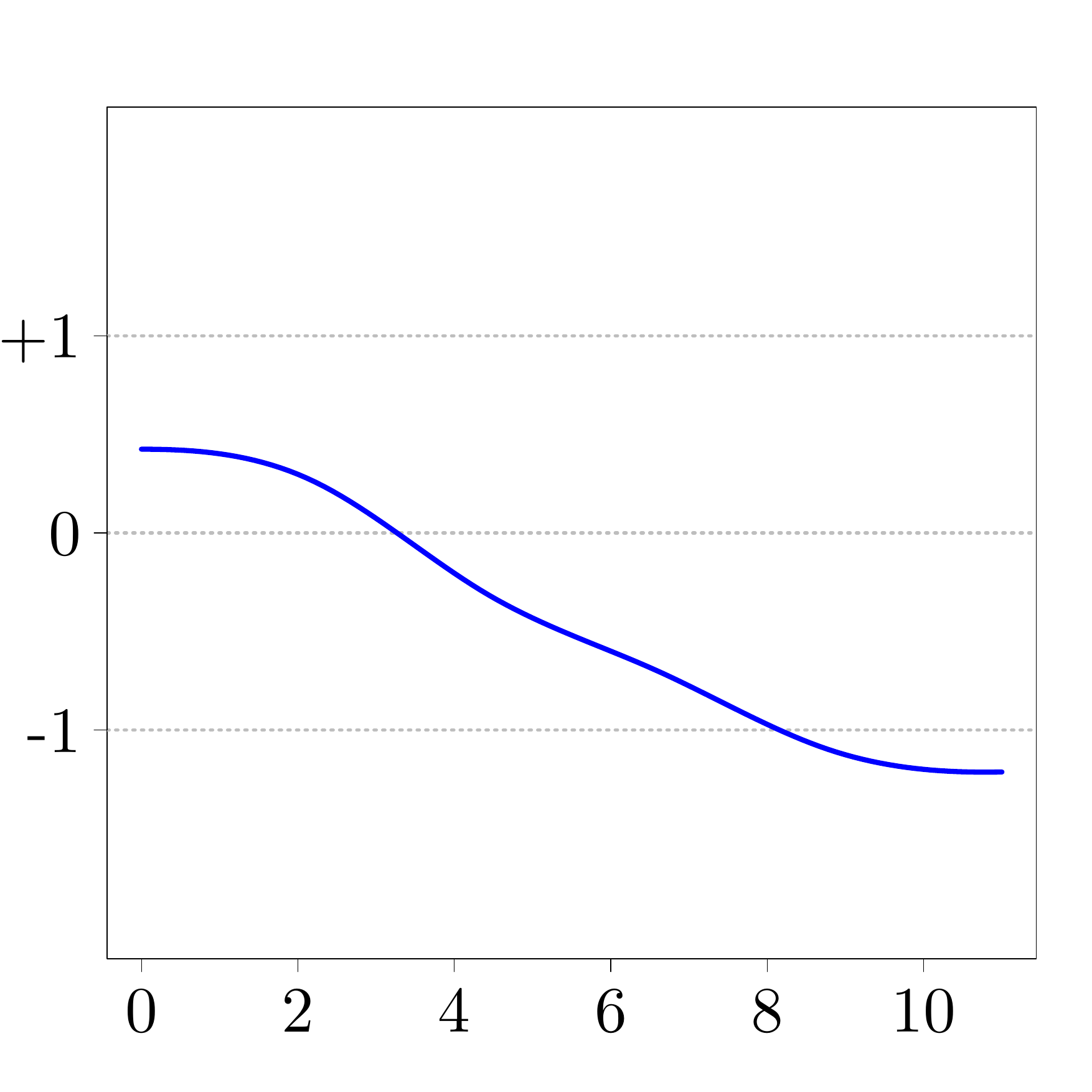} 
		\end{minipage}
		\vspace{-1em}
		\begin{minipage}[b]{1\linewidth}
			\centering
			\footnotesize{Month}
		\end{minipage}
		\vspace{-1em}
		
		\begin{minipage}[b]{1\linewidth}
			\begin{minipage}[b]{0.25\linewidth}
				\centering
				(a)
			\end{minipage}
			\begin{minipage}[b]{0.25\linewidth}
				\centering (b)
			\end{minipage}
			\begin{minipage}[b]{0.25\linewidth}
				\centering
				(c)
			\end{minipage}
			\begin{minipage}[b]{0.25\linewidth}
				\centering
				(d)
			\end{minipage}
		\end{minipage}
	\end{minipage}%
	\caption{estimated mean functions. Panel (a): females under 20 years old; panel (b): females aged 20 and over; panel (c): males under 20 years old; panel (d): males aged 20 and over.}
	
	\label{mad:mean}
\end{figure}

The estimated smooth mean functions for the four different combinations of covariates are depicted in Figure \ref{mad:mean}. In this figure, (a) is the mean function of the females aged under 20, i.e, $\beta_{0}+\beta_{1}+\beta_{2}+\beta_{3}$, (b) is the mean function of  the females aged 20 and over, i.e., $\beta_{0}+\beta_{2}$, (c) is the mean function of males aged under 20, i.e., $\beta_{0}+\beta_{1}$, and (d) is the mean function of males aged 20 and over, i.e., $\beta_{0}$.
By inspection of Figure \ref{mad:mean}, marked differences in the shapes of the four mean functions can be observed across all four groups. 
For instance, TDI is higher for younger females at the onset, although it is lower at the end of the course of illness; besides, the shape of the mean function is quite different in younger and older females, with the former showing a steady decline, and the latter instead a final increase in TDI at the end of the year. Moreover, the rate of reduction of TDI in older males is lower, meaning that TDI is less affected by hospitalization in this group. Another outstanding result that can be inferred from the estimated functional regression coefficients is that, while younger females experience a steady decrease in TDI, younger males show a quite different patter, with a stationary period followed by first an abrupt decrease in TDI in the middle of the year, and then by an increase in TDI towards the end.

\begin{figure}[tb!]
	\centering
	\begin{subfigure}{0.44\textwidth}
		\centering
		\includegraphics[width=.75\linewidth]{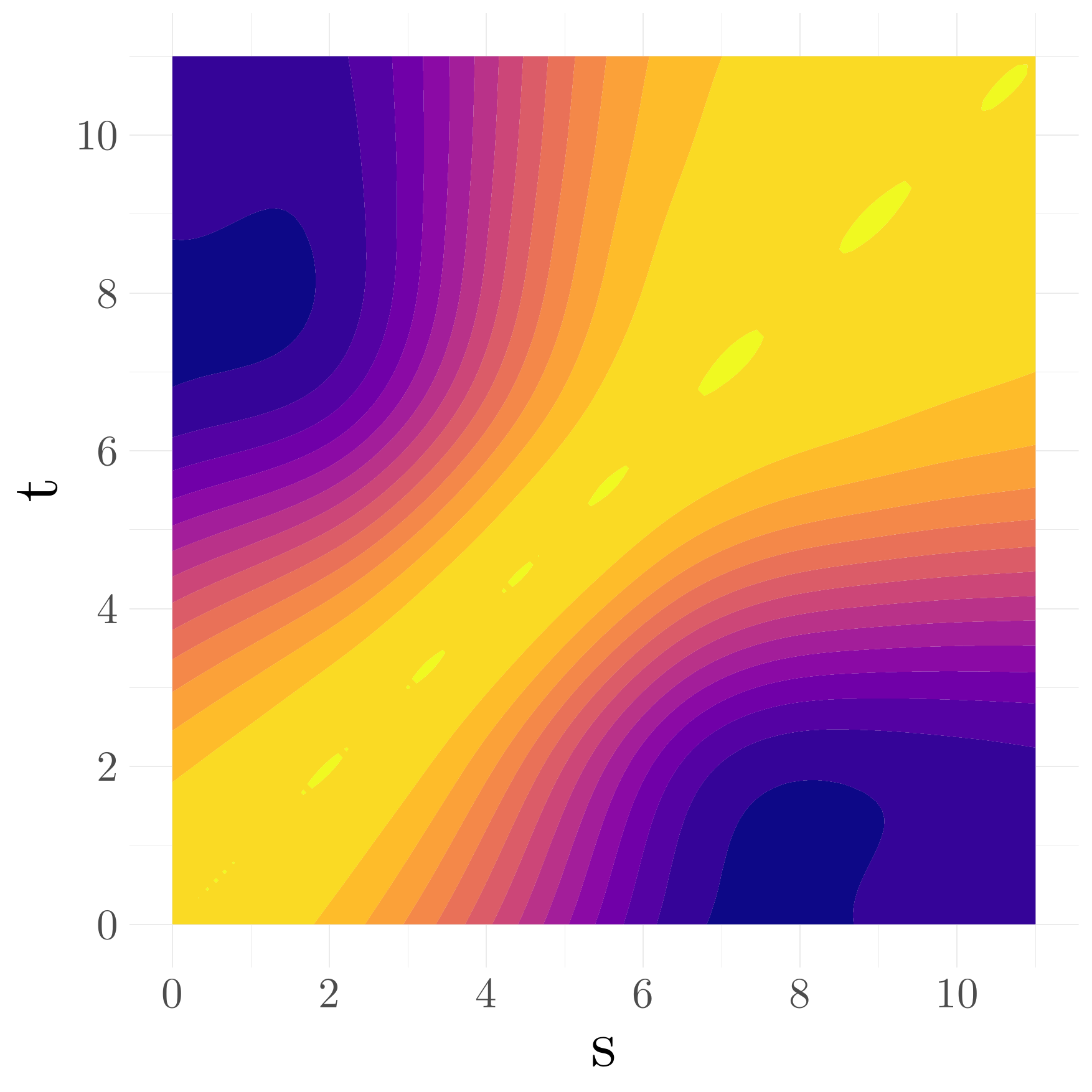}
		
	\end{subfigure}%
	\begin{subfigure}{0.14\textwidth}
		\centering
		
		
	\end{subfigure}%
	\begin{subfigure}{0.45\textwidth}
		\centering
		\includegraphics[width=1\linewidth]{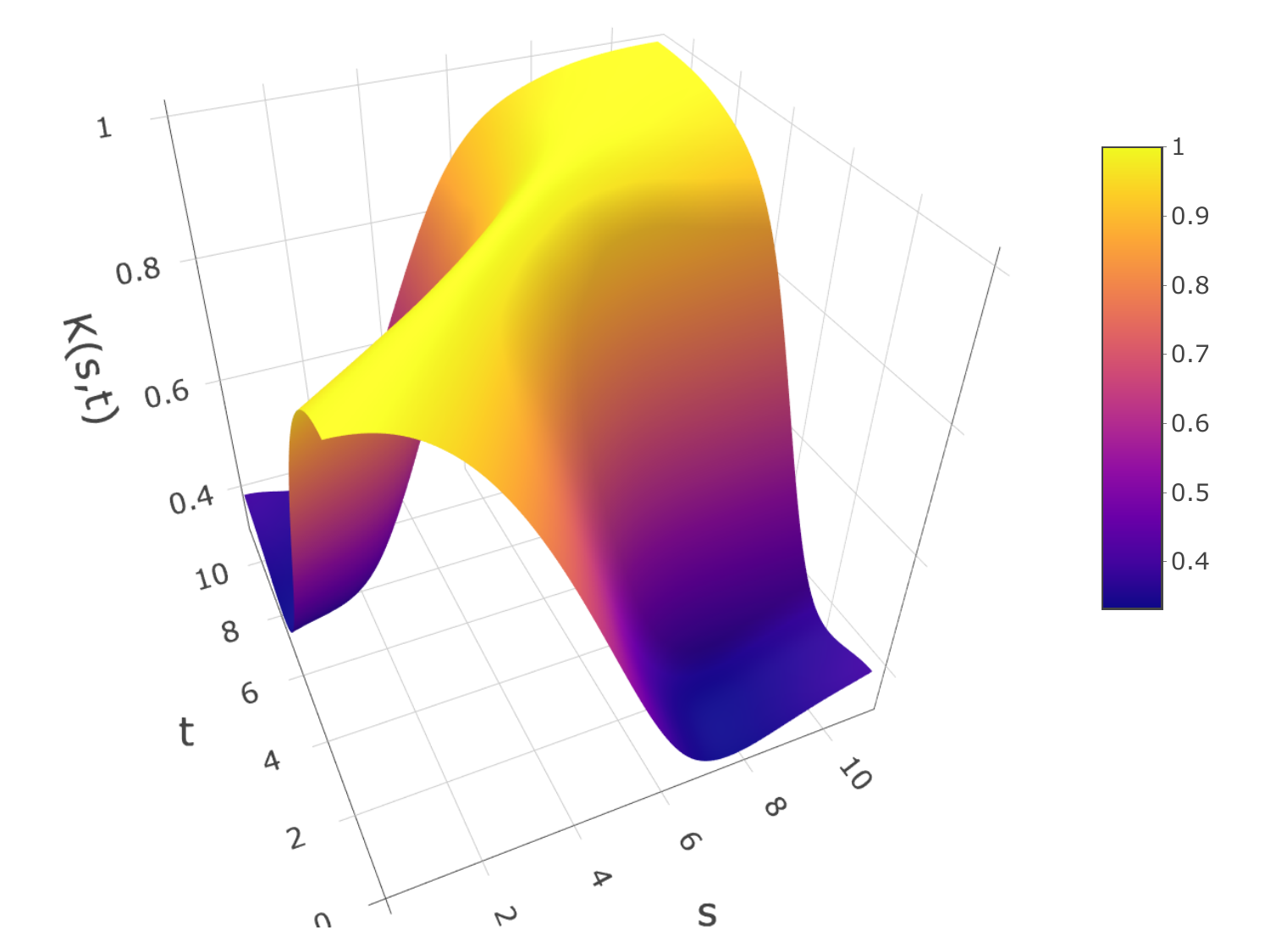}
	\end{subfigure}
	
	\caption{estimated covariance function. Left: countor plot; right: 3D surface plot. }
	
	\label{mad:kernel}

\end{figure}

\begin{figure}[tb!]
	
	\begin{minipage}{\dimexpr\linewidth-0cm\relax}%
		\begin{minipage}[b]{1\linewidth}
			\begin{minipage}[b]{0.25\linewidth}
				\centering
				\footnotesize{PC 1 (73.07\%)}
			\end{minipage}
			\begin{minipage}[b]{0.25\linewidth}
				\centering
				\footnotesize{PC 2 (21.95\%)}
			\end{minipage}
			\begin{minipage}[b]{0.25\linewidth}
				\centering
				\footnotesize{	PC 3 (3.79\%)}
			\end{minipage}
			\begin{minipage}[b]{0.25\linewidth}
				\centering
				\footnotesize{	PC 4 (0.94\%)}
			\end{minipage}
		\end{minipage}
		
		\begin{minipage}[b]{0.25\linewidth}
			\centering
			\includegraphics[width=1\linewidth]{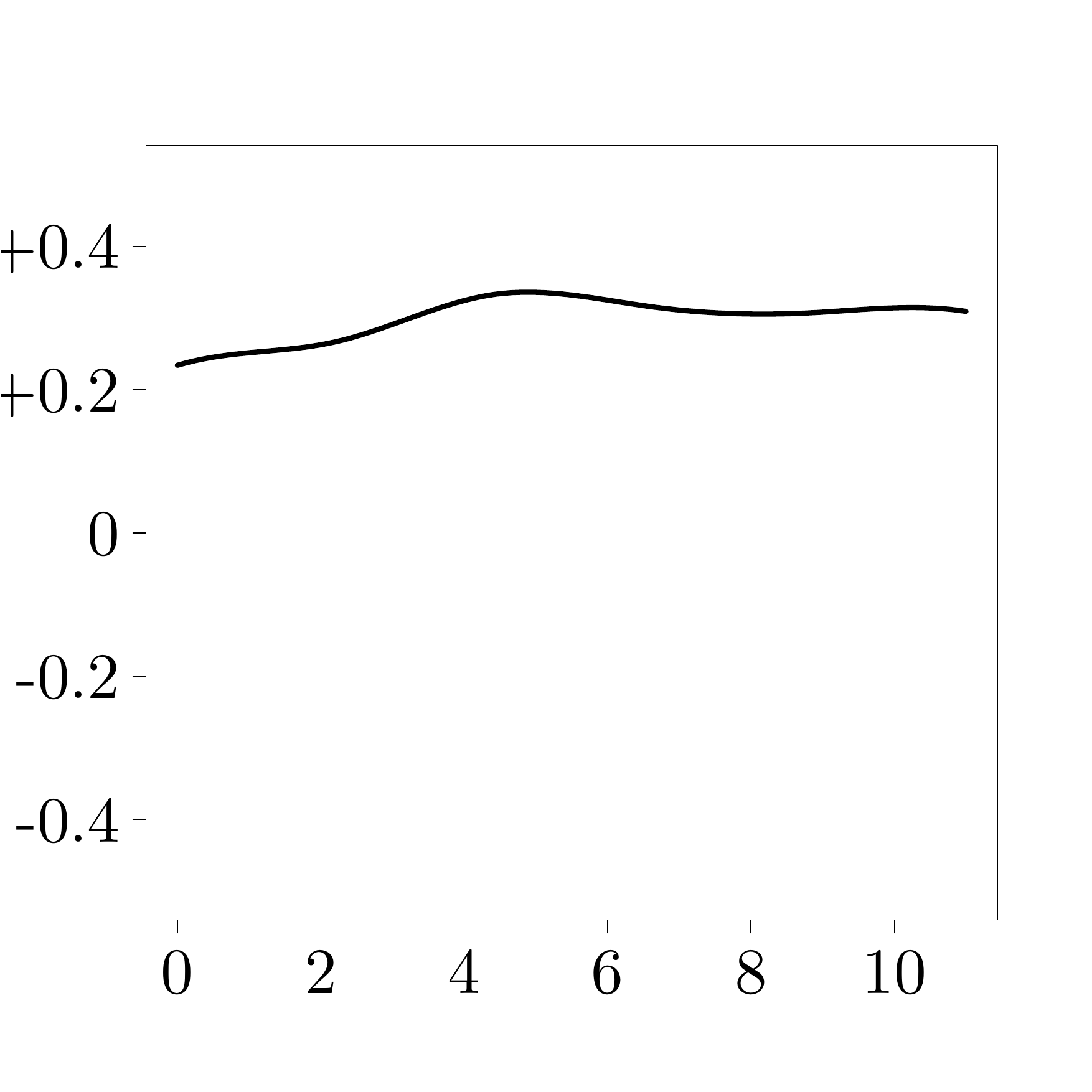} 
		\end{minipage}
		\begin{minipage}[b]{0.25\linewidth}
			\centering
			\includegraphics[width=1\linewidth]{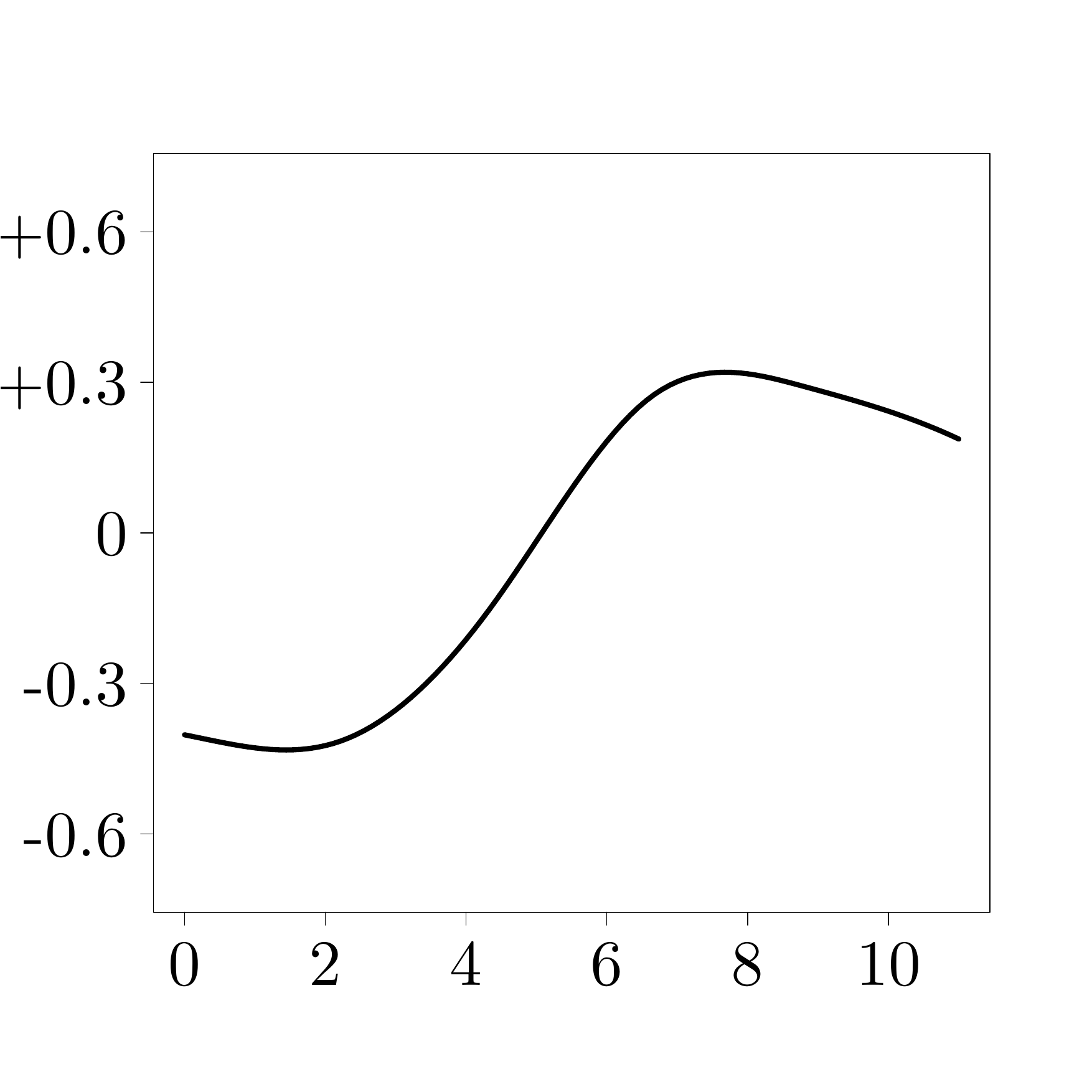} 
		\end{minipage}
		\begin{minipage}[b]{0.25\linewidth}
			\centering
			\includegraphics[width=1\linewidth]{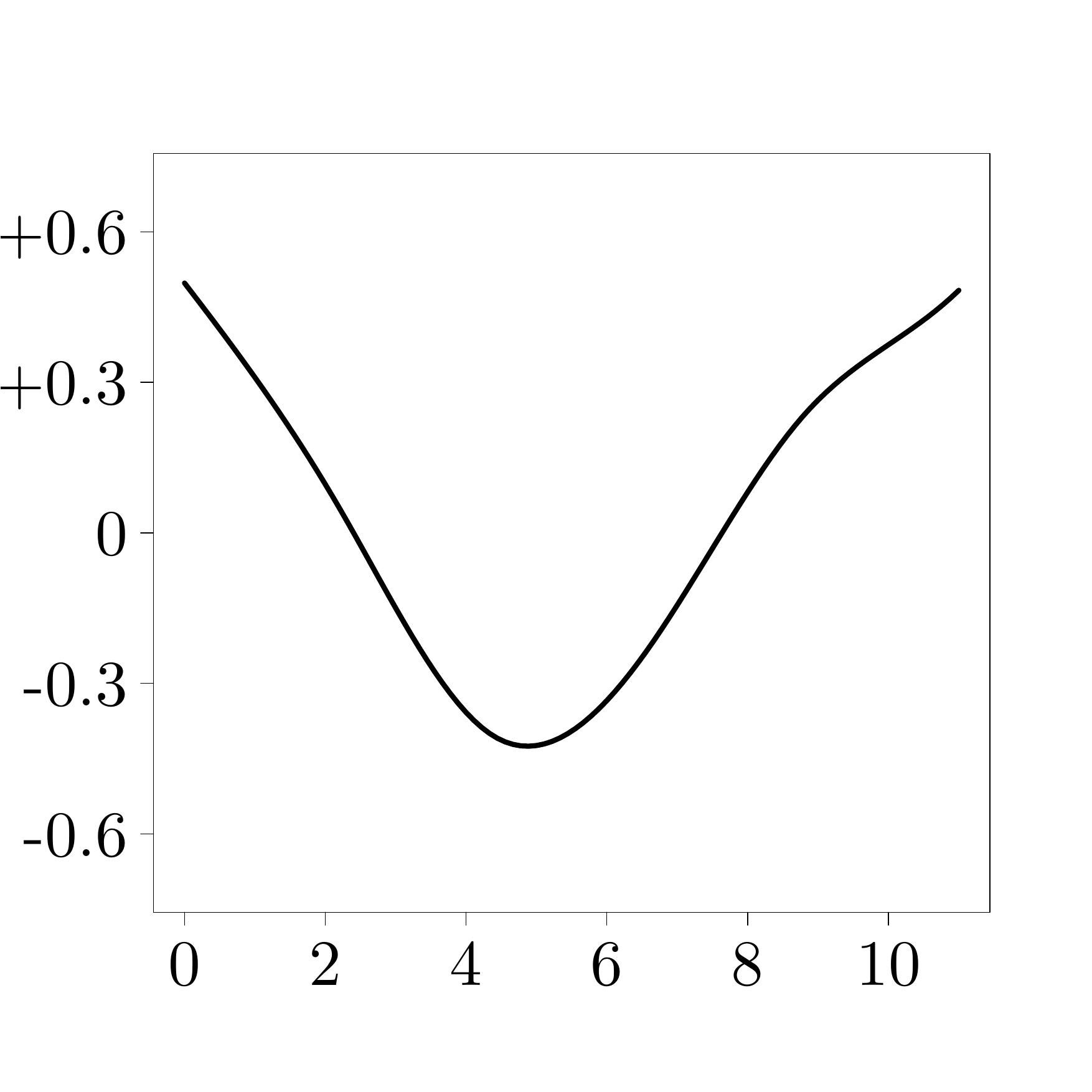} 
		\end{minipage}
		\begin{minipage}[b]{0.25\linewidth}
			\centering
			\includegraphics[width=1\linewidth]{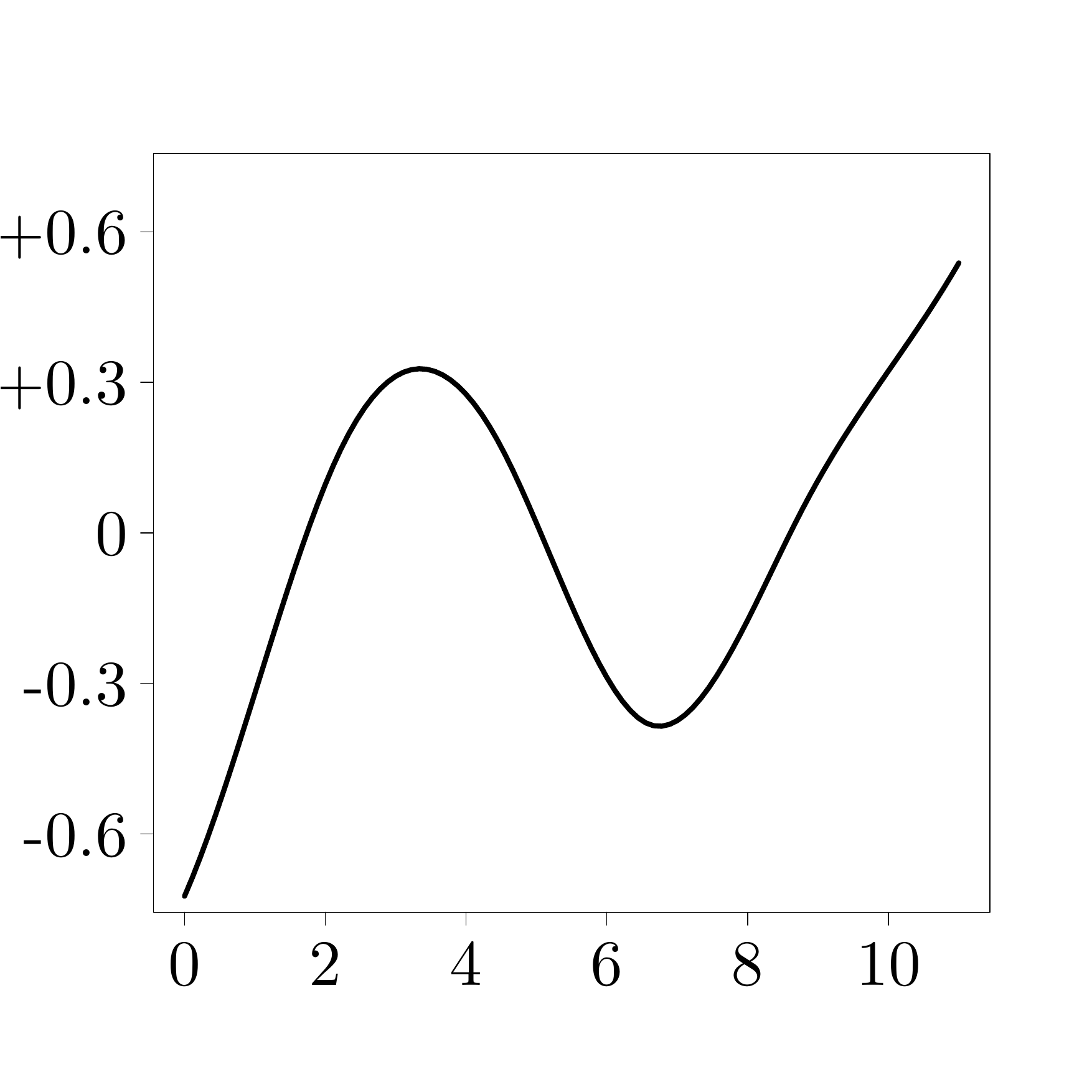} 
		\end{minipage}
		
		\begin{minipage}[b]{0.25\linewidth}
			\centering
			\includegraphics[width=1\linewidth]{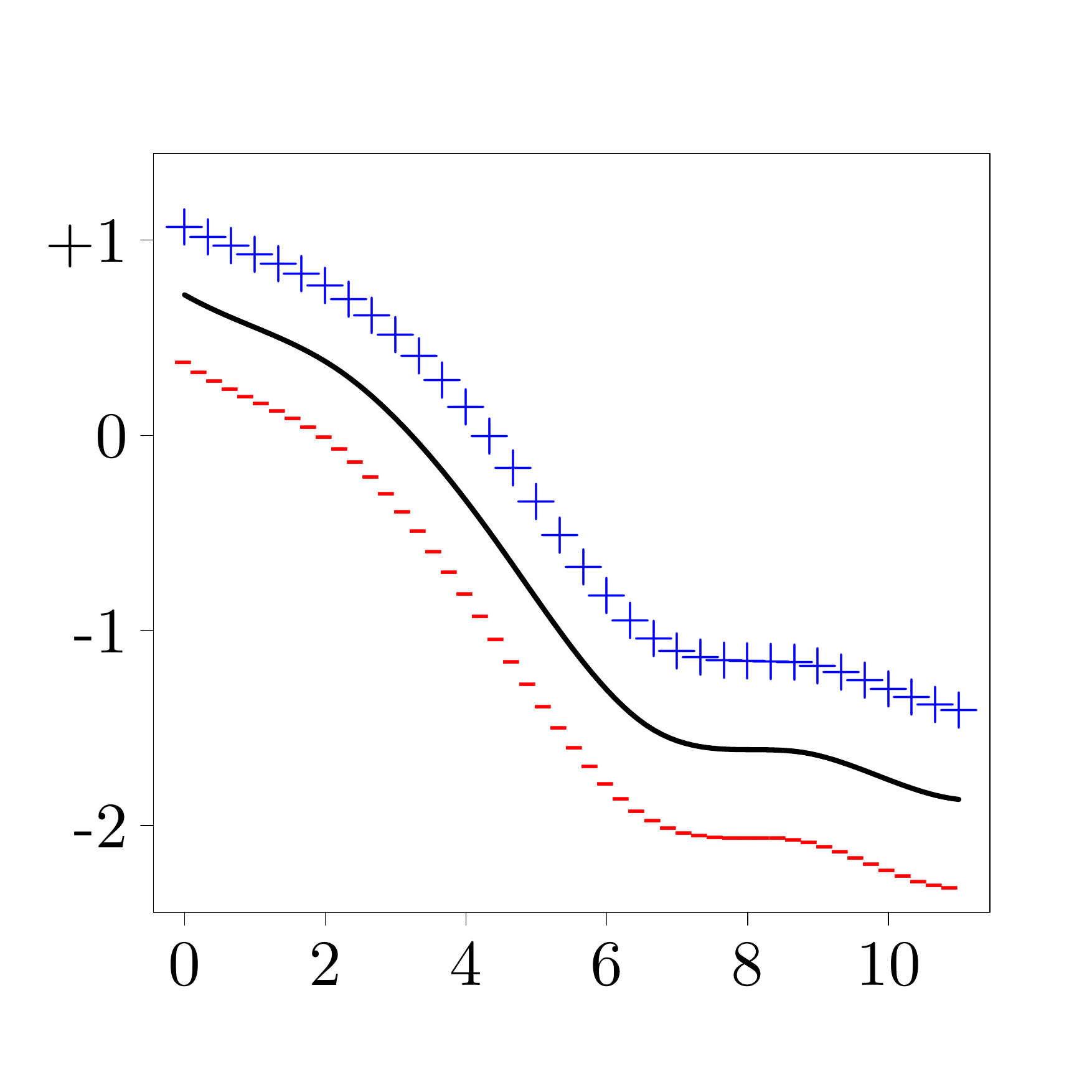} 
		\end{minipage}
		\begin{minipage}[b]{0.25\linewidth}
			\centering
			\includegraphics[width=1\linewidth]{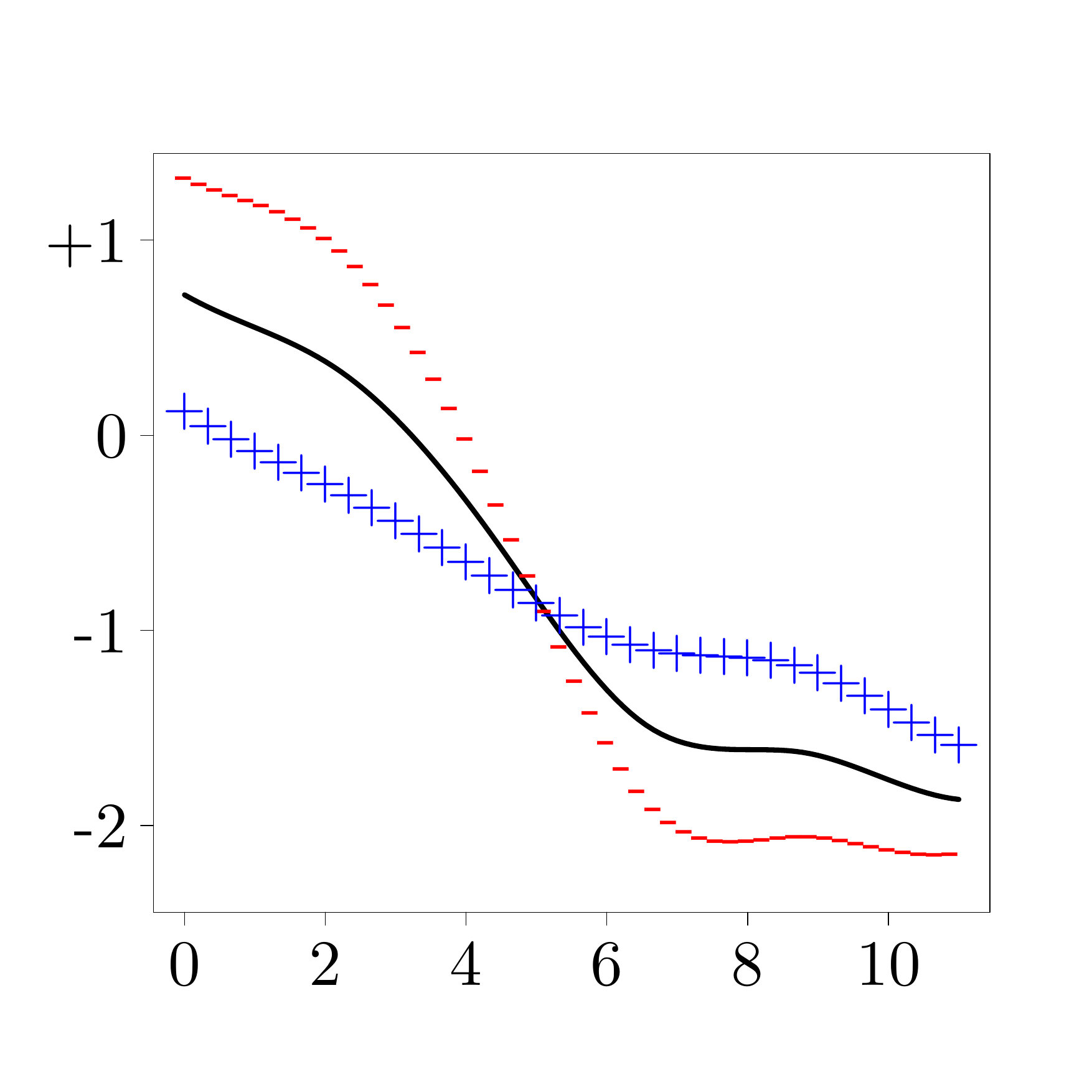} 
		\end{minipage}
		\begin{minipage}[b]{0.25\linewidth}
			\centering
			\includegraphics[width=1\linewidth]{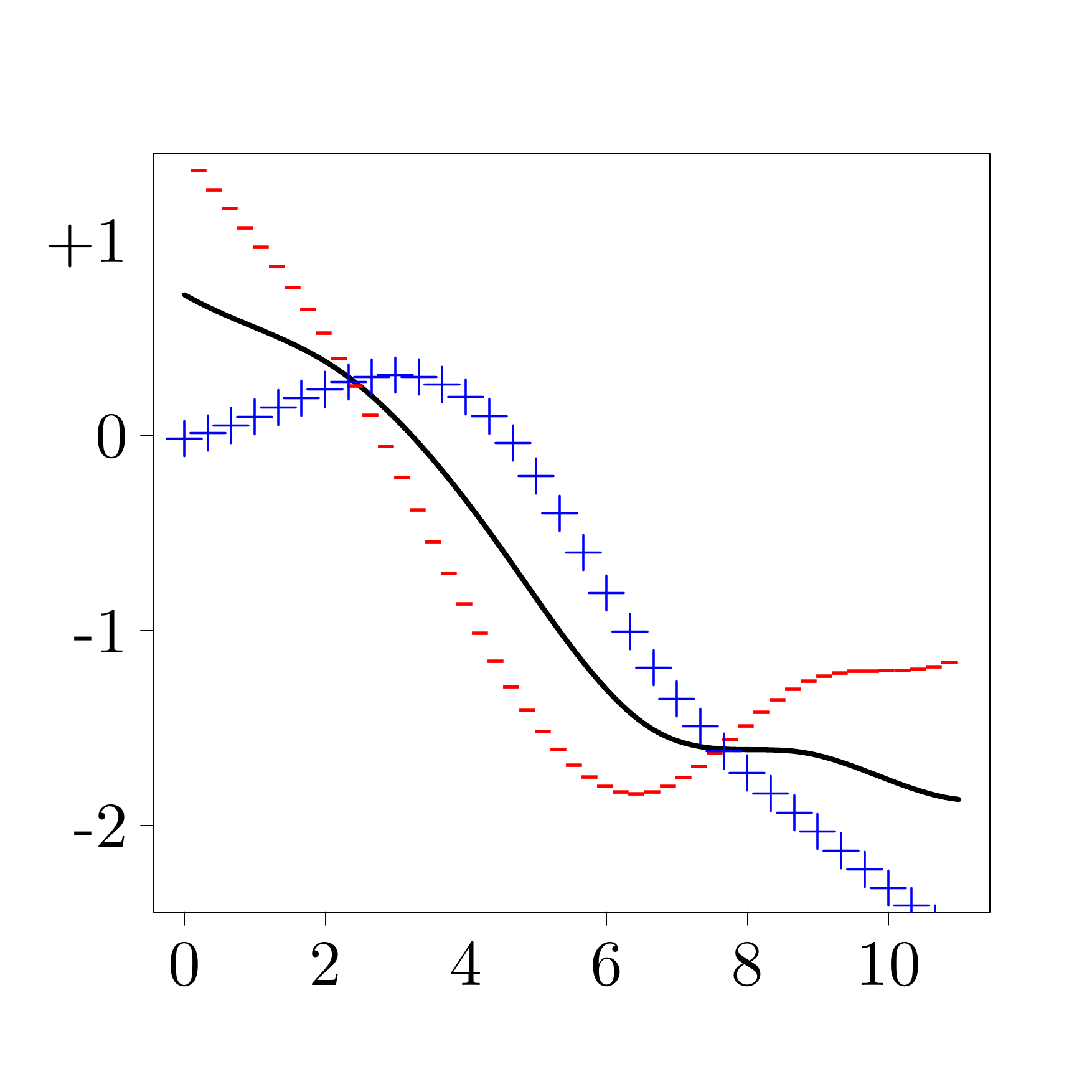} 
		\end{minipage}
		\begin{minipage}[b]{0.25\linewidth}
			\centering
			\includegraphics[width=1\linewidth]{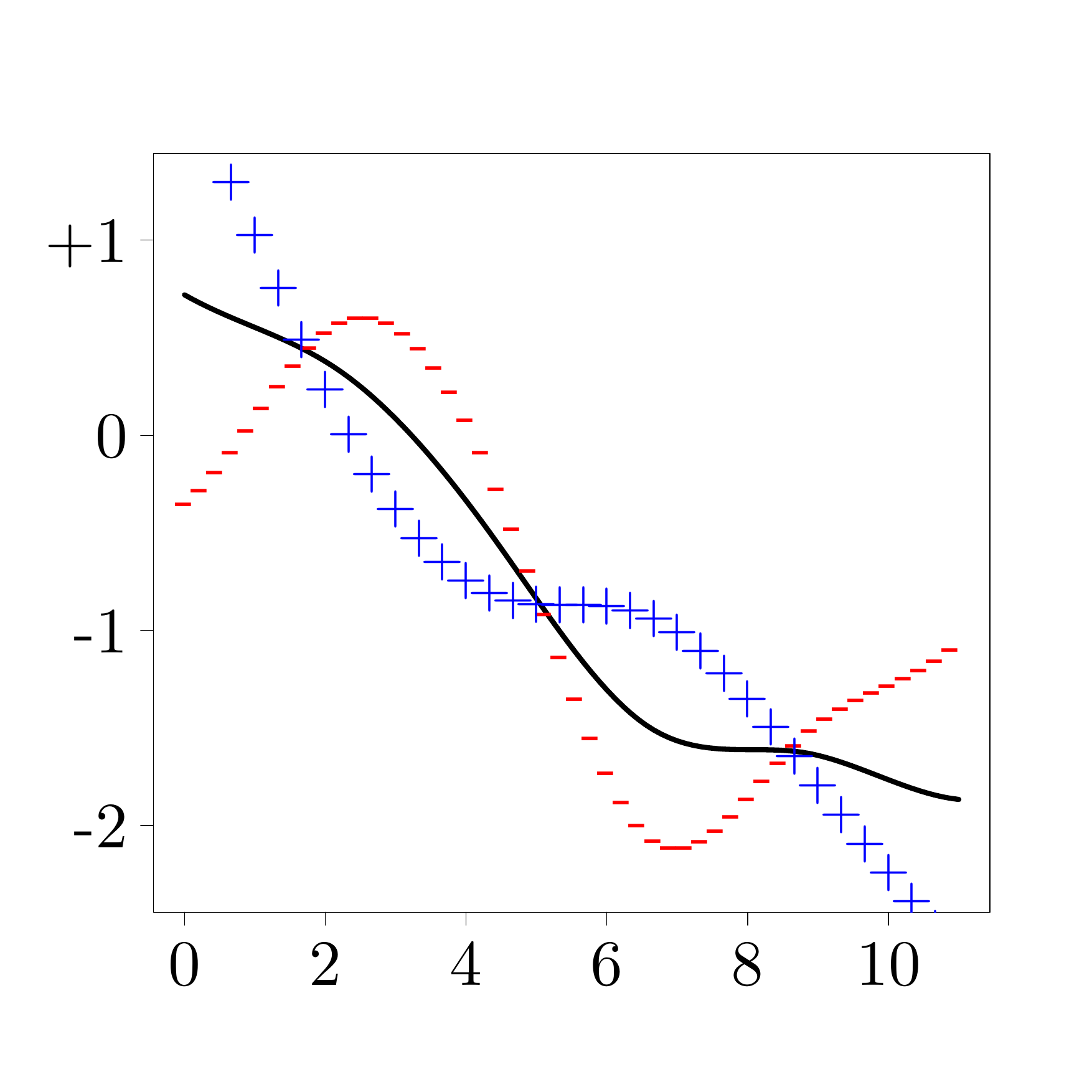} 
		\end{minipage}
		
		\vspace{-1em}
		\begin{minipage}[b]{1\linewidth}
			\centering
			\footnotesize{Month}
		\end{minipage}

	\end{minipage}%
	\caption{in the upper panels, estimated first four principal components of `thought disorder'; in the lower panels, mean function of TDI in younger females $\pm$ an appropriate multiple of the corresponding principal component. }
	
	\label{mad:eigen}
\end{figure}

The smooth standardized covariance function is illustrated in Figure \ref{mad:kernel}, which carries valuable information on the illness trajectories. The first four smooth principal components of the covariance function are shown in Figure \ref{mad:eigen}. The first two principal components account for 95.02\% of the variation in TDI. These principal components reveal meaningful additional information for the illness trajectories. PC1 indicates that 73.07\% of the variability of the trajectories is related to the variability of the TDI at baseline. 
%
%
PC2 indicates that 21.95\% of the variability in the trajectories is related to the variability in the rate of TDI reduction. According to PC1 and PC2, one could consider a mixed-effect regression model with both random intercept and random slope instead of FoSR, to account for this pattern of variation. 

\begin{figure}[tb!]
	
	\begin{minipage}{\dimexpr\linewidth-0cm\relax}%
		\begin{minipage}[b]{1\linewidth}

			\begin{minipage}[b]{0.25\linewidth}
				\centering
				\includegraphics[width=1\linewidth]{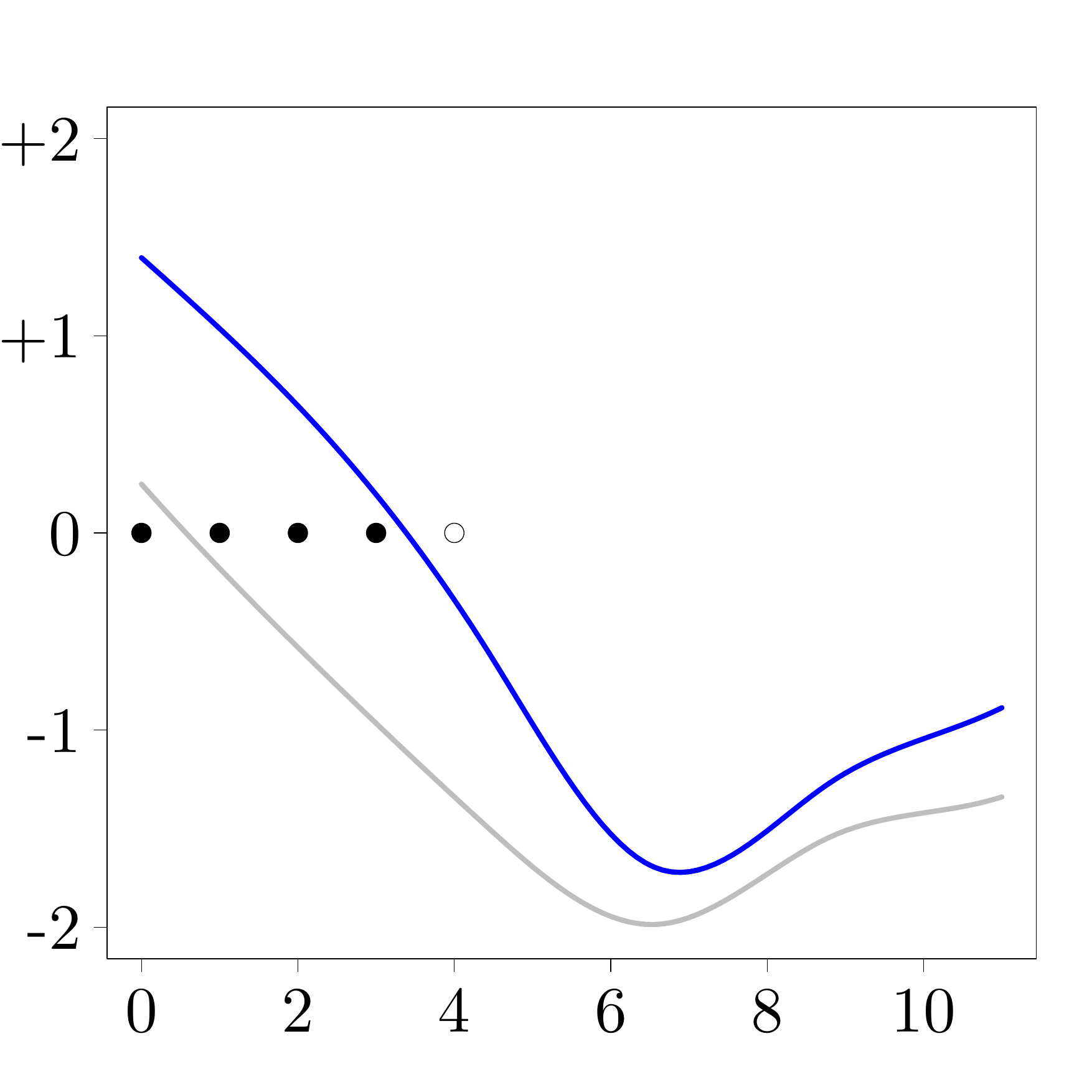} 
			\end{minipage}
			\begin{minipage}[b]{0.25\linewidth}
				\centering
				\includegraphics[width=1\linewidth]{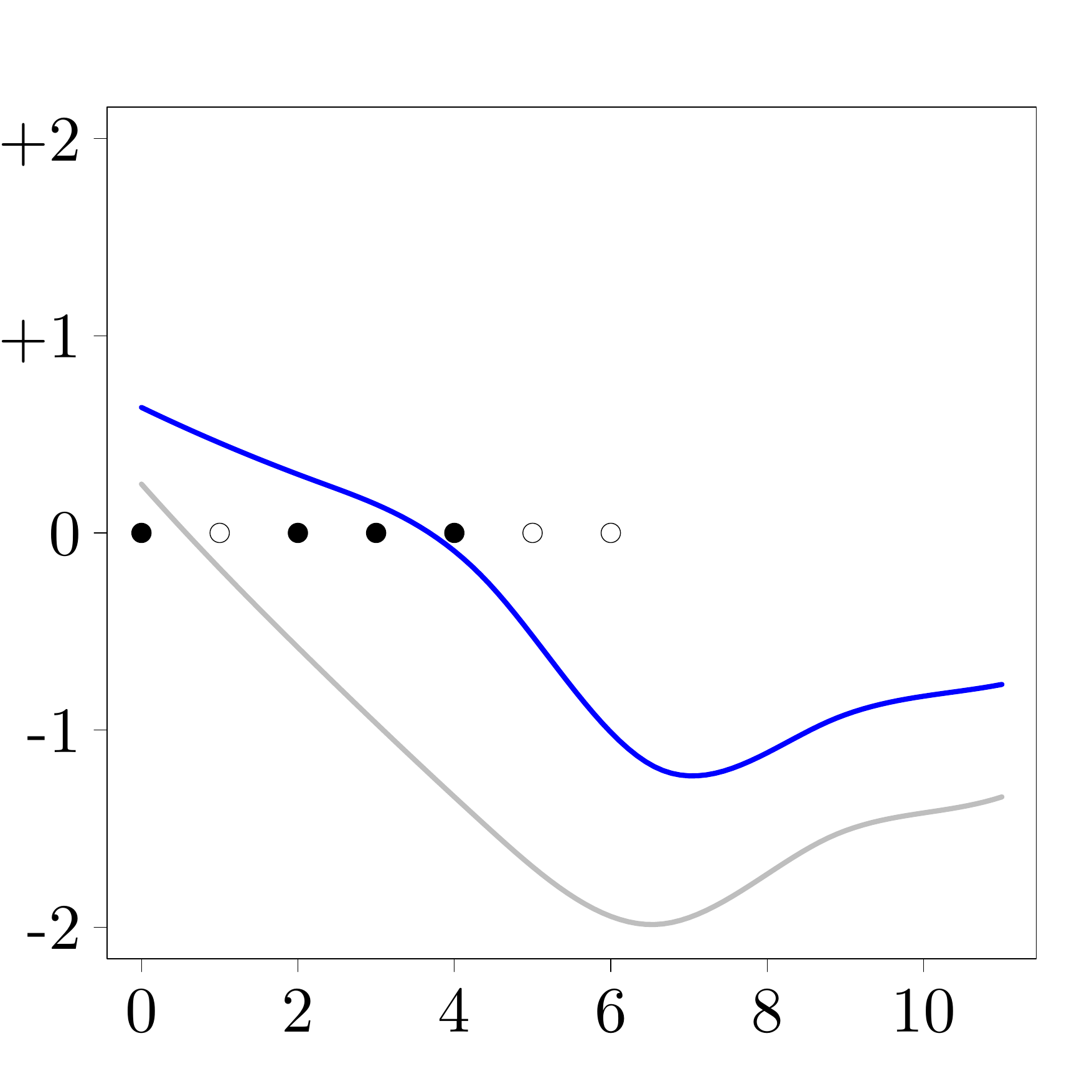} 
			\end{minipage}
			\begin{minipage}[b]{0.25\linewidth}
				\centering
				\includegraphics[width=1\linewidth]{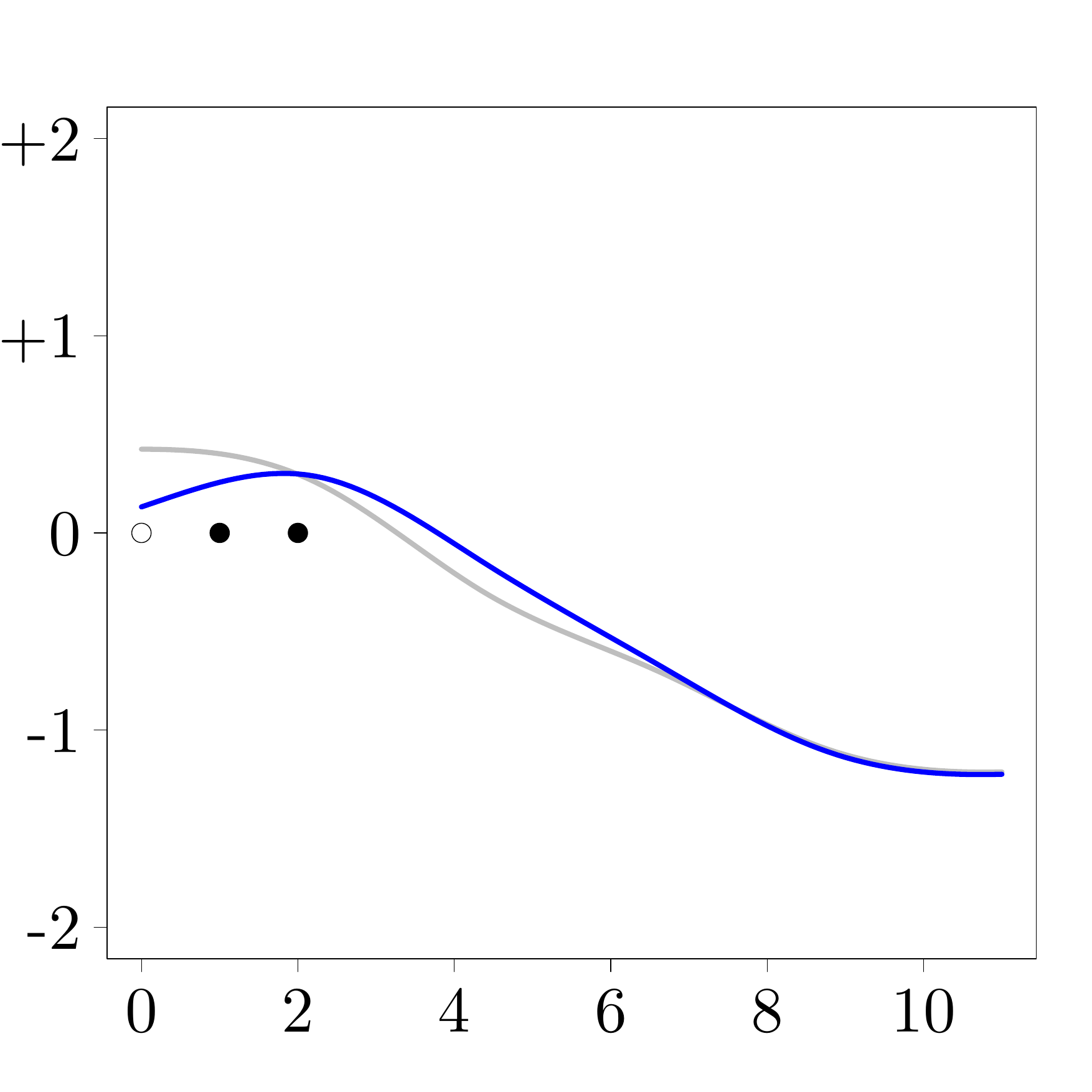} 
			\end{minipage}
			\begin{minipage}[b]{0.25\linewidth}
				\centering
				\includegraphics[width=1\linewidth]{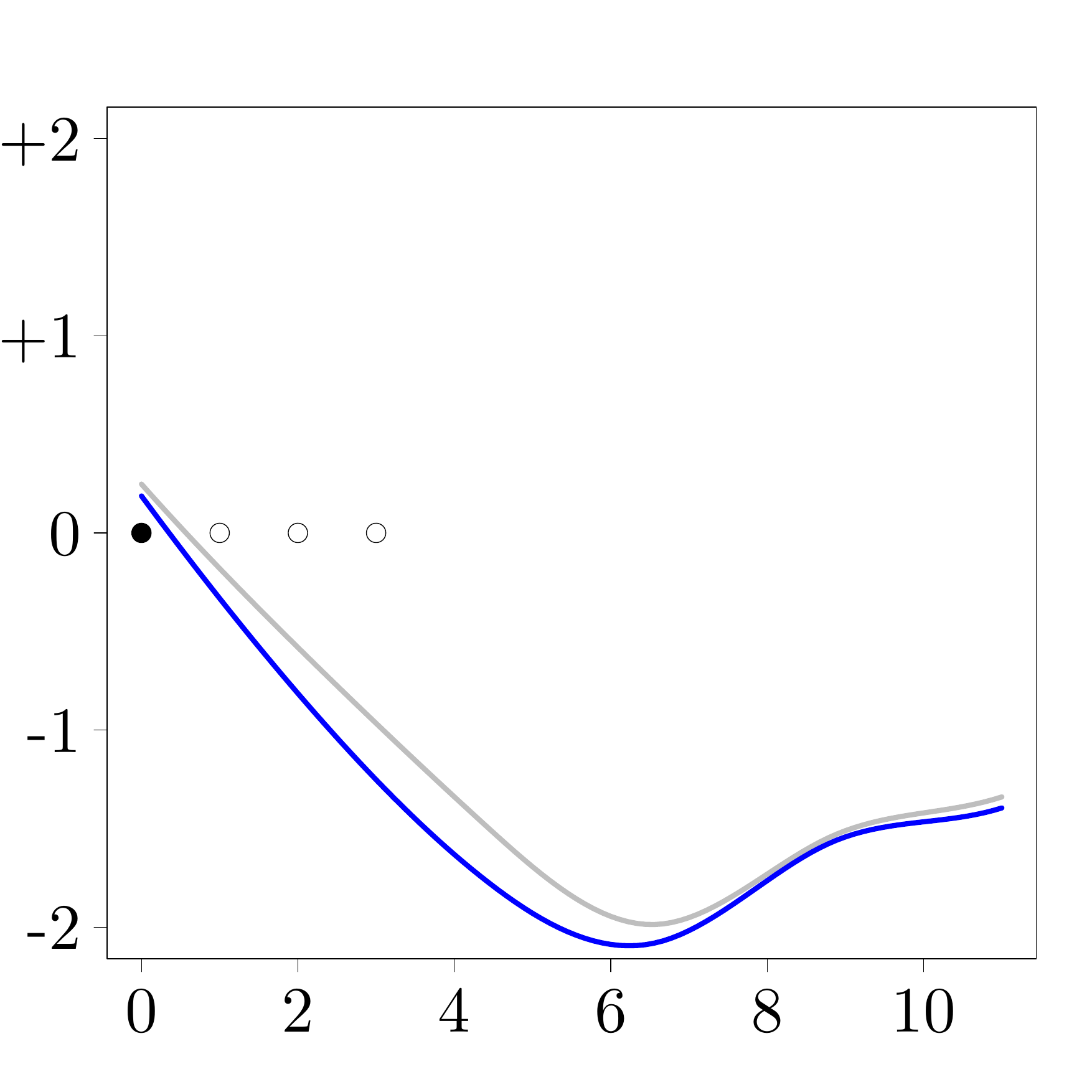} 
			\end{minipage}
			
			\vspace{-1em}
			\begin{minipage}[b]{1\linewidth}
				\centering
				\footnotesize{Month}
			\end{minipage}
		\end{minipage}
	\end{minipage}
	
	\caption{prediction of the subject-specific latent functions for four not fully observed binary sequences. 
		Dots in the plot indicate the sequence of observations of the `thought disorder' symptom for the subject: empty circles for absence, filled circles for presence. Solid lines show the predicted latent functions, given the covariates only (grey) or given the covariates and the sequence of observed binary response (blue).}
	\label{mad:prediction}
\end{figure}

In order to compare the two different prediction scenarios described in Sec. \ref{sec.4}, we consider four random samples with a few observed data. Predictions of the corresponding subject-specific trajectories given only the covariates, i.e., $\Bbb{E}(\text{TDI}\mid \mathbf{X}=\mathbf{x})$ , and given the covariates and the observed sequence $\Bbb{E}(\text{TDI}\mid \mathbf{X}=\mathbf{x},\mathbf{Y}=\mathbf{y}),$ are illustrated  in Figure \ref{mad:prediction}.

\section{Conclusions and Discussion}\label{sec.6}

We proposed a restricted maximum likelihood analysis framework for the FoSR model for the case in which a dichotomized version of the response curve is observed. A class of identifiable parameters was introduced, and a novel algorithm, namely the AMCEM algorithm, was proposed to provide smooth estimations of functional parameters. This approach does not rely on selecting hyperparameters, which is instead typical in penalized regression problems. In a quite extensive simulation study, we considered four different sampling designs including regular, truncated regular, regular with missing at random, and completely irregular designs for the observed timings of the observations. This simulation study demonstrated that the AMCEM algorithm provides satisfying results in all the aforementioned designs. We ran the simulation study for different combinations of sample sizes, number of points sampled per curve, magnitudes of the measurement error, and complexity of the covariance function. An \code{R} package named \code{dfrr} is available on CRAN for implementing our proposed method. In comparison to the \code{pffr}, the \code{R} package \code{dfrr} provides more accurate estimates; moreover, it also allows for the estimation of measurement error and covariance function, all giving useful information about the underlying process. 

Our proposed method was applied to the Madras longitudinal schizophrenia data to examine the effects of age and gender on the presence or absence of the `thought disorder' symptom during the first year of hospitalization. Our analysis concludes that the major variation of the binary sequences can be described by a FoSR model.
Comparing the estimated mean functions revealed that the younger females (older males) are more (less) affected by hospitalization with respect to the other groups. Another outstanding result was that the younger females experience a steady decrease in the TDI, while in younger males a quite different pattern is observed.
By inspection of the estimated principal components one can understand the main pattern of functional variation in the data: 73.07\% of the variation of the `thought disorder' intensity is due to differences in the baseline, while 21.95\% of the variation is due to differences in the rate of decrease of the `thought disorder' intensity during hospitalization.  

The current work can be extended in two directions. The covariates on the right-hand side of the model can be extended to functional covariates, and the left-hand side of the model can either be any discretized version of the functional response or censored from below or above. Thus, ordinal and Tobit function-on-function regression models are the natural extensions of this work.


\section*{Appendix}

\begin{proof}{\emph{Proposition \ref{t1}.}}
	
	To prove the proposition, it is sufficient to illustrate $\boldsymbol{\alpha}$ and $R$ are invariant under the transformation $\Ztipa_i^{'}=C\Ztipa_i$, where $C:L^2[0,1]\to L^2[0,1]~~g(\cdot)\mapsto f(\cdot)g(\cdot)$ for some postitive function $f(\cdot)$. 
	Let $\Ztipa$ be the Gaussian process given in (\ref{eq.2.2}) with location parameter  $\boldsymbol{\beta}$ and covariance operator $T$  with the kernel function $K$, i.e., $\left(Tg\right)(t)=\int K(s,t)g(s)ds$, and let $L_K:L^2[0,1]\to L^2[0,1]~~g(\cdot)\mapsto K(\cdot,\cdot)^{-\frac{1}{2}}g(\cdot)$ be the standardizer operator of $\Ztipa$ then $\boldsymbol{\alpha}=L_K\boldsymbol{\beta}$ and $R=L_KTL_K$.\newline
	If $\Ztipa^{'}=C\Ztipa$, then location parameter and covariance function of $\Ztipa^{'}$ are $C\boldsymbol{\beta}$ and $CTC$, respectively. From equation (\ref{p1.2}), the kernel function of $CTC$ equals to
	\[K^{'}(s,t)=f(s)K(s,t)f(t).\]
	Thus, using the standardizer operator $L_{K^{'}}:L^2[0,1]\to L^2[0,1]~~g(\cdot)\mapsto K^{'}(\cdot,\cdot)^{-\frac{1}{2}}g(\cdot)$, we have the following standardized parameters
	\begin{align*}
	\left(L_{K^{'}}C\boldsymbol{\beta}\right)(t)&=\left[f(t)K(t,t)f(t)\right]^{-\frac{1}{2}}f(t)\boldsymbol{\beta}(t)\\
	&=K(t,t)^{-\frac{1}{2}}\boldsymbol{\beta}(t)=\left(L_{K}\boldsymbol{\beta}\right)(t),
	\end{align*}
	and for any $g\in L^{2}[0,1]$
	\begin{align*}
	\left(L_{K^{'}}CTCL_{K^{'}}g\right)(t)&=K^{'}(t,t)^{-\frac{1}{2}}f(t)\int {K(s,t)f(s)K^{'}(s,s)^{-\frac{1}{2}}g(s)ds}\\
	&=\int{f(t)^{-1}f(t)K(t,t)^{-\frac{1}{2}}K(s,t)f(s)f(s)^{-1}K(s,s)^{-\frac{1}{2}}g(s)ds}\\
	&=\int {K(t,t)^{-\frac{1}{2}}K(s,t)K(s,s)^{-\frac{1}{2}}g(s)ds}=\left(Rg\right)(t).
	\end{align*}
\end{proof}

\begin{proof}{\emph{Theorem \ref{p2}.}}
	
	For $g\in L^{2}[0,1]$, we can write 
	\begin{align*}
	(Tg)(t)&=\int_{0}^{1}{K(s,t)g(s)ds}\\
	&=K(t,t)^{\frac{1}{2}}\int_{0}^{1}{K(t,t)^{-\frac{1}{2}}K(s,t)K(s,s)^{-\frac{1}{2}}K(s,s)^{\frac{1}{2}}g(s)ds}\\
	&=K(t,t)^{\frac{1}{2}}\int_{0}^{1}{K^{*}(s,t)K(s,s)^{\frac{1}{2}}g(s)ds}=(L^{'}RL^{'}g)(t).
	\end{align*}
	Using Cauchy-Schwarz inequality, we have
	\begin{align}\label{p2.2}\nonumber
	\parallel R\parallel &=\sup_{g\in L^2[0,1],\parallel g\parallel=1}\parallel Rg\parallel\\\nonumber
	&=\sup_{g\in L^2[0,1],\parallel g\parallel=1}\Big\{\int_{0}^{1}\Big[\int_{0}^{1}K^{*}(s,t)g(s)ds\Big]^{2}dt\Big\}^{\frac{1}{2}}\\\nonumber
	&\leq \sup_{g\in L^2[0,1],\parallel g\parallel=1}\Big\{\int_{0}^{1}\Big[\int_{0}^{1}K^{*}(s,t)^{2}ds\Big]^{2}\parallel g\parallel ^{2}dt\Big\}^{\frac{1}{2}}\\
	&=\Big\{\int_{0}^{1}\int_{0}^{1}K^{*}(s,t)^{2}dsdt\Big\}^{\frac{1}{2}}.
	\end{align}
	According to \citet[Lemma 4.6.6]{Hsing2015}, we have
	\begin{equation}\label{p2.3}
	\sqrt{K(s,s)K(t,t)}\geq \sum_{j\geq 1}\mid \nu_{j}\psi_{j}(s)\psi_{j}(t)\mid .
	\end{equation}
	On the other side, by (\ref{p2.3}), we get
	\begin{align}\label{p2.4}\nonumber
	\int_{0}^{1}\int_{0}^{1}K^{*}(s,t)^{2}dsdt&=\int_{0}^{1}\int_{0}^{1}\frac{K(s,t)^{2}}{K(s,s)K(t,t)}dsdt\\\nonumber
	&\leq \int_{0}^{1}\int_{0}^{1}\frac{[\sum_{j\geq 1} \nu_{j}\psi_{j}(s)\psi_{j}(t)]^2}{[\sum_{j\geq 1}\mid \nu_{j}\psi_{j}(s)\psi_{j}(t)\mid]^2}dsdt\\
	&\leq \int_{0}^{1}\int_{0}^{1}\frac{[\sum_{j\geq 1} \mid \nu_{j}\psi_{j}(s)\psi_{j}(t)\mid]^2}{[\sum_{j\geq 1}\mid \nu_{j}\psi_{j}(s)\psi_{j}(t)\mid]^2}dsdt=1 .
	\end{align}
	Thus,  (\ref{p2.2}) and (\ref{p2.4}) give $\parallel R\parallel\leq 1$ and the proof is finished.
\end{proof}

\begin{proof}{\emph{Proposition \ref{p3}.}}
	
	Let $X_1$ and $X_2$
	be independent variables. Suppose $f$ and $g$ are two functions in $H$ and $\mu_
	{X_1 ,X_2}$ is a borel probability measure on the Hilbert space $H^2=H\times H$, then
	\begin{align*}
	\left\langle C_{12}f,g\right\rangle &=\int_{H^{2}}{\left\langle x_1 -m_1 ,f\right\rangle\left\langle x_2-m_2,g\right\rangle \mu_{X_1,X_2}(dx_1,dx_2)}\\
	&=\int_{H_1}\int_{H_2}{\left\langle x_1 -m_1 ,f\right\rangle\left\langle x_2-m_2,g\right\rangle \mu_{X_1}(dx_1)\mu_{X_2}(dx_2)}\\
	&=\int_{H_1}{\left\langle x_1 -m_1 ,f\right\rangle \mu_{X_1}(dx_1)}\int_{H_2}{\left\langle x_2 -m_2 ,g\right\rangle \mu_{X_2}(dx_2)}.
	\end{align*}
	Therefore,  $ \left\langle C_{12}f,g\right\rangle =0$ which implies $ C_{12}f=0$ for all $f\in H$. With regard to $ C_{12}f=0$, we have   $C_{12}=0$. Similarly, we can prove $C_{21}$ is equal to zero. Conversly, we assume $C_{12}=0$. Thus, for all $f,g\in H$, we have $\left\langle C_{12}f,g\right\rangle=0$ which shows
	\begin{align}\label{eq.3.11}
	\text{cov}(\left\langle X_1 -m_1 ,f\right\rangle ,\left\langle X_2 -m_2 ,g\right\rangle)=\Bbb{E}\left\langle X_1 -m_1 ,f\right\rangle \left\langle X_2 -m_2 ,g\right\rangle =0.
	\end{align}
	Noting that $Y_1=\left\langle X_1 -m_1 ,f\right\rangle$ and $Y_2=\left\langle X_2 -m_2 ,g\right\rangle$ are jointly Gaussian, from (\ref{eq.3.11}), $Y_1$ and $Y_2$ are independent. We can write
	\begin{align*}
	\Bbb{E}\left\langle X_1 -m_1 ,f\right\rangle \left\langle X_2 -m_2 ,g\right\rangle &=\int_{H^{2}}{\left\langle x_1 -m_1 ,f\right\rangle\left\langle x_2-m_2,g\right\rangle \mu_{X_1,X_2}(dx_1,dx_2)}\\\nonumber
	&=\int_{R^2}{y_{1}y_{2} \mu_{Y_1,Y_{2}}(dy_1 ,dy_2)}\\\nonumber
	&=\int_{R}{y_{1} \mu_{Y_1}(dy_1)}\int_{R}{y_{2} \mu_{Y_2}(dy_2)}\\
	&=\int_{H_1}\hspace*{-0.4cm}{\left\langle x_1 -m_1 ,f\right\rangle \mu_{X_1}(dx_1)}\int_{H_2}\hspace*{-0.4cm}{\left\langle x_2 -m_2 ,g\right\rangle \mu_{X_2}(dx_2)}.
	\end{align*}
	As a result, $X_1$ and $X_2$ are independent.
\end{proof}

\begin{proof}{\emph{Proposition \ref{p4}.}}
	
	We know that $Y_i$ is Gaussian with the mean function  zero. Thus, if $f,g\in H$, we can write
	\begin{align*}
	\left\langle C_{ij}f,g\right\rangle =\Bbb{E}\left\langle Y_i ,f\right\rangle\left\langle Y_j ,g\right\rangle&=\Bbb{E}\left\langle \mathbf{q}_i^\top \mathbf{X} ,f\right\rangle\left\langle \mathbf{q}_j^\top \mathbf{X} ,g\right\rangle\\
	&=\sum_{k=1}^{n}\sum_{l=1}^{n}q_{ik}q_{jl}\Bbb{E}\left\langle X_{k},f\right\rangle\left\langle X_{l},g\right\rangle\\
	&=\sum_{k=1}^{n}q_{ik}q_{jk}\Bbb{E}\left\langle X_{k},f\right\rangle\left\langle X_{k},g\right\rangle\\
	&=\left\langle Cf,g\right\rangle \mathbf{q}^\top_{i}\mathbf{q}_j=\left\langle Cf,g\right\rangle \delta_{ij},
	\end{align*}
	which shows $C_{ii}=C$ and $C_{ij}$ are equal to zero,  for  all $i\neq j$, where $C_{ij}$ is the cross-covariance operator of $Y_i$ and $Y_j$. As a result, based on Proposition \ref{p3}, $Y_i$ for $i=1,2,\ldots,m$ are independent.
\end{proof}


\bibliography{article}

\end{document}